\documentclass[11pt]{article}
\usepackage{graphicx,ifthen,color,algorithm,palatino}
\usepackage{amsmath,amsthm,amssymb,amsfonts,verbatim}
\usepackage{mathrsfs,setspace}
\newboolean{ColourVersion}
\newboolean{UnBlinded}
\newboolean{DoubleSpaced}
\setboolean{ColourVersion}{true}
\setboolean{UnBlinded}{true}
\setboolean{DoubleSpaced}{false}
\def\bX{\boldsymbol{X}}
\def\bLambda{\boldsymbol{\Lambda}}
\def\Csc{{\mathcal C}}
\def\smhalf{{\textstyle{\frac{1}{2}}}}
\def\tr{\mbox{tr}}
\def\quarter{{\textstyle{1\over4}}}
\def\Qsc{{\mathcal Q}}
\def\argmindum{\mathop{\mbox{argmin}}}
\def\argmin#1{\argmindum_{#1}}
\def\infint{\int_{-\infty}^{\infty}}
\def\ba{\boldsymbol{a}}
\def\btau{\boldsymbol{\tau}}
\def\bdf{\boldsymbol{f}}
\def\by{\boldsymbol{y}}
\def\bbeta{\boldsymbol{\beta}}
\def\bI{\boldsymbol{I}}
\def\bzero{\boldsymbol{0}}
\def\bone{\boldsymbol{1}}
\def\bx{\boldsymbol{x}}
\def\btheta{\boldsymbol{\theta}}
\def\thickarrow{\longleftarrow}
\def\jump{\vskip3mm\noindent}
\def\vecof{\mbox{vec}}
\def\bA{\boldsymbol{A}}
\def\bmu{\boldsymbol{\mu}}
\def\bSigma{\boldsymbol{\Sigma}}
\def\bTheta{\boldsymbol{\Theta}}
\def\balpha{\boldsymbol{\alpha}}
\def\bo{\boldsymbol{o}}
\def\bomega{\boldsymbol{\omega}}
\def\bOmega{\boldsymbol{\Omega}}
\def\logit{\mbox{logit}}
\def\bu{\boldsymbol{u}}
\def\simind{\stackrel{{\tiny \mbox{ind.}}}{\sim}}
\def\bZ{\boldsymbol{Z}}
\def\blockdiagdum{\mathop{\mbox{blockdiag}}}
\def\blockdiag#1{\blockdiagdum_{#1}}
\def\relstack#1#2{\mathop{#1}\limits_{#2}}
\def\bC{\boldsymbol{C}}
\def\bc{\boldsymbol{c}}
\def\be{\boldsymbol{e}}
\def\bE{\boldsymbol{E}}
\def\utilde{{\widetilde u}}
\def\vech{\mbox{vech}}
\def\Bsc{{\mathcal B}}
\def\bib{\vskip12pt\par\noindent\hangindent=1 true cm\hangafter=1}
\def\Asc{{\mathcal A}}
\def\bb{\boldsymbol{b}}
\def\bv{\boldsymbol{v}}
\def\bz{\boldsymbol{z}}
\def\bL{\boldsymbol{L}}
\def\bm{\boldsymbol{m}}
\def\bB{\boldsymbol{B}}
\def\bD{\boldsymbol{D}}
\def\Psc{{\mathcal P}}
\def\bT{\boldsymbol{T}}
\def\bdeta{\boldsymbol{\eta}}
\def\myand{\&\ }
\def\thickarroweps{\stackrel{{\large{\varepsilon}}}{\thickarrow}}
\def\endproof{\rightline{\rule[.2ex]{1ex}{1.5ex}}}
\def\smallN{\mbox{\tiny$N$}}
\def\coeffmat{\boldsymbol{L}}
\def\coeffmatdagg{\boldsymbol{L}^{\dagger}}
\def\bvdagg{\bv^{\dagger}}
\def\LambdaTheta{\bLambda_{\mbox{\tiny $\bTheta$}}}
\def\dTheta{d^{\,\mbox{\tiny$\Theta$}}}
\def\kappaTheta{\kappa_{\mbox{\tiny $\bTheta$}}}
\def\bbetabu{
\left[
\begin{array}{c}
\bbeta\\
\bu
\end{array}
\right]
}
\def\bZgrp{\bZ_{\mbox{\tiny grp}}}
\def\bZspl{\bZ_{\mbox{\tiny spl}}}
\def\agrp{a_{\mbox{\tiny grp}}}
\def\aspl{a_{\mbox{\tiny spl}}}
\def\agrp{A_{\mbox{\tiny grp}}}
\def\Aspl{A_{\mbox{\tiny spl}}}
\def\bugrp{\bu_{\mbox{\tiny grp}}}
\def\sigmagrp{\sigma_{\mbox{\tiny grp}}}
\def\sigmaspl{\sigma_{\mbox{\tiny spl}}}
\def\ugrpi{u_{\mbox{\tiny{grp}},i}}
\def\utilgrpi{\utilde_{\mbox{\tiny{grp}},i}}
\def\utilsplk{\utilde_{\mbox{\tiny{spl}},k}}
\def\buspl{\bu_{\mbox{\tiny spl}}}
\def\usplk{u_{\mbox{\tiny{spl}},k}}
\def\betaspl{\beta_{\mbox{\tiny spl}}}
\def\KL{\mbox{KL}}
\def\proj{\mbox{proj}}
\def\projN{\proj_{\mbox{\tiny N}}}
\def\projICS{\proj_{\mbox{\tiny I$\chi^2$}}}
\def\projIW{\proj_{\mbox{\tiny IW}}}
\def\projMR{\proj_{\mbox{\tiny MR}}}
\def\AN{A_{\mbox{\tiny N}}}
\def\AICS{A_{\mbox{\tiny I$\chi^2$}}}
\def\AMR{A_{\mbox{\tiny MR}}}
\def\zerobtwo{\left[\begin{array}{c}   
0 \\
b_2
\end{array}
\right]}
\def\twozerozero{\left[\begin{array}{c}   
2 \\
0 \\
0
\end{array}
\right]}
\def\realnos{{\mathbb R}}
\def\biggerm{\mbox{\large $m$}}
\def\subscriptneighbours{\mbox{\scriptsize neighbours}}
\def\biggerf{\mbox{\large $f$}}
\def\dbu{d^{\bu}}

\def\dtheta{d^{\btheta}}
\def\dalpha{d^{\balpha}}
\def\biggerbdeta{\mbox{\Large $\bdeta$}}
\def\mSUBbalphaTObdelta{
\biggerm_{\mbox{\footnotesize$\balpha\to\delta(\balpha-\bA^T\btheta)$}}(\balpha)}
\def\etaSUBbalphaTObdelta{
\biggerbdeta_{\mbox{\footnotesize$\balpha\to\delta(\balpha-\bA^T\btheta)$}}}
\def\etaSUBalphaTOdelta{
\biggerbdeta_{\mbox{\footnotesize$\alpha\to\delta(\alpha-\ba^T\btheta)$}}}
\def\mSUBpnuTOnu{
\biggerm_{\mbox{\footnotesize$p(\nu)\to\nu$}}(\nu)}
\def\mSUBthetaTObdelta{
\biggerm_{\mbox{\footnotesize$\btheta\to\delta(\balpha-\bA^T\btheta)$}}(\btheta)}
\def\etaSUBthetaTObdelta{
\biggerbdeta_{\mbox{\footnotesize$\btheta\to\delta(\balpha-\bA^T\btheta)$}}}
\def\etaSUBthetaTOdelta{
\biggerbdeta_{\mbox{\footnotesize$\btheta\to\delta(\alpha-\ba^T\btheta)$}}}
\def\mSUBalphaTOpyalpha{
\biggerm_{\mbox{\footnotesize$\alpha\to p(y|\alpha)$}}(\alpha)}
\def\mSUBpyalphaTOalpha{
\biggerm_{\mbox{\footnotesize$p(y|\alpha)\to\alpha$}}(\alpha)}
\def\etaSUBpyalphaTOalpha{
\biggerbdeta_{\mbox{\footnotesize$p(y|\alpha)\to\alpha$}}}
\def\mSUBbdeltaTOtheta{
\biggerm_{\mbox{\footnotesize$\delta(\balpha-\bA^T\btheta)\to\btheta$}}(\btheta)}
\def\mSUBdeltaTOtheta{
\biggerm_{\mbox{\footnotesize$\delta(\alpha-\ba^T\btheta)\to\btheta$}}(\btheta)}

\def\mSUBthetaTOdelta{
\biggerm_{\mbox{\footnotesize$\btheta\to\delta(\alpha-\ba^T\btheta)$}}(\btheta)}
\def\mSUBdeltaTOalpha{
\biggerm_{\mbox{\footnotesize$\delta(\alpha-\ba^T\btheta)\to\alpha$}}(\alpha)}
\def\etaSUBdeltaTOalpha{
\biggerbdeta_{\mbox{\footnotesize$\delta(\alpha-\ba^T\btheta)\to\alpha$}}}
\def\mSUBbdeltaTObalpha{
\biggerbdeta_{\mbox{\footnotesize$\delta(\balpha-\bA^T\btheta)\to\balpha$}}(\balpha)}
\def\etaSUBbdeltaTObalpha{
\biggerbdeta_{\mbox{\footnotesize$\delta(\balpha-\bA^T\btheta)\to\balpha$}}}
\def\etaSUBbdeltaTOtheta{
\biggerbdeta_{\mbox{\footnotesize$\delta(\balpha-\bA^T\btheta)\to\btheta$}}}
\def\etaSUBdeltaTOtheta{
\biggerbdeta_{\mbox{\footnotesize$\delta(\alpha-\ba^T\btheta)\to\btheta$}}}
\def\omegaONE{\bomega}

\def\etaSUBpyalphaTOalpha{
\biggerbdeta_{\mbox{\footnotesize$p(y|\alpha)\to\alpha$}}}
\def\etaSUBalphaTOpyalpha{
\biggerbdeta_{\mbox{\footnotesize$\alpha\to p(y|\alpha)$}}}
\def\Hlogistic{H_{\tiny{\mbox{logistic}}}}
\def\HPoisson{H_{\tiny{\mbox{Poisson}}}}
\def\Hprobit{H_{\tiny{\mbox{probit}}}}
\def\mSUBalphaTOpyalphasigsq{
\biggerm_{\mbox{\footnotesize$\alpha\to p(y|\alpha,\sigma^2)$}}(\alpha)}

\def\etaSUBpyalphasigsqaTOalpha{
\biggerbdeta_{\mbox{\footnotesize$p(y|\alpha,\sigma^2,a)\to\alpha$}}}
\def\mSUBpyalphasigsqaTOalpha{
\biggerm_{\mbox{\footnotesize$p(y|\alpha,\sigma^2,a)\to\alpha$}}(\alpha)}

\def\mSUBpahalfnuTOa{
\biggerm_{\mbox{\footnotesize$p(a|\halfnu)\to a$}}(a)}

\def\mSUBhalfnuTOpahalfnu{
\biggerm_{\mbox{\footnotesize$\halfnu\to p(a|\halfnu)$}}(\halfnu)}

\def\mSUBpahalfnuTOhalfnu{
\biggerm_{\mbox{\footnotesize$p(a|\halfnu)\to\halfnu$}}(\halfnu)}

\def\etaSUBpahalfnuTOa{
\biggerbdeta_{\mbox{\footnotesize$p(a|\halfnu)\to a$}}}

\def\mSUBphalfnuTOhalfnu{
\biggerm_{\mbox{\footnotesize$p(\halfnu)\to\halfnu$}}(\halfnu)}

\def\etaSUBhalfnuTOpahalfnu{
\biggerbdeta_{\mbox{\footnotesize$\halfnu\to p(a|\halfnu)$}}}

\def\etaSUBpahalfnuTOhalfnu{
\biggerbdeta_{\mbox{\footnotesize$p(a|\halfnu)\to\halfnu$}}}

\def\etaSUBpyalphasigsqaTOsigsq{
\biggerbdeta_{\mbox{\footnotesize$p(y|\alpha,\sigma^2,a)\to\sigma^2$}}}
\def\mSUBpyalphasigsqaTOsigsq{
\biggerm_{\mbox{\footnotesize$p(y|\alpha,\sigma^2,a)\to\sigma^2$}}(\sigma^2)}

\def\etaSUBpyalphasigsqaTOa{
\biggerbdeta_{\mbox{\footnotesize$p(y|\alpha,\sigma^2,a)\to a$}}}
\def\mSUBpyalphasigsqaTOa{
\biggerm_{\mbox{\footnotesize$p(y|\alpha,\sigma^2,a)\to a$}}(a)}

\def\mSUBpyalphasigsqTOalpha{
\biggerm_{\mbox{\footnotesize$p(y|\alpha,\sigma^2)\to\alpha$}}(\alpha)}
\def\etaSUBpyalphasigsqTOalpha{
\biggerbdeta_{\mbox{\footnotesize$p(y|\alpha,\sigma^2)\to\alpha$}}}
\def\mSUBpyalphasigsqTOsigsq{
\biggerm_{\mbox{\footnotesize$p(by|\alpha,\sigma^2)\to\sigma^2$}}(\sigma^2)}
\def\etaSUBpyalphasigsqTOsigsq{
\biggerbdeta_{\mbox{\footnotesize$p(y|\alpha,\sigma^2)\to\sigma^2$}}}

\def\etaSUBalphaTOpyalphasigsq{
\biggerbdeta_{\mbox{\footnotesize$\alpha\to p(y|\alpha,\sigma^2)$}}}

\def\etaSUBpuSigmaTOSigma{
\biggerbdeta_{\mbox{\footnotesize$p(\bu|\bSigma)\to\bSigma$}}}

\def\mSUBpuSigmaTOSigma{
\biggerm_{\mbox{\footnotesize$p(\bu|\bSigma)\to\bSigma$}}(\bSigma)}

\def\etaSUBSigmaTOpuSigma{
\biggerbdeta_{\mbox{\footnotesize$\bSigma\to p(\bu|\bSigma)$}}}

\def\mSUBSigmaTOpuSigma{
\biggerm_{\mbox{\footnotesize$\bSigma\to p(\bu|\bSigma)$}}(\bSigma)}

\def\etaSUBuTOpuSigma{
\biggerbdeta_{\mbox{\footnotesize$\bu\to p(\bu|\bSigma)$}}}

\def\mSUBuTOpuSigma{
\biggerm_{\mbox{\footnotesize$\bu\to p(\bu|\bSigma)$}}(\bSigma)}

\def\mSUBuTOpuSigma{
\biggerm_{\mbox{\footnotesize$\bu\to p(\bu|\bSigma)$}}(\bu)}

\def\mSUBaTOpsigsqa{
\biggerm_{\mbox{\footnotesize$a\to p(\sigma^2|a)$}}(a)}

\def\mSUBpsigsqaTOa{
\biggerm_{\mbox{\footnotesize$p(\sigma^2|a)\to a$}}(a)}
\def\mSUBpsigsqaTOsigsq{
\biggerm_{\mbox{\footnotesize$p(\sigma^2|a)\to\sigma^2$}}(\sigma^2)}
\def\mSUBsigsqTOpsigsqa{
\biggerm_{\mbox{\footnotesize$\sigma^2\to p(\sigma^2|a)$}}(\sigma^2)}
\def\mSUBsigsqTOpyalphasigsq{
\biggerm_{\mbox{\footnotesize$\sigma^2\to p(y|\alpha,\sigma^2)$}}(\sigma^2)}
\def\etaSUBsigsqTOpyalphasigsq{
\biggerbdeta_{\mbox{\footnotesize$\sigma^2\to p(y|\alpha,\sigma^2)$}}}
\def\etaSUBpsigsqaTOsigsq{
\biggerbdeta_{\mbox{\footnotesize$p(\sigma^2|a)\to\sigma^2$}}}
\def\etaSUBpsigsqaTOa{
\biggerbdeta_{\mbox{\footnotesize$p(\sigma^2|a)\to a$}}}
\def\etaSUBsigsqTOpsigsqa{
\biggerbdeta_{\mbox{\footnotesize$\sigma^2\to p(\sigma^2|a)$}}}
\def\etaSUBaTOpsigsqa{
\biggerbdeta_{\mbox{\footnotesize$a\to p(\sigma^2|a)$}}}
\def\mSUBpThetaTOTheta{
\biggerm_{\mbox{\footnotesize$p(\bTheta)\to\bTheta$}}(\bTheta)}
\def\etaSUBpThetaTOTheta{
\biggerbdeta_{\mbox{\footnotesize$p(\bTheta)\to\bTheta$}}}
\def\etaSUBpthetaTOtheta{
\biggerbdeta_{\mbox{\footnotesize$p(\btheta)\to\btheta$}}}
\def\mSUBpthetaTOtheta{
\biggerm_{\mbox{\footnotesize$p(\btheta)\to\btheta$}}(\btheta)}
\def\mSUBthetaTOptheta{
\biggerm_{\mbox{\footnotesize$\theta\to p(\btheta)$}}(\btheta)}
\def\sigmabeta{\sigma_{\footnotesize{\bbeta}}}
\def\StarOfDavid{\bigtriangleup\mkern-16mu\bigtriangledown}

\begin{document}
\ifthenelse{\boolean{DoubleSpaced}}{\setstretch{1.5}}{}

\thispagestyle{empty}

\centerline{\LARGE\bf Factor Graph Fragmentization of Expectation Propagation}
\vskip7mm
\ifthenelse{\boolean{UnBlinded}}{
\centerline{\large\sc By Wilson Y. Chen and Matt P. Wand}
\vskip6mm
\centerline{\textit{University of Technology Sydney}}
\vskip6mm
\centerline{15th January, 2018}
}{\null}

\vskip6mm

\centerline{\large\bf Abstract}
\vskip2mm

Expectation propagation is a general approach to fast approximate inference
for graphical models. The existing literature treats models separately when 
it comes to deriving and coding expectation propagation inference algorithms. 
This comes at the cost of similar, long-winded algebraic steps being repeated
and slowing down algorithmic development. We demonstrate how 
\emph{factor graph fragmentization} can overcome this impediment. This involves 
adoption of the message passing on a factor graph approach to expectation propagation 
and identification of factor graph sub-graphs, which we call \emph{fragments}, 
that are common to wide classes of models. Key fragments and their corresponding 
messages are catalogued which means that their algebra does not need to be repeated.
This allows compartmentalization of coding and efficient software development.


\section{Introduction}\label{sec:intro}

\emph{Expectation propagation} (e.g.\ Minka, 2005) is gaining popularity as a general
approach to fitting and inference for large graphical models, including those that
arise in statistical contexts such as Bayesian generalized linear mixed models 
(e.g.\ Gelman \textit{et al.}, 2014; Kim \myand Wand, 2017). Compared
with Markov chain Monte Carlo approaches, expectation propagation has the 
attractions of speed and parallelizability of the computing across
multiple processors making it more amenable to high volume/velocity data
applications. One price to be paid is inferential accuracy since expectation
propagation uses product density simplifications of joint posterior density
functions. Another is algebraic overhead: as demonstrated by Kim \myand Wand (2016)
several pages of algebra are required to derive explicit programmable expectation
propagation algorithms for even very simple Bayesian models. This article
alleviates the latter cost. Using the notions of \emph{message passing} and
\emph{factor graph fragments} we demonstrate the compartmentalization of expectation
propagation algebra and coding. The resultant infrastructure and updating
formulae lead to much more efficient expectation propagation fitting and
inference and allows extension to arbitrarily large Bayesian models.

  Expectation propagation and \emph{mean field variational Bayes} are the
two most common paradigms for obtaining fast approximate inference
algorithms for graphical models (e.g.\ Bishop, 2006; Wainwright \myand Jordan, 2008;
Murphy, 2012). Each is driven by minimum Kullback-Leibler divergence considerations.
As explained in Minka (2005), they can both be expressed as message passing algorithms
on factor graphs. The alternative appellation \emph{variational message passing}
is used for mean field variational Bayes when such an approach is used.
The software platform \textsf{Infer.NET} (Minka \textit{et al.}, 2014)
uses both expectation propagation and variational message passing to perform
fast approximate inference for graphical models. Recently Wand (2017) introduced 
factor graph fragmentization to streamline variational message passing 
for semiparametric regression analysis.
Semiparametric regression (e.g.\ Ruppert \textit{et al.}, 2009) is a big
class of flexible regression models that includes generalized linear mixed
models, generalized additive models and varying-coefficient models as special
cases. Nolan \myand Wand (2017) and McLean \myand Wand (2018) built on 
Wand (2017) for more elaborate likelihood fragments.

The crux of this article is to show how the factor graph fragment idea also can
be used to streamline expectation propagation. We focus on semiparametric 
regression models. However, the approach is quite general and applies to 
other graphical models for which expectation propagation is feasible.
The fragment updating algorithms presented and derived here cover a wide
range of semiparametric models and pave the way for future derivations 
of the same type.

Section \ref{sec:background} provides the background material needed for 
factor graph fragmentization of expectation propagation. This includes
exponential family and Kullback-Leibler projection theory, as well
as the notions of factor graphs and their fragment sub-graphs.
The article's centerpiece is Section \ref{sec:fragGLMM} in which
several key fragments are identified and have message updates
derived and catalogued. Such cataloguing implies that updates
for a particular fragment never have to be derived again and only 
need to be implemented once in an expectation propagation software suite. 
An illustration involving generalized additive mixed model analysis
of data from a longitudinal public health study is provided
in Section \ref{sec:illustration}. Section \ref{sec:elaborate}
contains some commentary of fragmentization of 
expectation propagation for more elaborate models.
	
\section{Background Material}\label{sec:background}

Factor graph fragmentization of expectation propagation relies on
definitions and results concerning both distribution theory and 
graph theory, not all of which are commonplace in the statistics 
literature. We provide the necessary background material in this 
section. 

\subsection{Exponential Family Distributions}

A random variable $x$ has an exponential family distribution if its
probability mass function or density function admits the form
$$p(x)=\exp\{\bT(x)^T\bdeta-A(\bdeta)\}h(x),\quad x\in\realnos,\ \bdeta\in H.$$
The vectors $\bT(x)$ and $\bdeta$ are called, respectively, the
\emph{sufficient statistic} and \emph{natural parameter}. 
The set $H$ is the space of allowable natural parameter values.
The function $A(\bdeta)$ is called the \emph{log-partition function} 
and $h(x)$ is the \emph{base measure}. 
A key exponential family distributional result is that 
\begin{equation}
E\{\bT(x)\}=\nabla A(\bdeta)
\label{eq:delAres}
\end{equation}
where $\nabla A(\bdeta)$ is the column vector of partial derivatives of $A(\bdeta)$
with respect to each of the components of $\bdeta$.

Table \ref{tab:exponFam} lists each of the exponential families distributions 
arising in this article, along with their defining functions and parameter spaces.
The Normal and Inverse Chi-Squared exponential families are well known. 
The Moon Rock exponential family is less established, and is given this name
in McLean \myand Wand (2018). In Table \ref{tab:exponFam} and elsewhere
we use the following indicator function notation: $I(\Psc)=1$ if the proposition
$\Psc$ is true and $I(\Psc)=0$ if $\Psc$ is false.

Note also that the Inverse Chi-Squared exponential family is equivalent to
the \emph{Inverse Gamma} exponential family. The two families differ in their
common parametrizations as explained in, for example, Section S.1.3 of the 
online supplement of Wand (2017). The Inverse Chi-Squared distribution has the 
advantage of being the special case of the Inverse Wishart distribution for
$1\times1$ random matrices. Throughout this article we write 
$$\bX\sim\mbox{Inverse-Wishart}(\kappa,\bLambda)$$
to denote a $d\times d$ random matrix $\bX$ having density function
$$p(\bX)=\Csc_{d,\kappa}^{-1}|\bLambda|^{\kappa/2}\,
|\bX|^{-(\kappa+d+1)/2}
\exp\{-\smhalf\tr(\bLambda\bX^{-1} )\}\,I(\mbox{$\bX$ symmetric and positive definite})
$$
where $\kappa>d-1$, $\bLambda$ is a $d\times d$ symmetric positive definite matrix
and
\begin{equation}
\Csc_{d,\kappa}\equiv2^{d\kappa/2}\pi^{d(d-1)/4}
\prod_{j=1}^d\Gamma\left(\frac{\kappa+1-j}{2}\right).
\label{eq:CscDefn}
\end{equation}
For the special case of $d=1$ we write 
$$x\sim\mbox{Inverse-$\chi^2$}(\kappa,\lambda).$$

\begin{table}[!t]
\begin{center}
{\setlength{\tabcolsep}{6pt}
\begin{tabular}{lclcl}
name    & $\bT(x)$ & $A(\bdeta)$ & $h(x)$ & $H$ \\[0.1ex]
\hline\\[-0.9ex]
Normal               & $\left[{\setlength{\arraycolsep}{0pt}\begin{array}{c}x\\x^2\end{array}}\right]$   &  
$-\quarter(\eta_1^2/\eta_2)$  & $1$                    &  $\{(\eta_1,\eta_2):$ \\[0ex] 
&           &  $\quad-\smhalf\,\log(-2\eta_2)$  &            & $\eta_1\in\realnos,\eta_2<0\}$\\[4ex]
Inverse Chi-Squared & $\left[{\setlength{\arraycolsep}{0pt}\begin{array}{c}\log(x)\\1/x\end{array}}\right]$      
&  $(\eta_1+1)\log(-\eta_2)$ & $I(x>0)$   & $\{(\eta_1,\eta_2):$      \\[0ex]
&              &  $\quad+\log\Gamma(-\eta_1-1)$  &            &  $\eta_1<-1,\eta_2<0\}$ \\[4ex]
Moon Rock             & $\left[{\setlength{\arraycolsep}{0pt}\begin{array}{c}\{x\log(x)\\-\log\Gamma(x)\}\\[1.5ex]x\end{array}}\right]$ 
&$\log\Big[\int_0^{\infty}\{t^t/\Gamma(t)\}^{\eta_1}$
&$I(x>0)$&$\{(\eta_1,\eta_2):$\\[-2ex] 
&  &$\qquad\times\exp(\eta_2\,t)\,dt\Big]$&&$\eta_1>0,\eta_1+\eta_2<0\}$ \\[2ex]
\hline
\end{tabular}
}
\caption{\textit{Sufficient statistics, log-partition functions, base measures and natural parameter spaces
of three exponential families.}
}
\label{tab:exponFam}
\end{center}
\end{table}

\subsection{Kullback-Leibler Projection}

If $p_1$ and $p_2$ are two univariate density functions then the 
\emph{Kullback-Leibler divergence} of $p_2$ from $p_1$ is 
$$\KL(p_1\Vert p_2)\equiv\int_{-\infty}^{\infty}p_1(x)\log\{p_1(x)/p_2(x)\}\,dx.$$
If $\Qsc$ is a family of univariate density functions then the projection of the
univariate density function $p$ onto $\Qsc$ is 
\begin{equation}
\proj_{\Qsc}[p]\equiv\argmin{q\in\Qsc}\,\KL(p\Vert q).
\label{eq:projQsc}
\end{equation}
A core aspect of expectation propagation is projection of an 
arbitrary \emph{input} density function onto a particular exponential
family. This corresponds to (\ref{eq:projQsc}) with 
$$\Qsc=\big\{q(\,\cdot\,;\bdeta):q(x;\bdeta)=\exp\{\bT(x)^T\bdeta-A(\bdeta)\}h(x),\ \ \bdeta\in H\big\}.$$
As explained in Section 2.3 of Kim \myand Wand (2016), the exponential
family Kullback-Leibler problem 
$$\bdeta^*=\argmin{\bdeta\in H}\KL\big(p\Vert q(\cdot;\bdeta)\big)$$
is equivalent to the sufficient statistic moment matching problem
\begin{equation}
\infint\bT(x)\,\exp\{\bT(x)^T\bdeta^*-A(\bdeta^*)\}h(x)\,dx=\infint \bT(x)\,p(x)\,dx.
\label{eq:momentMatch}
\end{equation}
Because of (\ref{eq:delAres}) we can re-write (\ref{eq:momentMatch}) as 
$$(\nabla A)(\bdeta^*)=\infint \bT(x)\,p(x)\,dx.$$
Then, assuming that the inverse of $\nabla A$ is well-defined, 
\begin{equation}
\bdeta^*=(\nabla A)^{-1}\left(\infint \bT(x)\,p(x)\,dx\right).
\label{eq:delAinv}
\end{equation}
Hence, given the $\bT$ moments, Kullback-Leibler projection of
a density function $p$ onto an exponential family boils down to 
inversion of $\nabla A$. Section 3 of Wainwright \myand Jordan (2008)
provides a detailed study of exponential families including
properties of $A$ and $\nabla A$. An exponential family 
distribution with the sufficient statistic $\bT(x)$ being a $d\times 1$ vector is said to be 
\emph{regular} if $H$ is an open set in $\realnos^d$ and \emph{minimal} 
if there is no $d\times1$ vector $\ba$ and constant $b\in\realnos$ such
that $\ba^T\bT(x)=b$ almost surely. Each of the exponential
families in Table \ref{tab:exponFam} are regular and minimal.
Result 1 provides a summary of results from Section 3 of Wainwright \myand Jordan (2008) 
that is relevant to (\ref{eq:delAinv}). It depends on: 

\vskip3mm
\noindent
{\bf Definition 1.}
\textit{Consider a function $\bdf:\realnos\to\realnos^{d}$. Then the set of realizable 
expectations of $\bdf$ is the set of points $[\tau_1,\ldots,\tau_d]^T\in\realnos^d$ 
such that there exists a univariate random variable $x$ for which 
$E\{\bdf(x)\}=[\tau_1,\ldots,\tau_d]^T$.}

\begin{figure}[h!]
\begin{center}
\includegraphics[width=0.45\textwidth]{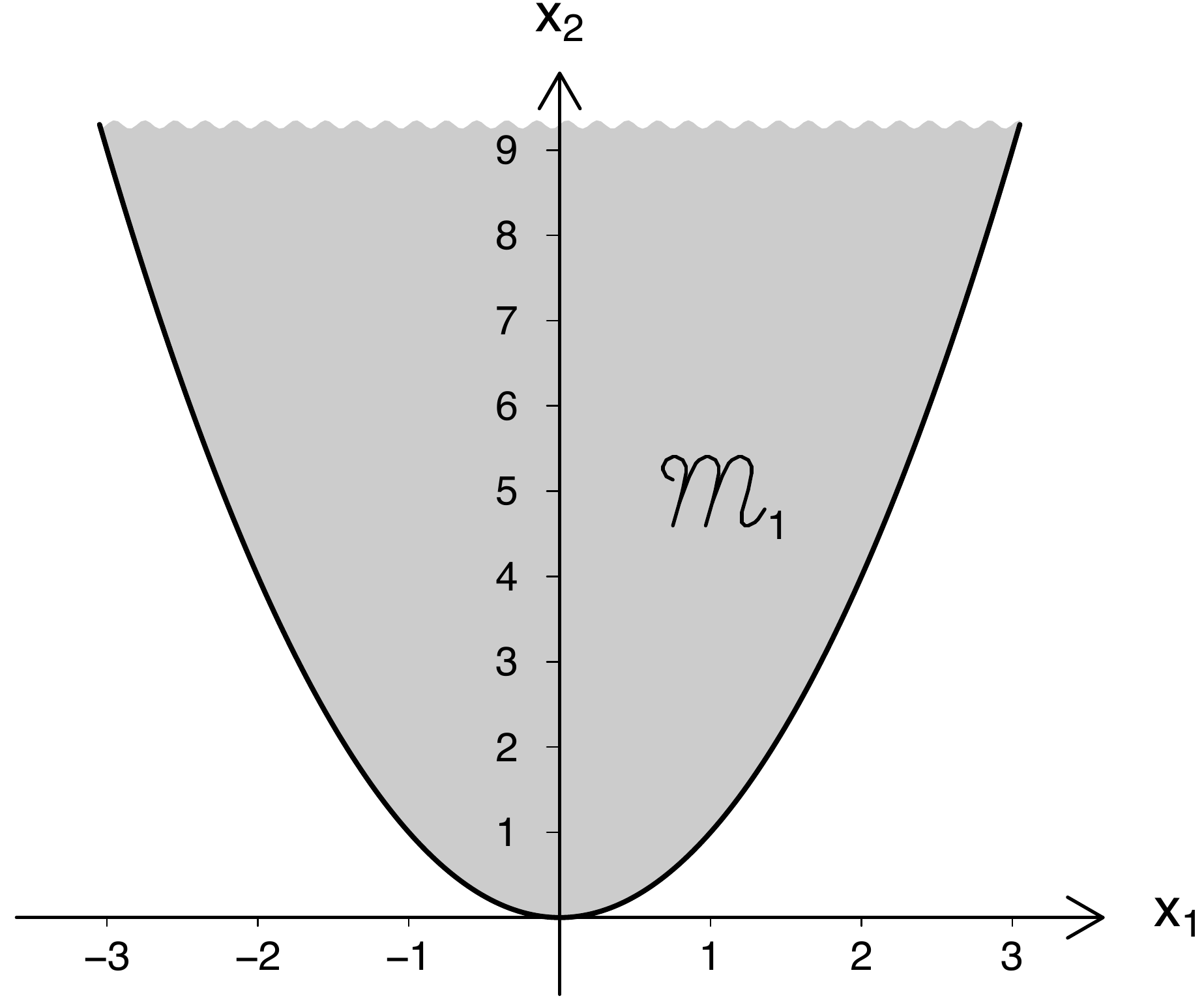}  
\includegraphics[width=0.45\textwidth]{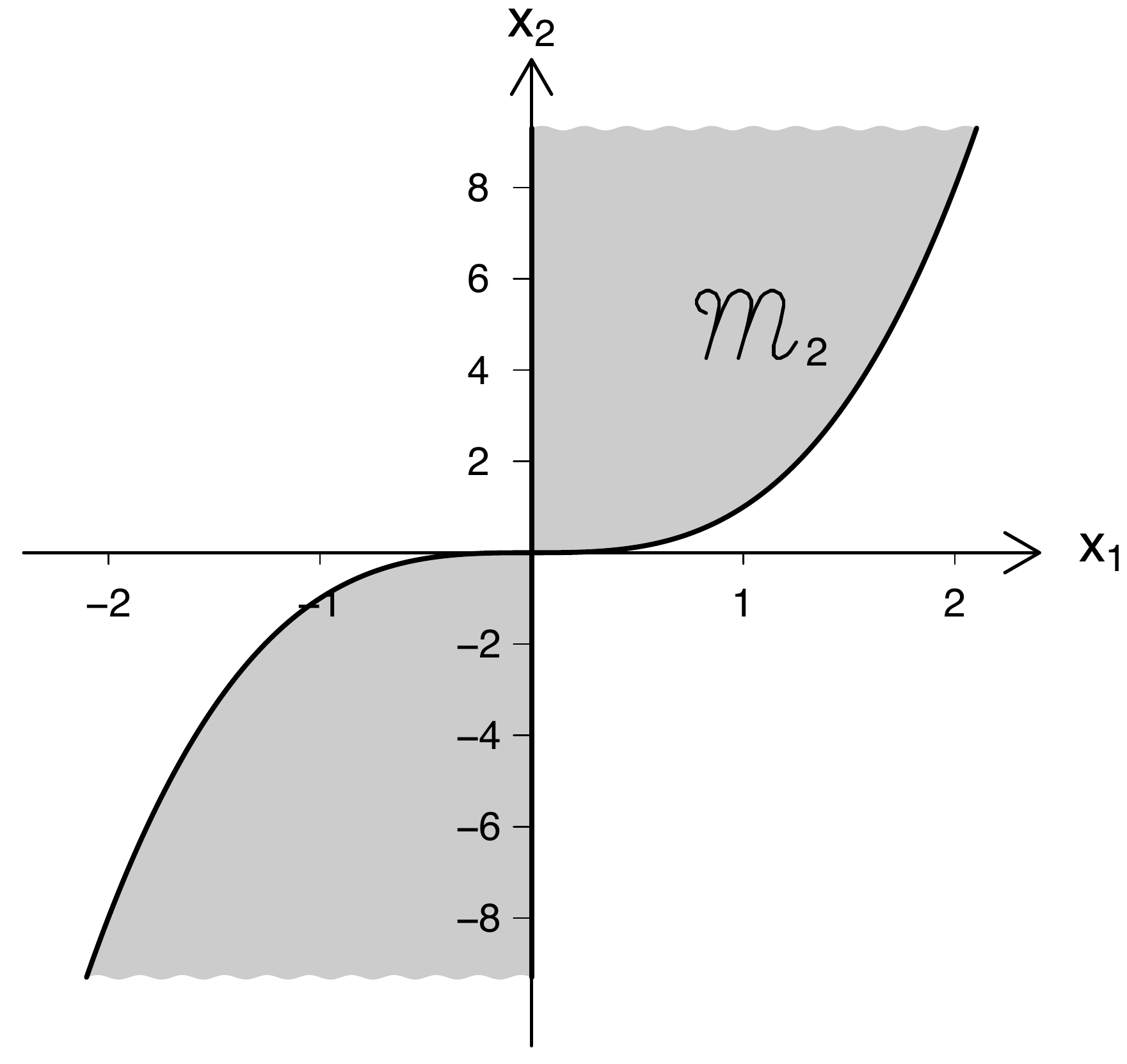}  
\caption{\it Left panel: The shaded region is $\mathfrak{M}_1$, the set of realizable expectations
of $\bdf_1$. Right panel: The shaded region is $\mathfrak{M}_2$, the set of realizable expectations
of $\bdf_2$.}
\label{fig:F1andF2realisExps}
\end{center}
\end{figure}

\vskip3mm
\noindent
To illustrate the notion of the set of realizable expectations,
consider the functions $\bdf_1:\realnos\to\realnos^2$ and $\bdf_2:\realnos\to\realnos^2$ given by 
$$
\bdf_1(x)=\left[
\begin{array}{c}
x\\[1ex]
x^2
\end{array}
\right]   
\quad\mbox{and}\quad
\bdf_2(x)=\left[
\begin{array}{c}
x\\[1ex]
x^3
\end{array}
\right].   
$$
The sets of all realizable expectations of $\bdf_1$ and $\bdf_2$ are, respectively
$$\mathfrak{M}_1\equiv\left\{\left[
\begin{array}{c}
x_1\\[1ex]
x_2
\end{array}
\right]:x_2\ge x_1^2
\right\}\quad\mbox{and}\quad
\mathfrak{M}_2\equiv
\left\{\left[
\begin{array}{c}
x_1\\[1ex]
x_2
\end{array}
\right]:\,\mbox{sign}(x_1)\,x_2\ge|x_1|^3\right\}.
$$
To show that $\mathfrak{M}_1$ is the set of all realizable expectations of $\bdf_1$ note
that
$\mathfrak{M}_1=\mathfrak{M}_{11}\cup\mathfrak{M}_{12}$ where $\mathfrak{M}_{11}\equiv\{[x_1\ x_2]^T:x_2=x_1^2\}$
and $\mathfrak{M}_{12}\equiv\{[x_1\ x_2]^T:x_2>x_1^2\}$. Then for any $[x_1\ x_2]\in\mathfrak{M}_{11}$
we can take $x$ to be the degenerate random variable  with probability mass function
$p(x)=I(x=x_1)$. For such $x$, $E\{\bdf_1(x)\}=[x_1\ x_1^2]^T=[x_1\ x_2]^T\in\mathfrak{M}_{11}$ which shows that all
elements of $\mathfrak{M}_{11}$ are realizable expectations of $\bdf_1$. For any $[x_1\ x_2]\in\mathfrak{M}_{12}$
taking $x\sim N(x_1,x_2-x_1^2)$ leads to $E\{\bdf_1(x)\}=[x_1\ x_2]^T$ verifying that
all elements of $\mathfrak{M}_{12}$ are realizable by $E\{\bdf_1(x)\}$ for some $x$.
Hence, all entries of $\mathfrak{M}_1$ are realizable by $E\{\bdf_1(x)\}$ for some $x$.. Values $[x_1\ x_2]^T\notin\mathfrak{M}_1$
are not realizable because Jensen's inequality implies that $E(x^2)\ge\{E(x)\}^2$ for any
random variable $x$. Similar arguments can be used to establish that $\mathfrak{M}_2$ is the 
set of all realizable expectations of $\bdf_2$. Figure \ref{fig:F1andF2realisExps} shows
the sets $\mathfrak{M}_1$ and $\mathfrak{M}_2$.

\vskip3mm\noindent
We are now ready to give the pivotal:
\vskip2mm
\noindent
{\bf Result 1 (Wainwright \myand Jordan, 2008).}
\textit{Consider a regular and minimal exponential family with $d$-dimensional
sufficient statistic $\bT(x)$ and corresponding natural parameter vector $\bdeta$. Then}
\begin{itemize}
\item[(a)] \textit{$H$ is a strictly convex subset of $\realnos^d$.}
\item[(b)] \textit{$A$ is a strictly convex and infinitely differentiable function on $H$.} 
\item[(c)] \textit{$\nabla A$ is a one-to-one function.}
\item[(d)] \textit{The image of $\nabla A$, which we denote by $T$, is the 
interior of the set of all realizable expectations of $\bT$.} 
\end{itemize}

\vskip5mm\noindent
Result 1 guarantees that 
$\nabla A:H\to T$
is a bijective map and that 
$(\nabla A)^{-1}:T\to H$
is well-defined.

\subsubsection{Normal Distribution Special Case}

The Normal distribution is the one of simplest exponential families since
$\nabla A$ and $(\nabla A)^{-1}$ admit simple closed forms. Firstly, we have
$$\nabla A(\bdeta)=\left[
\begin{array}{c}
-\eta_1/(2\eta_2)\\[1ex]
(\eta_1^2-2\eta_2)/(4\eta_2^2)
\end{array}
\right].
$$
It is straightforward to show that
the image of $H$ under $\nabla A$ is
$$T=\{(\tau_1,\tau_2):\tau_2>\tau_1^2\}$$
and the inverse of $\nabla A$ is
$$(\nabla A)^{-1}(\btau)=\left[
\begin{array}{c}
\tau_1/(\tau_2-\tau_1^2)\\
-1/\{2(\tau_2-\tau_1^2)\}
\end{array}
\right].
$$

\subsubsection{Inverse Chi-Squared Distribution Special Case}

For the Inverse Chi-Squared distribution we have
$$\nabla A(\bdeta)=\left[
\begin{array}{c}
\log(-\eta_2)-\mbox{digamma}(-\eta_1-1)\\[1ex]
(\eta_1+1)/\eta_2
\end{array}
\right]
$$
where $\mbox{digamma}(x)\equiv\frac{d}{dx}\log\Gamma(x)$.
Determination of the image of $H$ under $\nabla A$ is 
more challenging for the Inverse Chi-Squared distribution.
It is aided by Theorem 1 of Kim \myand Wand (2016) 
which establishes that $\log-\mbox{digamma}$ is a bijective
map between $\realnos_+$ and $\realnos_+$. This leads to 
$$T=\{(\tau_1,\tau_2):\tau_2>e^{-\tau_1}\}.$$
The inverse or $\nabla A$ is
$$(\nabla A)^{-1}(\btau)=\left[
\begin{array}{c}
-(\log-\mbox{digamma})^{-1}\big(\tau_1+\log(\tau_2)\big)-1\\[1ex]
-(\log-\mbox{digamma})^{-1}\big(\tau_1+\log(\tau_2)\big)/\tau_2
\end{array}
\right].
$$
Theorem 1 of Kim \myand Wand (2016) implies that $(\log-\mbox{digamma})^{-1}$
is well-defined.

\begin{figure}[!ht]
\centering
{\includegraphics[width=0.98\textwidth]{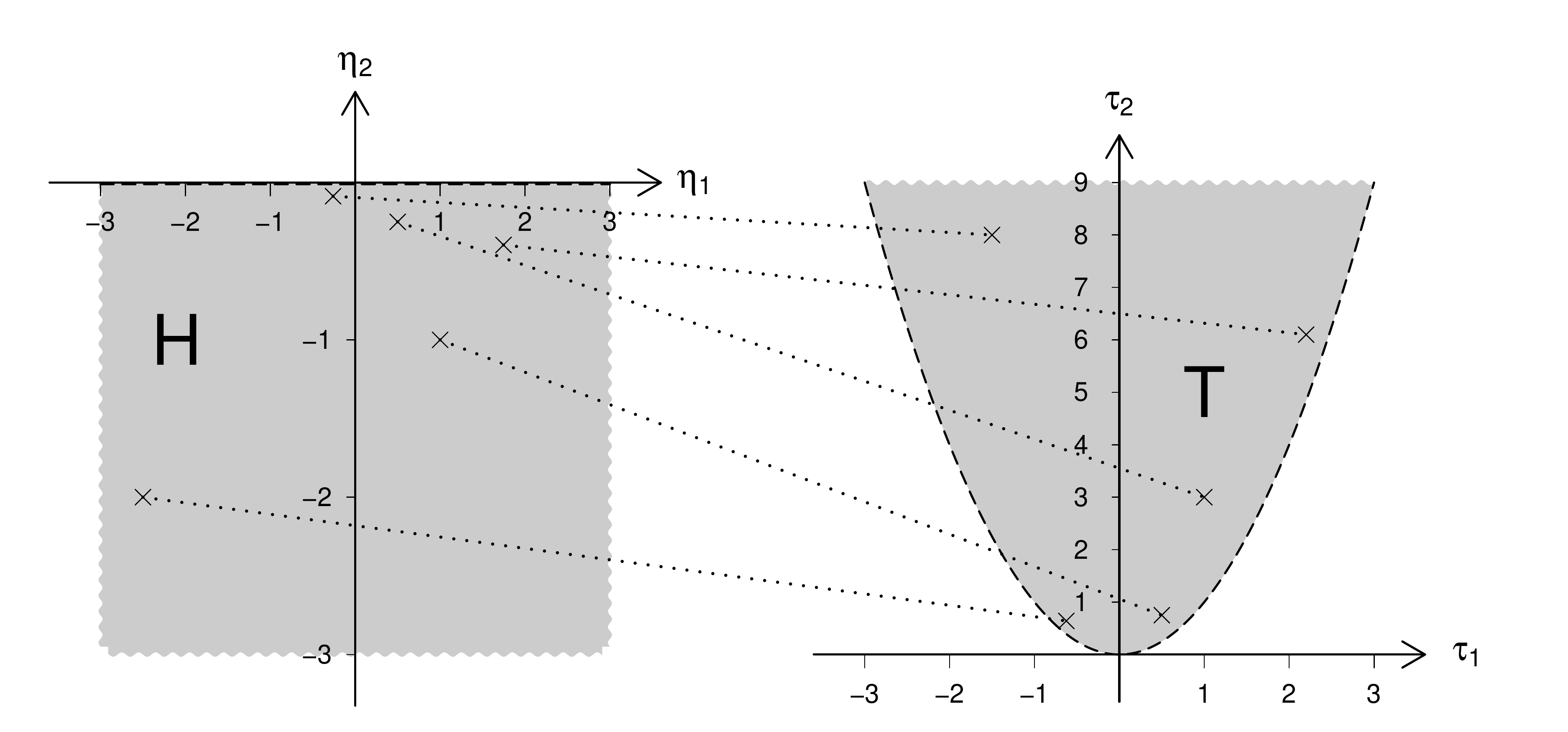}}
{\includegraphics[width=0.98\textwidth]{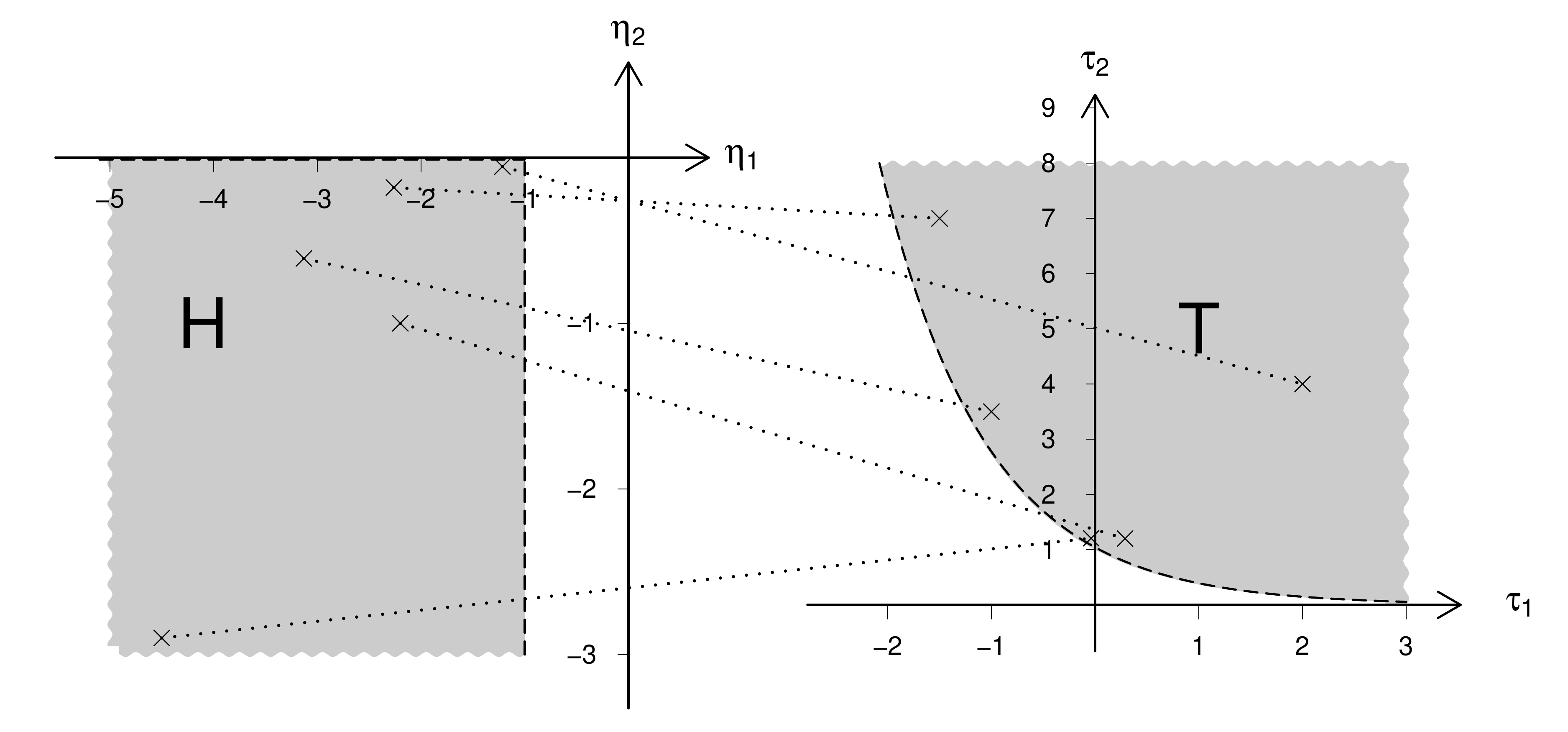}}
\caption{\it Upper panel: Illustration of the bijective maps between $H$ and $T$ for
the Normal exponential family. The crosses and dotted lines depict five example $\bdeta\in H$ and 
$\btau=\nabla A(\bdeta)\in T$ pairs. Since $\nabla A$ is a bijective map, the crosses
and dotted lines equivalently depict five example $\btau\in T$ and $\bdeta=(\nabla A)^{-1}(\btau)\in H$
pairs. Lower panel: Similar illustration for the Inverse Chi-Squared 
exponential family.}
\label{fig:HTmaps} 
\end{figure}

\vskip8mm
Figure \ref{fig:HTmaps} depicts the $\nabla A$ and $(\nabla A)^{-1}$ bijective maps between $H$ and $T$
for both the Normal and Inverse Chi-Squared exponential family distributions.

\subsection{Factor Graphs and Factor Graph Fragments}

\begin{figure}[!ht]
\centering
{\includegraphics[width=0.88\textwidth]{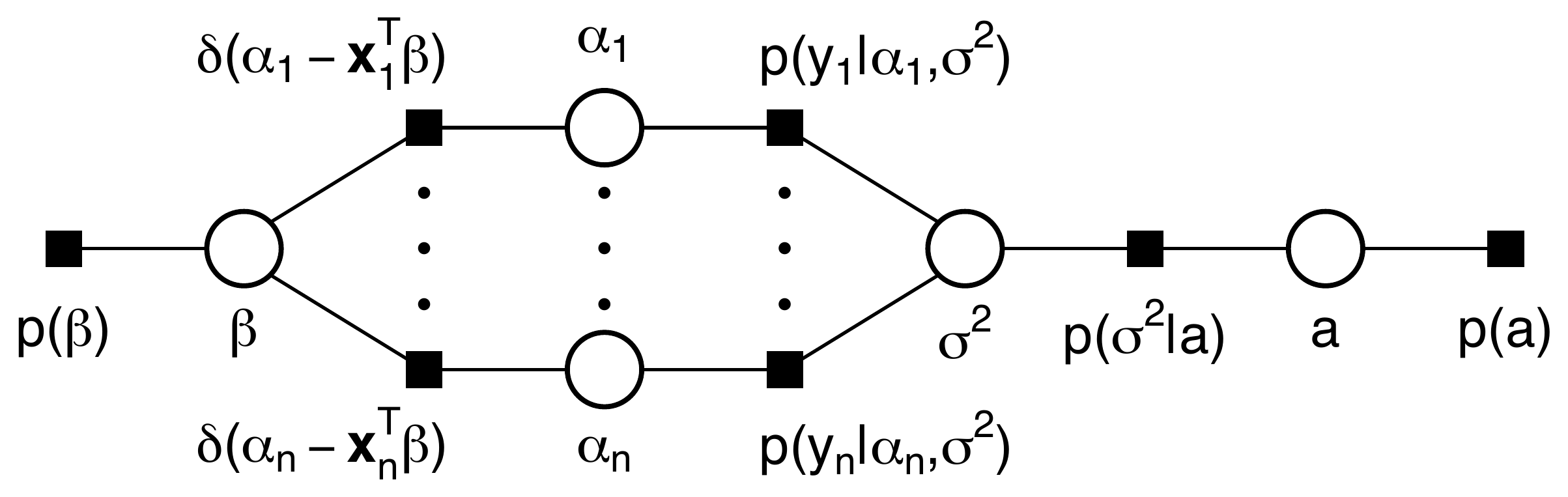}}
\caption{\it Factor graph representation of (\ref{eq:linModDerivVar}).}
\label{fig:linModFacGraph} 
\end{figure}

A \emph{factor graph} is a graphical representation of the factor/argument
dependencies of a multivariate function. Even though the concept applies to functions
in general, the relevant functions are joint density functions in the context of
expectation propagation. As an illustration, consider the Bayesian linear model
$$\by|\bbeta,\sigma^2\sim N(\bX\bbeta,\sigma^2\,\bI),$$ 
where $\by$ is an $n\times1$ vector of responses, with the following prior distributions 
on the regression coefficients and error standard deviation:
$$\bbeta\sim N(\bzero,\sigmabeta^2\bI)\quad\mbox{and}\quad\sigma\sim\mbox{Half-Cauchy}(A),$$
The second prior specification means that $\sigma$ has prior density function 
$p(\sigma)=2/[A\pi\{1+(\sigma/A)^2\}]$ for $\sigma>0$. An equivalent representation of the
model, involving the auxiliary variable $a$, is
\begin{equation}
\begin{array}{c}
\by|\bbeta,\sigma^2\sim N(\bX\bbeta,\sigma^2\bI),\quad\bbeta\sim N(\bzero,\sigmabeta^2\bI),\\[2ex]
\sigma^2|a\sim\mbox{Inverse-$\chi^2$}(1,1/a),\quad a\sim\mbox{Inverse-$\chi^2$}(1,1/A^2).
\end{array}
\label{eq:linModel}
\end{equation}
We work with this auxiliary variable representation since it aids tractability of expectation
propagation. The joint density function of the random variables and random vectors in (\ref{eq:linModel})
admits the following factorized form:
$$p(\by,\bbeta,\sigma^2,a)=p(\bbeta)p(\by|\bbeta,\sigma^2)p(\sigma^2|a)p(a).$$
Now let $\bx_i^T$ be the $i$th row of $\bX$ for $1\le i\le n$. Then a further breakdown of 
$p(\by,\bbeta,\sigma^2,a)$ is 
\begin{equation}
p(\by,\bbeta,\sigma^2,a)=p(\bbeta)\,\left\{\displaystyle{\prod_{i=1}^n}
\infint\delta(\alpha_i-\bx_i^T\bbeta)\,p(y_i|\alpha_i,\sigma^2) 
\,d\alpha_i\right\}\,p(\sigma^2|\,a)\,p(a)
\label{eq:linModDerivVar}
\end{equation}
where $\delta$ is the Dirac delta function and 
$p(y_i|\alpha_i,\sigma^2)\equiv(2\pi\sigma^2)^{-1/2}\exp\{-(y_i-\alpha_i)^2/(2\sigma^2)\}$. 
Figure \ref{fig:linModFacGraph} is a factor graph representation of 
$p(\by,\bbeta,\sigma^2,a)$ according to the factors 
that appear in (\ref{eq:linModDerivVar}). At this point we note that we are not using
the conventional factor graph definition here since some of the factors appear
inside the integrals in (\ref{eq:linModDerivVar}). Kim \myand Wand (2017) introduced
the term \emph{derived variable factor graph} to make this distinction. 
We will simply call it a \emph{factor graph} from now onwards.
The circles are called \emph{stochastic nodes} and the rectangles are called \emph{factors}.
Both circles and rectangles are \emph{nodes} of the factor graph. We say that a two nodes
are \emph{neighbors} of each other if they are joined by an edge.

\begin{figure}[!ht]
\centering
{\includegraphics[width=0.88\textwidth]{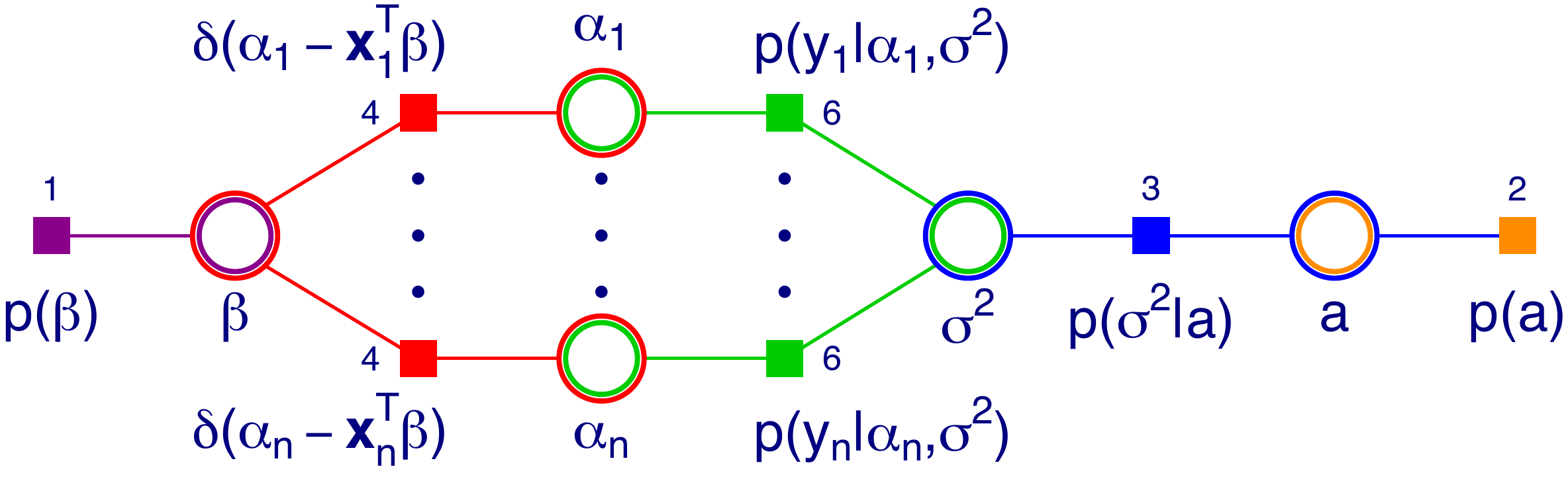}}
\caption{\it Fragmentization of the Figure \ref{fig:linModFacGraph} factor graph.
Different colors signify fragments of the same type, and 
are included in Table \ref{tab:GLMMfrags}.} 
\label{fig:linModFacGraphFrag} 
\end{figure}

Figure \ref{fig:linModFacGraphFrag} is a representation of Figure \ref{fig:linModFacGraph} 
with factor graph \emph{fragments} of the same type identified via
\ifthenelse{\boolean{ColourVersion}}{color-coding and numbering of the factors.}
{numbering of the factors.}
As defined in Wand (2017), a fragment is a sub-graph of a factor graph consisting
of a factor and each of its neighboring stochastic nodes. 

The five different colors in Figure \ref{fig:linModFacGraphFrag} correspond to 
five different fragment types. Some of the fragment types, such as that corresponding
to the $p(\bbeta)$ factor, only appear once in this factor graph. Other types,
such as those corresponding to $\delta(\alpha_i-\bx_i^T\bbeta)$, $1\le i\le n$,
appear multiple times. Recognition of the recurrence of fragments of the same
type in this factor graph and factor graphs for other models is at the core
of extension to arbitrarily large models. Wand (2017) demonstrated factor
graph fragmentization of variational message passing. Our goal here is
to do the same for expectation propagation.

\subsection{Expectation Propagation}\label{sec:expecPropag}

\begin{figure}[!ht]
\centering
{\includegraphics[width=0.475\textwidth]{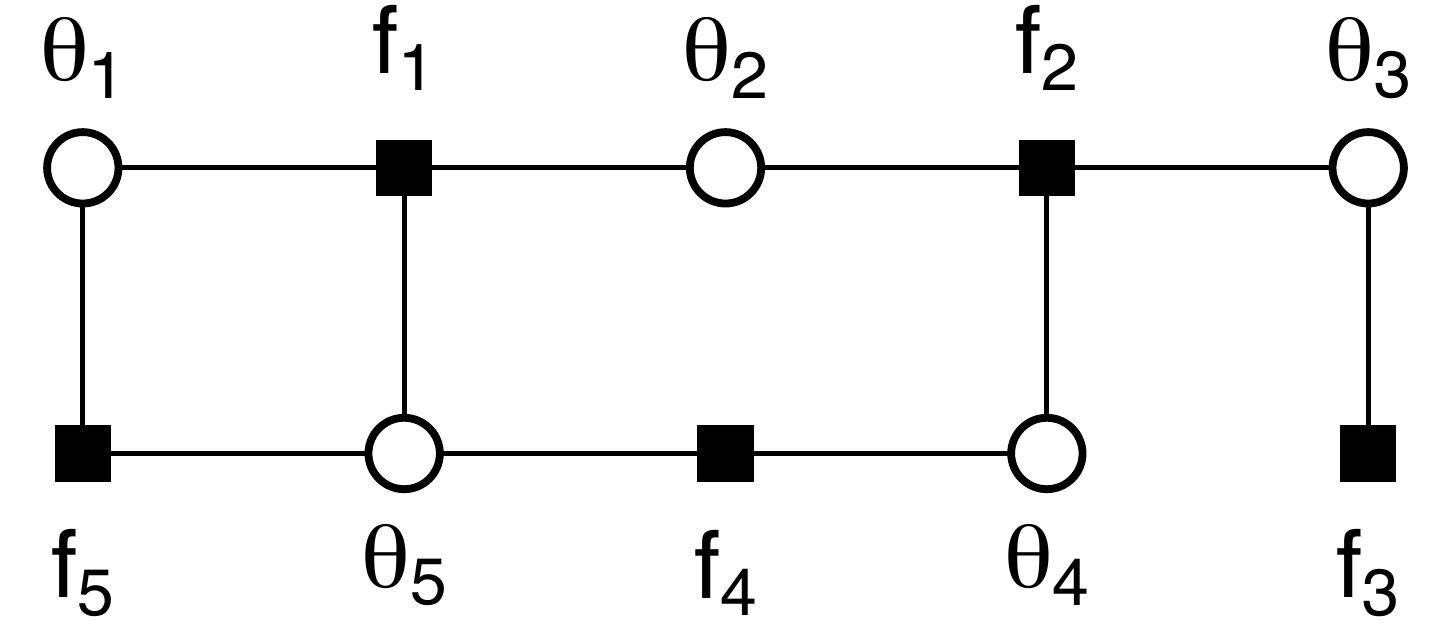}}
\caption{\textit{An illustrative generic form factor graph.}}
\label{fig:simpFacGraph} 
\end{figure}

Recent summaries of expectation propagation are provided in Kim \myand Wand (2016, 2017).
We briefly cover the main points here. The function $\mbox{neighbors}(\cdot)$ 
plays an important role in the algebraic description of the message updates.
Consider the illustrative generic form factor graph shown in Figure \ref{fig:simpFacGraph},
corresponding to the joint density function of random vectors 
$\btheta_1,\ldots,\btheta_5$ according to a particular Bayesian model.
Then $\mbox{neighbours}(1)=\{1,2,5\}$
since the factor $f_1$ is connected by edges to each of $\btheta_1$,
$\btheta_2$ and $\btheta_5$. Similarly, $\mbox{neighbours}(2)=\{2,3,4\}$,
$\mbox{neighbours}(3)=\{3\}$, $\mbox{neighbours}(4)=\{4,5\}$ and
$\mbox{neighbours}(5)=\{1,5\}$. For general factor graphs with the
$\btheta_i$ and $f_j$ labeling, $\mbox{neighbours}(j)$ 
is the set of indices of the $\btheta_i$ that are connected to $f_j$ by an edge.
%
%
Based on (54) of Minka (2005), the stochastic node
to factor messages are updated according to 
\begin{equation}
\biggerm_{\btheta_i\to\ f_j}(\btheta_i)\thickarrow 
\prod_{j'\ne j:\,i\in\subscriptneighbours(j')}
\biggerm_{f_{j'}\to\btheta_i}(\btheta_i)
\label{eq:stochToFac}
\end{equation}
and, based on (83) of Minka (2005), the factor to stochastic node 
messages updates are
\begin{equation}
\begin{array}{l}
\biggerm_{f_j\to\btheta_i}(\btheta_i)\thickarrow\\[1ex]
\displaystyle{\frac{\proj\Bigg[\biggerm_{\btheta_i\to\,f_j}(\btheta_i)
{\displaystyle\int}\biggerf_j(\btheta_{\subscriptneighbours(j)}) 
\displaystyle{\prod_{i'\in\subscriptneighbours(j)\backslash\{i\}}}
\biggerm_{\btheta_{i'}\to f_j}(\btheta_{i'})
\,d\btheta_{\subscriptneighbours(j)\backslash\{i\}}/Z\Bigg]}
{\biggerm_{\btheta_i\to\,f_j}(\btheta_i)}},
\end{array}
\label{eq:facToStoch}
\end{equation}
where $Z$ is the normalizing factor that ensures that the 
function of $\btheta_i$ inside the $\proj[\ \cdot\ ]$ is a density
function. The normalizing factor in (\ref{eq:facToStoch}) involves
summation if some of the $\btheta_{i'}$ have discrete components.
The $\proj[\ \cdot\ ]$ in (\ref{eq:facToStoch}) denotes
Kullback-Leibler projection onto an appropriate exponential family of density
functions. However, in Kim \myand Wand (2016) illustration was done 
only via a simple example in which all of the stochastic nodes were 
univariate. In the case of linear models, in which vector parameters 
are present, some adjustments are necessary to avoid intractable
multivariate integrals. The first one is an intrinsically important 
convention and is now spelt out:

\jump
\noindent
\textbf{Convention 1.} \textsl{Derived variable factor graphs are treated 
as ordinary factor graphs when it comes to applying the message passing
expressions (\ref{eq:stochToFac}) and (\ref{eq:facToStoch}).}
\jump

In practice, iteration involving (\ref{eq:stochToFac}) and (\ref{eq:facToStoch})
may require some tweaking to achieve convergence. Minka (2005) recommends
the \emph{damping} adjusment 
\begin{equation}
\biggerm_{f_j\to\btheta_i}(\btheta_i)\thickarrow
\biggerm_{f_j\to\btheta_i}(\btheta_i)^{\varepsilon}\times
\{\mbox{right-hand side of (\ref{eq:facToStoch})}\}^{1-\varepsilon}.
\label{eq:epsAdjust}
\end{equation}
for some $0\le\varepsilon<1$. Kim and Wand (2017) noted that setting $\varepsilon$ 
to a small positive number such as $\varepsilon=0.1$ aided convergence
for their expectation propagation algorithms for fitting linear models.
Therefore, we build this adjustment into the fragment updates
in the next section.

The full expectation propagation iterative algorithm is:

\begin{minipage}[t]{134mm}
\hrule
\vskip3mm
\begin{itemize}
\item[] Initialize all factor to stochastic node messages.
\item[] Cycle until all factor to stochastic node messages converge:
\begin{itemize}
\item[]For each factor:
\begin{itemize}
\item[] Compute the messages passed to the factor using (\ref{eq:stochToFac}).
\item[] Compute the messages passed from the factor using (\ref{eq:facToStoch})
and (\ref{eq:epsAdjust}).
\end{itemize}
\end{itemize}
\end{itemize}
\hrule
\end{minipage}

\vskip5mm
\noindent
Upon convergence the expectation propagation-approximate posterior 
density function of $\btheta_i$ is obtained from
$$q^*(\btheta_i)\propto\prod_{j:i\in\subscriptneighbours(j)}
\biggerm_{f_j\to\btheta_i}(\btheta_i).$$

\section{Fragmentization for Generalized, Linear and Mixed Models}\label{sec:fragGLMM}

Each of the generalized, linear and mixed models dealt with in Kim \myand Wand (2017)
can be handled with nine distinct fragment types, which are listed in Table \ref{tab:GLMMfrags}.
The message updates for each fragment type only needs to be derived once. Each subsection
deals with the required derivation and summarizes the updates as an algorithm.
For a software suite that uses expectation propagation to fit generalized, linear and 
mixed models the fragment only needs to be implemented once. We now work through
each of the Table \ref{tab:GLMMfrags} fragments in turn.

The algorithms use the matrix functions $\vecof$ and its inverse $\vecof^{-1}$
which we define here. If $\bA$ is $d\times d$ matrix then $\vecof(\bA)$ is
the $d^2\times 1$ vector obtained by stacking the columns
of $\bA$ underneath each other in order from left to right.
if $\ba$ is a $d^2\times1$ vector then $\vecof^{-1}(\ba)$ is
the $d\times d$ matrix formed from listing the entries of $\ba$
in a column-wise fashion in order from left to right and
is the usual function inverse when the domain of $\vecof$ 
is restricted to square matrices.

The following shorthand is used throughout this section:
$$a\thickarroweps b\quad\mbox{denotes}\quad a\thickarrow \varepsilon\,a+(1-\varepsilon)\,b. $$

\begin{table}[!ht]
\begin{center}
\begin{footnotesize}
\begin{tabular}{lcl}
\hline\\[-1.3ex]
Fragment name $\qquad\qquad$        & Diagram    &$\qquad\qquad$ Distributional statement    \\[0.1ex]
\hline\\[-0.9ex]
1. Gaussian prior             &   \includegraphics[width=0.15\textwidth, trim=0 8mm 0 0]{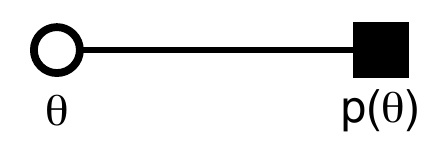} &  
$\qquad\qquad\btheta\sim N(\bmu_{\btheta},\bSigma_{\btheta})$\\[3ex]
2. Inverse Wishart          & \includegraphics[width=0.15\textwidth, trim=0 8mm 0 0]{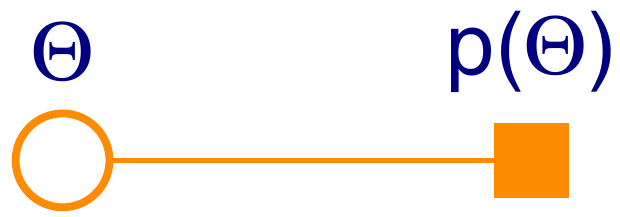}     &  
$\qquad\qquad\bTheta\sim\mbox{Inverse-Wishart}(\kappaTheta,\bLambda_{\bTheta})$  \\
\ \ \ \ prior                     &            &    \\[3ex]
3. Iterated Inverse Chi-Squared     &    \includegraphics[width=0.20\textwidth, trim=0 8mm 0 0]{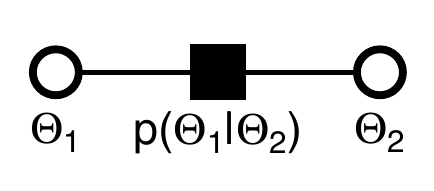}   & 
$\qquad\qquad\sigma^2|a\sim\mbox{Inverse-$\chi^2$}(\nu,1/a)$  \\
\ \ \ \             &        &                                                \\[3ex] 
4. Linear combination  &  \includegraphics[width=0.19\textwidth, trim=0 8mm 0 0]{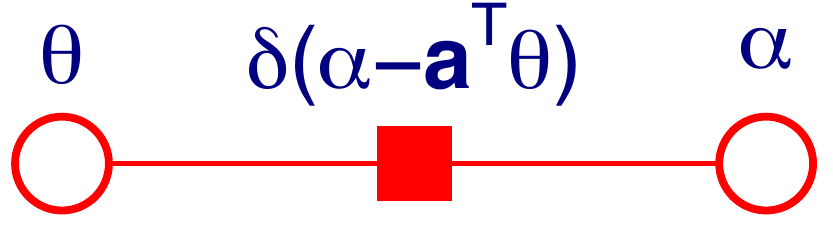}     &    
$\qquad\qquad\alpha\equiv\ba^T\btheta$   \\
\ \ \ \ derived variable       &          &                                              \\[3ex]
5. Multivariate linear combi-  &  \includegraphics[width=0.19\textwidth, trim=0 8mm 0 0]{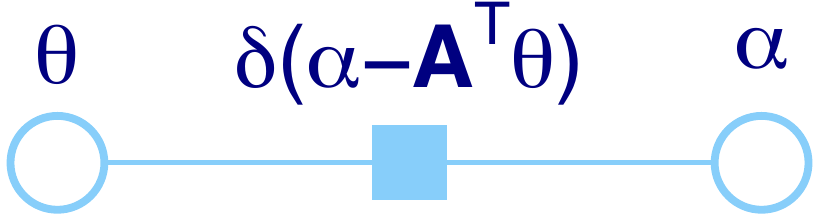}     &    
$\qquad\qquad\balpha\equiv\bA^T\btheta$   \\
\ \ \ \ nation derived variable       &          &                                              \\[3ex]
6. Gaussian      &  \includegraphics[width=0.19\textwidth, trim=0 8mm 0 0]{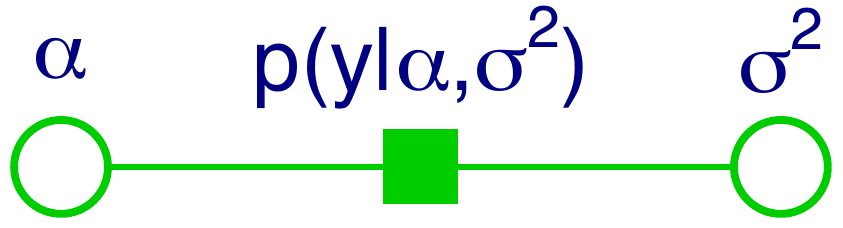}     &    
$\qquad\qquad y|\,\alpha,\sigma^2\sim N(\alpha,\sigma^2)$   \\[5ex]
7. Logistic likelihood       &  \includegraphics[width=0.15\textwidth, trim=0 8mm 0 0]{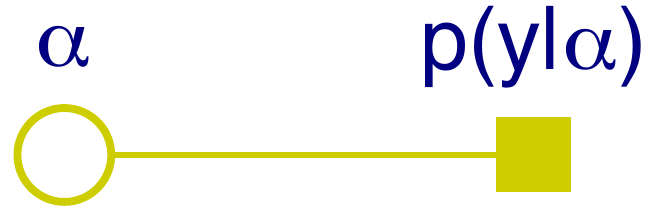}     &    
$\qquad\qquad y|\,\alpha\sim \mbox{Bernoulli}(\mbox{logit}^{-1}(\alpha))$   \\[5ex]
8. Probit likelihood       &  \includegraphics[width=0.15\textwidth, trim=0 8mm 0 0]{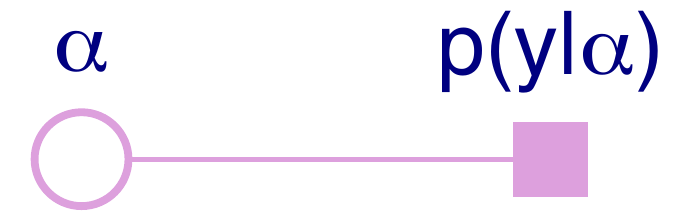}     &    
$\qquad\qquad y|\,\alpha\sim \mbox{Bernoulli}(\Phi(\alpha))$   \\[5ex]
9. Poisson likelihood       &  \includegraphics[width=0.15\textwidth, trim=0 8mm 0 0]{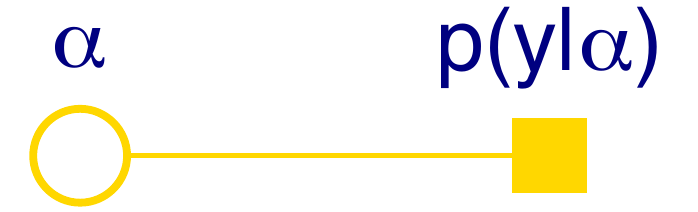}     &    
$\qquad\qquad y|\,\alpha\sim \mbox{Poisson}(\exp(\alpha))$   \\[5ex]
\hline
\end{tabular}
\end{footnotesize}
\end{center}
\caption{\it Fundamental factor graph fragments for expectation propagation fitting
of generalized, linear and mixed models.}
\label{tab:GLMMfrags} 
\end{table}

\subsection{Gaussian Prior Fragment}

The Gaussian prior fragment arises from the following prior distribution specification:
$$\btheta\sim N(\bmu_{\btheta},\bSigma_{\btheta})$$
for user-specified hyperparameters $\bmu_{\btheta}$ and $\bSigma_{\btheta}$.
The fragment factor is
$$p(\btheta)=(2\pi)^{-d_{\theta}/1}|\bSigma_{\btheta}|^{-1/2}
\exp\big\{-\smhalf(\btheta-\bmu_{\btheta})^T\bSigma_{\btheta}^{-1}
(\btheta-\bmu_{\btheta})\big\}.
$$
We assume that 
\begin{equation}
\begin{array}{l}
\mbox{all messages passed to $\btheta$ from factors outside of the}\\[0.1ex] 
\mbox{fragment are in the Multivariate Normal family.}\\[0.1ex] 
\end{array}
\label{eq:conjugacyGaussPrior}
\end{equation}
The message from $p(\btheta)$ to $\btheta$ takes the form
$$\mSUBpthetaTOtheta=\exp\left\{
\left[\begin{array}{c}
\btheta\\
\vecof(\btheta\btheta^T)
\end{array}
\right]^T
\etaSUBpthetaTOtheta\right\}.
$$
Algorithm \ref{alg:GaussianPriorFrag} provides the natural parameter update for
this simple fragment. The derivation of Algorithm \ref{alg:GaussianPriorFrag}
is given in Section \ref{sec:GaussPriorFragDeriv} of the online supplement.

\begin{algorithm}[!th]
\begin{center}
\begin{minipage}[t]{80mm}
\hrule
\jump
\textbf{Hyperparameter Inputs:} $\bmu_{\btheta},\bSigma_{\btheta}$.
\jump\noindent
\textbf{Update:}
\begin{itemize}
\item[] $\etaSUBpthetaTOtheta\thickarrow
\left[
\begin{array}{c}
\bSigma_{\btheta}^{-1}\bmu_{\btheta}\\[2ex]
-\smhalf\vecof(\bSigma_{\btheta}^{-1})
\end{array}
\right]
$
\end{itemize}
\jump
\textbf{Parameter Output:} $\etaSUBpthetaTOtheta$.
\jump
\hrule
\end{minipage}
\end{center}
\caption{\it The input, update and output of the Gaussian prior fragment.}
\label{alg:GaussianPriorFrag} 
\end{algorithm}

\subsection{Inverse Wishart Prior Fragment}

Let $\bTheta$ be a $\dTheta\times \dTheta$ symmetric
positive definite random matrix. The prior specification 
$$\bTheta\sim\mbox{Inverse-Wishart}\big(\kappaTheta,\LambdaTheta\big)$$
leads to a factor graph fragment with factor
$$p(\bTheta)=\Csc_{\dTheta,\kappaTheta}^{-1}|\LambdaTheta|^{\kappaTheta/2}\,
|\bTheta|^{-(\kappaTheta+\dTheta+1)/2}
\exp\{-\smhalf\tr(\LambdaTheta\bTheta^{-1} )\}\,I(\mbox{$\bTheta$ symmetric and positive definite}).
$$
where $\Csc_{\dTheta,\kappaTheta}$ is defined via (\ref{eq:CscDefn}).
The message from $p(\bTheta)$ to $\bTheta$ takes the form
$$\mSUBpThetaTOTheta=\exp\left\{      
\left[
\begin{array}{c}
\log|\bTheta|\\[1ex]
\vecof(\bTheta^{-1})
\end{array}
\right]^T\etaSUBpThetaTOTheta
\right\}.$$
Algorithm \ref{alg:InvWishartPriorFrag} gives the 
$\etaSUBpThetaTOTheta$ update based on hyperparameter
inputs $\kappaTheta$ and $\LambdaTheta$.

\begin{algorithm}[!th]
\begin{center}
\begin{minipage}[t]{80mm}
\hrule
\jump
\textbf{Hyperparameter Inputs:} $\kappaTheta,\LambdaTheta$.
\jump\noindent
\textbf{Update:}
\begin{itemize}
\item[] $\etaSUBpThetaTOTheta\thickarrow
\left[
\begin{array}{c}
-\smhalf(\kappaTheta+\dTheta+1)\\[2ex]
-\smhalf\vecof(\LambdaTheta)
\end{array}
\right]
$
\end{itemize}
\jump
\textbf{Parameter Output:} $\etaSUBpThetaTOTheta$.
\jump
\hrule
\end{minipage}
\end{center}
\caption{\it The input, update and output of the Inverse Wishart prior fragment.}
\label{alg:InvWishartPriorFrag} 
\end{algorithm}

A derivation of Algorithm \ref{alg:InvWishartPriorFrag} is given in 
Section \ref{sec:InvWishartFragDeriv} of the online supplement.

\subsection{Iterated Inverse Chi-Squared Fragment}

This fragment arises from the following distributional fact
(e.g. Wand {\it et al.} (2011), Result 5):
\begin{equation}
\begin{array}{c}
\sigma^2|a\sim\mbox{Inverse-$\chi^2$}(\nu,\nu/a)\quad\mbox{and}\quad
a\sim\mbox{Inverse-$\chi^2$}(1,1/A^2)\\[1ex]
\mbox{implies}\quad\sigma\sim\mbox{Half-t}(A,\nu)
\end{array}
\label{eq:Halft}
\end{equation}
where $x\sim\mbox{Half-t}(A,\nu)$ if and only if 
$$p(x)=\frac{2\Gamma\left(\frac{\nu+1}{2}\right)I(x>0)}{\sqrt{\pi\nu}\,\Gamma(\nu/2)A\{1+(x/A)^2/\nu\}^{(\nu+1)/2}}.$$
The advantage of fact (\ref{eq:Halft}) is that non-informative priors within the
Half-$t$ family can be imposed on standard deviation parameters using messages
within the Inverse Chi-Squared family.

The fragment factor is
$$p(\sigma^2|a)=\frac{\{\nu/(2a)\}^{\nu/2}}{\Gamma(\nu/2)}\,(\sigma^2)^{-(\nu/2)-1}\exp\{-\nu/(2a\sigma^2)\}\,
I(\sigma^2>0)\,I(a>0)$$
and it is assumed that:
\begin{equation}
\begin{array}{l}
\mbox{all messages passed to $\sigma^2$ from factors outside of}\\[0.1ex] 
\mbox{the fragment are in the Inverse Chi-Squared family and}\\[0.1ex] 
\mbox{all messages passed to $a$ from factors outside of}\\[0.1ex] 
\mbox{the fragment are also in the Inverse Chi-Squared family.} 
\end{array}
\label{eq:conjugacyIterIG}
\end{equation}
The messages from the factor to its neighboring stochastic nodes are
$$\mSUBpsigsqaTOsigsq=\exp\left\{\left[    
\begin{array}{c}
\log(\sigma^2)\\[1ex]
1/\sigma^2
\end{array}
\right]^T
\etaSUBpsigsqaTOsigsq
\right\}
$$
and
$$\mSUBpsigsqaTOa=\exp\left\{\left[    
\begin{array}{c}
\log(a)\\[1ex]
1/a
\end{array}
\right]^T
\etaSUBpsigsqaTOa
\right\}.
$$
Algorithm \ref{alg:iterInvChisqFrag} provides the updates of the 
natural parameters of these messages given messages from outside
the fragment. The function $G^{\mbox{\tiny IG3}}$ is defined
in Section \ref{sec:explicitFns}.

\begin{algorithm}[!th]
\begin{center}
\begin{minipage}[t]{150mm}
\hrule
\jump
\textbf{Data Input:} $\nu>0$,\ \ \ $0\le\varepsilon<1$.
\jump\noindent
\textbf{Parameter Inputs:} $\etaSUBpsigsqaTOsigsq$, $\etaSUBpsigsqaTOa$,
$\etaSUBsigsqTOpsigsqa$, $\etaSUBaTOpsigsqa$.
\jump\noindent
\textbf{Updates:}
\begin{itemize}
\item[] $\etaSUBpsigsqaTOsigsq\thickarroweps G^{\mbox{\tiny IG3}}
\left(\etaSUBsigsqTOpsigsqa,\etaSUBaTOpsigsqa;\nu+2,\nu\right)$
\item[] $\etaSUBpsigsqaTOa\thickarroweps G^{\mbox{\tiny IG3}}
\left(\etaSUBaTOpsigsqa,\etaSUBsigsqTOpsigsqa;\nu,\nu\right)$
\end{itemize}
\jump
\textbf{Parameter Outputs:} $\etaSUBpsigsqaTOsigsq$, $\etaSUBpsigsqaTOa$.
\jump
\end{minipage}
\hrule
\end{center}
\caption{\it The inputs, updates and outputs of the iterated Inverse Chi-Squared fragment.}
\label{alg:iterInvChisqFrag} 
\end{algorithm}

\subsection{Linear Combination Derived Variable Fragment}

The  linear combination derived variable fragment corresponds to equating
a scalar variable $\alpha$ with a linear combination $\ba^T\btheta$.
If $g$ is a general function that depends on the linear combination form
$\ba^T\btheta$ and other variables, denoted by $\bo$, then the derived
variable $\alpha$ arises from the equality:
\begin{equation}
g(\ba^T\btheta;\bo)=\infint\delta(\alpha-\ba^T\btheta)\,g(\alpha;\bo)\,d\alpha.
\label{eq:gLinComb}
\end{equation}
where $\delta$ is the Dirac delta function.
Under Convention 1 given in Section \ref{sec:expecPropag}, the integral sign is ignored 
when it comes to applying the expectation propagation updates (\ref{eq:stochToFac})
and (\ref{eq:facToStoch}). We assume that: 
\begin{equation}
\begin{array}{l}
\mbox{all messages passed to $\alpha$ from factors outside of}\\[0.1ex] 
\mbox{the fragment are in the Univariate Normal family}\\[0.1ex] 
\mbox{and all messages passed to $\btheta$ from factors outside}\\[0.1ex] 
\mbox{of the fragment are in the Multivariate Normal family.} 
\end{array}
\label{eq:conjugacyLinComb}
\end{equation}
The function $\delta(\alpha-\ba^T\btheta)$ is the factor for this fragment.
According to conjugacy restrictions, messages passed from $\delta(\alpha-\ba^T\btheta)$
to $\alpha$ and $\btheta$ take the forms
\begin{equation}
\mSUBdeltaTOalpha=\exp\left\{
\left[\begin{array}{c}
\alpha\\
\alpha^2
\end{array}
\right]^T\etaSUBdeltaTOalpha
\right\}
\label{eq:alphaMsgScalar}
\end{equation}
and
$$\mSUBdeltaTOtheta=\exp\left\{
\left[\begin{array}{c}
\btheta\\
\vecof(\btheta\btheta^T)
\end{array}
\right]^T\etaSUBdeltaTOtheta
\right\}.
$$
Algorithm \ref{alg:LinCombFrag} provides the updates to the natural parameter
vectors
$$\etaSUBdeltaTOalpha\quad\mbox{and}\quad\etaSUBdeltaTOtheta$$
given inputs
$$\etaSUBalphaTOdelta\quad\mbox{and}\quad\etaSUBthetaTOdelta.$$
It uses the notation
{\setlength{\arraycolsep}{1pt}
\begin{eqnarray*}
(\etaSUBthetaTOdelta\Big)_1&\equiv&\mbox{vector containing the first $d^{\btheta}$ entries of $\etaSUBthetaTOdelta$}\\
\mbox{and}\ 
(\etaSUBthetaTOdelta\Big)_2&\equiv&\mbox{vector containing the remaining $(d^{\btheta})^2$ entries of $\etaSUBthetaTOdelta$}\\
\end{eqnarray*}
}
where $d^{\btheta}$ is the number of entries in $\btheta$.
The derivations of these updates are given in Section \ref{sec:MultLinCombFragDeriv}
of the online supplement.

\begin{algorithm}[!th]
\begin{center}
\begin{minipage}[t]{150mm}
\hrule
\jump
\textbf{Data Input:} $\ba$ (vector having the same dimension as $\btheta$),\ \ \ $0\le \varepsilon<1$.
\jump\noindent
\textbf{Parameter Inputs:} $\etaSUBdeltaTOalpha$, $\etaSUBdeltaTOtheta$, $\etaSUBalphaTOdelta$, $\etaSUBthetaTOdelta$.
\jump\noindent
\textbf{Updates:}
\begin{itemize}
\item[]$\omegaONE\thickarrow\,\Big\{\vecof^{-1}\Big(\Big(\etaSUBthetaTOdelta\Big)_2\Big)\Big\}^{-1}\ba$
\item[]$\etaSUBdeltaTOalpha\thickarroweps\displaystyle{\frac{1}{\omegaONE^T\ba}}\left[
\begin{array}{c}   
\omegaONE^T\Big(\etaSUBthetaTOdelta\Big)_1\\[2ex]
1
\end{array}
\right]$
\item[]$\etaSUBdeltaTOtheta\thickarroweps
\left[
\begin{array}{c}
\ba\,\Big(\etaSUBalphaTOdelta\Big)_1\\[2ex]
\vecof(\ba\ba^T)\,\Big(\etaSUBalphaTOdelta\Big)_2
\end{array}
\right]$
\end{itemize}
\jump
\textbf{Parameter Outputs:} $\etaSUBdeltaTOalpha,\etaSUBdeltaTOtheta$.
\jump
\end{minipage}
\hrule
\end{center}
\caption{\it The inputs, updates and outputs of the linear combination derived variable fragment.}
\label{alg:LinCombFrag} 
\end{algorithm}

\subsection{Multivariate Linear Combination Derived Variable Fragment}

Now consider the following bivariate extension of (\ref{eq:gLinComb}):
\begin{equation}
g\big(\ba_1^T\btheta,\ba_2^T\btheta;\bo\big)=
\infint\infint\delta(\alpha_1-\ba_1^T\btheta)\delta(\alpha_2-\ba_2^T\btheta)
\,g(\alpha_1,\alpha_2;\bo)\,d\alpha_1\,d\alpha_2,
\label{eq:coppDing}
\end{equation}
where the primary argument of the function $g$ is now bivariate. 
The established result for the Dirac delta function
applied to bivariate arguments leads to the equivalent form
for the right-hand side of (\ref{eq:coppDing}) taking the form:
$$\infint\infint\delta\left(\left[\begin{array}{c}
\alpha_1\\
\alpha_2
\end{array}
\right]-\bA^T\btheta\right)
\,g(\alpha_1,\alpha_2;\bo)\,d\alpha_1\,d\alpha_2
\quad\mbox{where}\quad\bA\equiv[\ba_1\ \ba_2].$$
It follows that
$$\balpha\equiv\left[   
\begin{array}{c}
\alpha_1\\
\alpha_2
\end{array}
\right]
$$
is a bivariate derived variable corresponding to the multivariate linear 
combination $\bA^T\btheta$.

In the most general case, $\btheta$ and $\balpha$ are, respectively, $\dtheta\times1$
and $\dalpha\times1$ vectors and $\bA$ is a $\dtheta\times\dalpha$
matrix. The fragment factor is $\delta(\balpha-\bA^T\btheta)$ and the 
message given in (\ref{eq:alphaMsgScalar}) generalizes to 
$$
\mSUBbdeltaTObalpha=\exp\left\{
\left[\begin{array}{c}
\balpha\\
\vecof(\balpha\balpha^T)
\end{array}
\right]^T\etaSUBbdeltaTObalpha
\right\}.
$$
Algorithm \ref{alg:MultLinCombFrag} lists the natural parameter updates.
Their derivations are given in Section \ref{sec:MultLinCombFragDeriv}
of the online supplement.

\begin{algorithm}[!th]
\begin{center}
\begin{minipage}[t]{150mm}
\hrule
\jump
\textbf{Data Input:} $\bA$ (matrix with number of columns matching the dimension of $\btheta$),\ \ \ 
$0\le\varepsilon<1$.
\jump\noindent
\textbf{Parameter Inputs:} $\etaSUBbdeltaTObalpha$,$\,\etaSUBbdeltaTOtheta$, 
$\,\etaSUBbalphaTObdelta$,$\,\etaSUBthetaTObdelta$.
\jump\noindent
\textbf{Updates:}
\begin{itemize}
\item[]$\bOmega\thickarrow\,\Big\{\vecof^{-1}\Big(\Big(\etaSUBthetaTObdelta\Big)_2\Big)\Big\}^{-1}\bA$
\item[]$\etaSUBbdeltaTObalpha\thickarroweps\left[
\begin{array}{c}   
(\bOmega^T\bA)^{-1}\bOmega^T\Big(\etaSUBthetaTObdelta\Big)_1\\[2ex]
\vecof\big((\bOmega^T\bA)^{-1}\big)
\end{array}
\right]$
\item[]$\etaSUBbdeltaTOtheta\thickarroweps
\left[
\begin{array}{c}
\bA\,\Big(\etaSUBbalphaTObdelta\Big)_1\\[2ex]
(\bA\otimes\bA)\,\Big(\etaSUBbalphaTObdelta\Big)_2
\end{array}
\right]$
\end{itemize}
\jump
\textbf{Parameter Outputs:} $\etaSUBbdeltaTObalpha,\etaSUBbdeltaTOtheta$.
\jump
\end{minipage}
\hrule
\end{center}
\caption{\it The inputs, updates and outputs of the multivariate
linear combination derived variable fragment.}
\label{alg:MultLinCombFrag} 
\end{algorithm}

Note that Algorithm \ref{alg:MultLinCombFrag} is a generalization of Algorithm \ref{alg:LinCombFrag}.
Therefore, from a strict mathematical standpoint, Algorithm \ref{alg:LinCombFrag}. However,
since ordinary linear combinations are common in expectation propagation fitting of 
linear models we feel that it is worth having a separate fragment and algorithm for this special case.

\subsection{Gaussian Fragment}\label{sec:GaussLik}

The Gaussian fragment corresponds to the
specification
$$y|\alpha,\sigma^2\sim N(\alpha,\sigma^2).$$
The fragment's factor is 
$$p(y|\alpha,\sigma^2)=(2\pi\sigma^2)^{-1/2}\,\exp\{-(y-\alpha)^2/(2\sigma^2)\}$$
which, as a function of $\alpha$, is in the Normal family and, as a function of $\sigma^2$, is
in the Inverse Chi-Squared family. Exponential family constraint considerations
then lead to the following assumption for the Gaussian fragment:
\begin{equation}
\begin{array}{l}
\mbox{all messages passed to $\alpha$ from factors outside of}\\[0.1ex] 
\mbox{the fragment are in the Univariate Normal family}\\[0.1ex] 
\mbox{and all messages passed to $\sigma^2$ from factors outside}\\[0.1ex] 
\mbox{of the fragment are in the Inverse Chi-Squared family.} 
\end{array}
\label{eq:conjugacyGaussLik}
\end{equation}
The messages from $p(y|\alpha,\sigma^2)$ take the forms 
$$\mSUBpyalphasigsqTOalpha=
\exp\left\{
\left[   
\begin{array}{c}
\alpha\\[1ex]
\alpha^2
\end{array}
\right]^T
\etaSUBpyalphasigsqTOalpha       
\right\}
$$
and
$$
\mSUBpyalphasigsqTOsigsq=
\exp\left\{
\left[   
\begin{array}{c}
\log(\sigma^2)\\[1ex]
1/\sigma^2
\end{array}
\right]^T
\etaSUBpyalphasigsqTOsigsq       
\right\}
$$
with natural parameters updated according to Algorithm \ref{alg:GaussFrag}.
The functions $G^{\mbox{\tiny N}}$ and $G^{\mbox{\tiny IG3}}$ are defined in Section \ref{sec:explicitFns}.
Algorithm \ref{alg:GaussFrag}'s derivation is given in Section \ref{sec:GaussLikFragDeriv}.

\begin{algorithm}[!th]
\begin{center}
\begin{minipage}[t]{150mm}
\hrule
\jump
\textbf{Data Input:} $y\in\realnos$,\ \ \ $0\le\varepsilon<1$.
\jump\noindent
\textbf{Parameter Inputs:} $\etaSUBpyalphasigsqTOalpha$,
$\etaSUBpyalphasigsqTOsigsq$, $\etaSUBalphaTOpyalphasigsq$, $\etaSUBsigsqTOpyalphasigsq$.
\jump\noindent
\textbf{Update:}
\begin{itemize}
\item[]$\etaSUBpyalphasigsqTOalpha\thickarroweps G^{\mbox{\tiny N}}
\left(\etaSUBalphaTOpyalphasigsq,\etaSUBsigsqTOpyalphasigsq;
       \left[\begin{array}{l}1\\ y \\ y^2 \end{array}\right]\right)$
\item[]$\etaSUBpyalphasigsqTOsigsq\thickarroweps G^{\mbox{\tiny IG1}}
\left(\etaSUBsigsqTOpyalphasigsq,\etaSUBalphaTOpyalphasigsq;
       \left[\begin{array}{l}1\\ y \\ y^2 \end{array}\right]\right)$
\end{itemize}
\jump
\textbf{Parameter Outputs:} $\etaSUBpyalphasigsqTOalpha$, $\etaSUBpyalphasigsqTOsigsq$.
\jump
\end{minipage}
\hrule
\end{center}
\caption{\it The inputs, updates and outputs of the Gaussian fragment.}
\label{alg:GaussFrag} 
\end{algorithm}

\subsection{Logistic Likelihood Fragment}

The logistic likelihood fragment corresponds to the specification
$$y|\alpha\sim\mbox{Bernoulli}\big\{\logit^{-1}(\alpha)\big\}.$$
The factor of the fragment is 
$$p(y|\alpha)=\exp\{y\alpha-\log(1+e^{\alpha})\}.$$
We assume that: 
\begin{equation}
\begin{array}{l}
\mbox{all messages passed to $\alpha$ from other factors are}\\[0.1ex]
\mbox{within the Univariate Normal exponential family.} 
\end{array}
\label{eq:logistConjugAss}
\end{equation}
Conjugacy then dictates that 
\begin{equation}
\mSUBpyalphaTOalpha=\exp\left\{
\left[
\begin{array}{c}
\alpha\\
\alpha^2
\end{array}
\right]^T
\etaSUBpyalphaTOalpha
\right\}.
\label{eq:msgFormLogist}
\end{equation}
Algorithm \ref{alg:logisticFrag} provides the update to the natural parameter vector
$$\etaSUBpyalphaTOalpha\quad\mbox{based on input}\quad\etaSUBalphaTOpyalpha$$
and depends on the function $\Hlogistic$ defined at (\ref{eq:HlogPoiDefns})
in the online supplement.

Its derivation is given in Section \ref{sec:logisticFragDeriv} of the
online supplement.

\begin{algorithm}[!th]
\begin{center}
\begin{minipage}[t]{80mm}
\hrule
\jump
\textbf{Data Input:} $y\in\{0,1\}$,\ \ \ $0\le\varepsilon<1$.
\jump\noindent
\textbf{Parameter Inputs:} $\etaSUBpyalphaTOalpha$, $\etaSUBalphaTOpyalpha$.
\jump\noindent
\textbf{Update:}
\begin{itemize}
\item[]$\etaSUBpyalphaTOalpha\thickarroweps\Hlogistic(\etaSUBalphaTOpyalpha;y)$
\end{itemize}
\jump
\textbf{Parameter Output:} $\etaSUBpyalphaTOalpha$.
\jump
\hrule
\end{minipage}
\end{center}
\caption{\it The inputs, updates and outputs of the logistic likelihood fragment.}
\label{alg:logisticFrag} 
\end{algorithm}

\subsection{Probit Likelihood Fragment}

The probit likelihood fragment corresponds to the specification
$$y|\alpha\sim\mbox{Bernoulli}\big(\Phi(\alpha)\big).$$
The factor of the fragment is 
$$p(y|\alpha)=\exp\big[y\log\{\Phi(\alpha)\}+(1-y)\log\{1-\Phi(\alpha)\}\big].$$
As for the logistic likelihood fragment, we also assume (\ref{eq:logistConjugAss})
which implies that $\mSUBpyalphaTOalpha$ also takes the form (\ref{eq:msgFormLogist}).
The fragment update is given in Algorithm \ref{alg:probitFrag}, with
justification deferred to Section \ref{sec:probitFragDeriv} of the online supplement.
The function $\Hprobit$ is defined in Section \ref{sec:explicitFns} of the online
supplement. Note that $\Hprobit$ has the advantage of admitting 
a closed form expression. This is not the case for $\Hlogistic$ and numerical integration
is required for its evaluation.

\begin{algorithm}[!th]
\begin{center}
\begin{minipage}[t]{80mm}
\hrule
\jump
\textbf{Data Input:} $y\in\{0,1\}$,\ \ \ $0\le\varepsilon<1$.
\jump\noindent
\textbf{Parameter Inputs:} $\etaSUBpyalphaTOalpha$, $\etaSUBalphaTOpyalpha$.
\jump\noindent
\textbf{Update:}
\begin{itemize}
\item[]$\etaSUBpyalphaTOalpha\thickarroweps\Hprobit(\etaSUBalphaTOpyalpha;y)$
\end{itemize}
\jump
\textbf{Parameter Outputs:} $\etaSUBpyalphaTOalpha$.
\jump
\hrule
\end{minipage}
\end{center}
\caption{\it The inputs, updates and outputs of the probit likelihood fragment.}
\label{alg:probitFrag} 
\end{algorithm}

\subsection{Poisson Likelihood Fragment}

The Poisson likelihood fragment matches
$$y|\alpha\sim\mbox{Poisson}\big\{\exp(\alpha)\big\}$$
and the factor of the fragment is 
$$p(y|\alpha)=\exp\{y\alpha-e^{\alpha}-\log(y!)\}.$$
As for the logistic and Poisson likelihood fragments, we also assume (\ref{eq:logistConjugAss})
which implies that $\mSUBpyalphaTOalpha$ also takes the form (\ref{eq:msgFormLogist}).
The fragment update is given in Algorithm \ref{alg:PoissonFrag} with the
$\HPoisson$ function defined at (\ref{eq:HlogPoiDefns})

Section \ref{sec:PoissonFragDeriv} of the online supplement 
contains justification of Algorithm \ref{alg:PoissonFrag}.

\begin{algorithm}[!th]
\begin{center}
\begin{minipage}[t]{80mm}
\hrule
\jump
\textbf{Data Input:} $y\in\{0,1,2,\ldots\}$,\ \ \ $0\le\varepsilon<1$.
\jump\noindent
\textbf{Parameter Inputs:} $\etaSUBpyalphaTOalpha$, $\etaSUBalphaTOpyalpha$.
\jump\noindent
\textbf{Update:}
\begin{itemize}
\item[]$\etaSUBpyalphaTOalpha\thickarroweps\HPoisson(\etaSUBalphaTOpyalpha;y)$
\end{itemize}
\jump
\textbf{Parameter Outputs:} $\etaSUBpyalphaTOalpha$.
\jump
\hrule
\end{minipage}
\end{center}
\caption{\it The inputs, updates and outputs of the Poisson likelihood fragment.}
\label{alg:PoissonFrag} 
\end{algorithm}

\section{Illustration}\label{sec:illustration}

We now provide illustration via a generalized additive mixed model analysis.
The data are from the Indonesian Children's Health Study (Sommer, 1982),
corresponding to a cohort of $275$ Indonesian children who are repeatedly
examined. The response variable is 
\begin{equation}
y_{ij}=\left\{\begin{array}{l}
1,\quad\mbox{if respiratory infection present in the $i$th child at the $j$th examination,}\\
0,\quad\mbox{otherwise}
\end{array}
\right.
\label{eq:yDefnIndon}
\end{equation}
for $1\le i\le m$ and $1\le j\le n_i$. For these data note that $m=275$
and the $n_i\in\{1,\ldots,6\}$.
Potential predictor variables are age, indicator of vitamin A deficiency,
indicator of being female, height, indicator of being stunted and 
indicators for the number of clinic visits for each child.
We let $a_{ij}$ denote the age in years of the $i$th child at the
$j$th examination. 
Consider the following Bayesian generalized additive mixed model:
\begin{equation}
\begin{array}{c}
y_{ij}|\beta_0,\bbeta_{\bx},\betaspl,\bugrp,\buspl\simind\mbox{Bernoulli}\Big(
\mbox{logit}^{-1}\big(\beta_0+\ugrpi+\bbeta_{\bx}^T\bx_{ij}+f(a_{ij})\big)\Big),\\[3ex]
f(a_{ij})\equiv\betaspl\,a_{ij}+{\displaystyle\sum_{k=1}^K}\,\usplk\,z_k(a_{ij})
\ \mbox{is a low-rank smoothing spline in $a_{ij}$,}\\[2ex]
\mbox{where $\{z_k(\cdot):1\le k\le K\}$ is a suitable spline basis},\\[2ex]
\bbeta\equiv\left[
\begin{array}{c}   
\beta_0\\
\bbeta_{\bx}\\
\betaspl
\end{array}
\right]\sim N(\bmu_{\bbeta},\bSigma_{\bbeta}),\quad
\bu\equiv\left[
\begin{array}{c}   
\bugrp\\
\buspl
\end{array}
\right]
\Bigg|\sigmagrp^2,\sigmaspl^2
\sim
N\left(
\left[
\begin{array}{c}   
\bzero\\
\bzero
\end{array}
\right],
\left[
\begin{array}{cc}   
\sigmagrp^2\bI_m& \bzero\\
\bzero          & \sigmaspl^2\bI_K
\end{array}
\right]
\right),\\[4ex]
\sigmagrp^2|\agrp\sim\mbox{Inverse-$\chi^2$}(1,1/\agrp),\quad
\sigmaspl^2|\aspl\sim\mbox{Inverse-$\chi^2$}(1,1/\aspl),\\[3ex]
\agrp\sim\mbox{Inverse-$\chi^2$}(1,1/\agrp^2),\quad
\aspl\sim\mbox{Inverse-$\chi^2$}(1,1/\Aspl^2).
\end{array}
\label{eq:indonModel}
\end{equation}
The `grp' and `spl' subscripting indicates whether the random effect vector
and corresponding variance parameter is for the random subject intercept
or for spline coefficients in the non-linear function of age. 
Let $\by$ denote the $N\times1$ vector containing the $y_{ij}$, where
$N\equiv\sum_{i=1}^mn_i$. Despite the common use of double subscript notation
as in (\ref{eq:yDefnIndon}), it is more convenient 
to label the entries of $\by$ with a single subscript when it comes
to fitting via expectation propagation. To avoid a notational clash we use 
$y^s_{\ell}$, $1\le\ell\le N$, to denote the
$\ell$th entry of $\by$. Let $d^{\bbeta}$ be the number of rows in $\bbeta$.
For the Indonesian Children's Health Study Data application $d^{\bbeta}=11$.
Then let $\bX$ by the $N\times d^{\bbeta}$ matrix containing the 
predictor data. The random effects design matrix is $\bZ=[\bZgrp\ \bZspl]$
where 
$$\bZgrp\equiv\blockdiag{1\le i\le m}(\bone_{n_i})\quad\mbox{and}\quad
\bZspl\equiv\Big[\relstack{z_k(a_{ij})}{1\le k\le K}\Big]_{1\le j\le n_i,\ 1\le i\le m}.$$
Then the likelihood can be written as 
$$y^s_{\ell}|\bbeta,\bu\simind\mbox{Bernoulli}\Big(\mbox{logit}^{-1}\big((\bX\bbeta+\bZ\bu)_{\ell}\big)\Big),
\quad 1\le\ell\le N.$$
Next, let $\bC\equiv[\bX\ \bZ]$ so that 
$$\bX\bbeta+\bZ\bu=\bC\bbetabu$$
and let $\bc_{\ell}^T$ be the $\ell$th row of $\bC$. Let $\be_r$ be the $(m+K)\times1$
vector with $r$th entry equal to $1$ and zeroes elsewhere for $1\le r\le m+K$. 
Lastly, let $\bE_{d^{\bbeta}}$ be the $(d^{\bbeta}+m+K)\times d^{\bbeta}$ matrix with the 
$d^{\bbeta}\times d^{\bbeta}$ identity matrix at the top and all other entries equal to zero.
The joint density function of all random variables in the model is
\begin{equation}
{\setlength{\arraycolsep}{1pt}
\begin{array}{rcl}
&&p(\by,\bbeta,\bu,\sigmagrp^2,\sigmaspl^2,\agrp,\aspl)\\[2ex]
&&\quad=p(\by|\bbeta,\bu)p(\bbeta)p(\bu|\sigmagrp^2,\sigmaspl^2)p(\sigmagrp^2|\agrp)
p(\sigmaspl^2|\aspl)p(\agrp)p(\aspl)\\[2ex]
&&\quad=p(\bbeta)\left\{{\displaystyle\prod_{\ell=1}^N}
p(y_\ell^s|\bbeta,\bu)\right\}\left\{{\displaystyle\prod_{i=1}^m}
p(\ugrpi|\sigmagrp^2)\right\}
\left\{{\displaystyle\prod_{k=1}^K}p(\usplk|\sigmaspl^2)\right\}p(\sigmagrp^2|\agrp)\\[0ex]
&&\qquad\qquad\times\,p(\sigmaspl^2|\aspl)p(\agrp)p(\aspl)\\[2ex]
&&\quad=\left\{{\displaystyle\int_{\realnos^{d^{\beta}}}}
p({\widetilde\bbeta})\,
\delta\left({\widetilde\bbeta}-\bE_{d^{\beta}}^T\bbetabu\right)\,
d{\widetilde\bbeta}\right\}
\left\{\displaystyle{\prod_{\ell=1}^N}\displaystyle{\infint} 
p(y_{\ell}^s|\,\alpha_{\ell})\delta\left(\alpha_{\ell}-\bc_{\ell}^T\bbetabu\right)\,d\alpha_{\ell}\right\}\\[0ex]
&&\qquad\times\left\{\displaystyle{\prod_{i=1}^m}\displaystyle{\infint} 
p(\utilgrpi|\,\sigmagrp^2)\delta\left(\utilgrpi-\be_{d^{\bbeta}+i}^T\bbetabu\right)\,d\utilgrpi\right\}
p(\sigmagrp^2|\agrp)p(\agrp)\\[3ex]
&&\qquad\times\left\{\displaystyle{\prod_{k=1}^K}\displaystyle{\infint} 
p(\utilsplk|\,\sigmagrp^2)\delta\left(\utilsplk-\be_{d^{\bbeta}+m+k}^T\bbetabu\right)\,d\utilsplk\right\}
p(\sigmaspl^2|\aspl)p(\aspl).
\end{array}
}
\label{eq:indonFactn}
\end{equation}
Figure \ref{fig:indonRespirFacGraph} is the derived variable factor graph corresponding to the
representation of the joint density function given in (\ref{eq:indonFactn}). 
All of the fragments in Figure \ref{fig:indonRespirFacGraph}
are versions of fundamental fragments listed in Table \ref{tab:GLMMfrags}
\ifthenelse{\boolean{ColourVersion}}{and are color-coded and numbered accordingly.}{and are numbered accordingly.}
Expectation propagation inference for this model and data involves iteratively passing messages between neighboring
nodes on the Figure \ref{fig:indonRespirFacGraph} factor graph. The parameter updates
for the factor to stochastic node messages  are given by the relevant algorithms 
in Section \ref{sec:fragGLMM}. The stochastic node to factor message parameter updates
are a simple consequence of (\ref{eq:stochToFac}).

\ifthenelse{\boolean{ColourVersion}}
{
\begin{figure}[!ht]
\centering
{\includegraphics[width=0.88\textwidth]{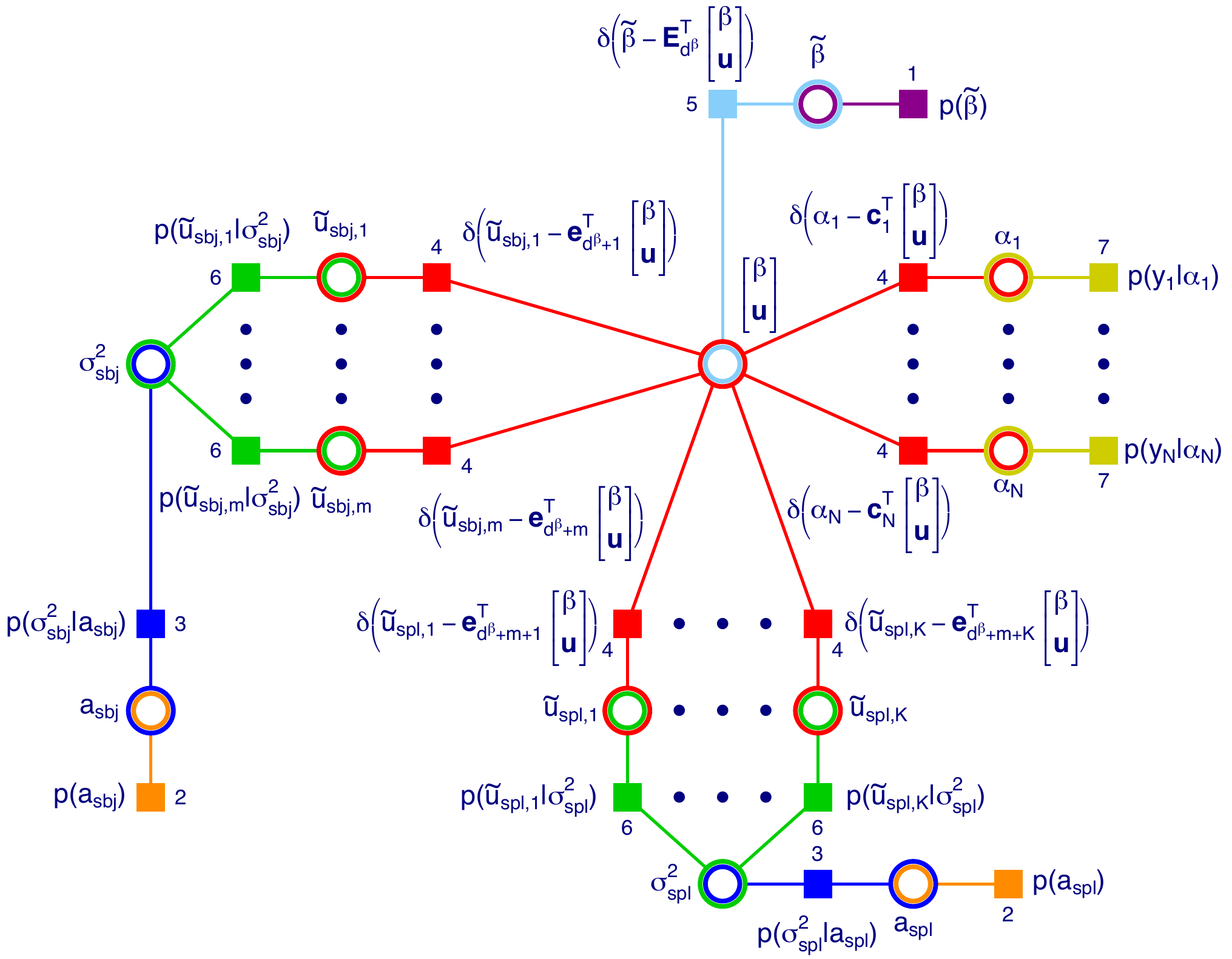}}
\caption{\it Derived variable factor graph corresponding to the representation of 
the joint density function of random variables in the generalized additive mixed model
(\ref{eq:indonModel}) given by (\ref{eq:indonFactn}).}
\label{fig:indonRespirFacGraph} 
\end{figure}
}
{
\begin{figure}[!ht]
\centering
{\includegraphics[width=0.88\textwidth]{indonRespirFacGraphBaW.pdf}}
\caption{\it Derived variable factor graph corresponding to the representation of 
the joint density function of random variables in the generalized additive mixed model
(\ref{eq:indonModel}) given by (\ref{eq:indonFactn}).}
\label{fig:indonRespirFacGraph} 
\end{figure}
}

\ifthenelse{\boolean{ColourVersion}}
{
\begin{figure}[!ht]
\centering
{\includegraphics[width=\textwidth]{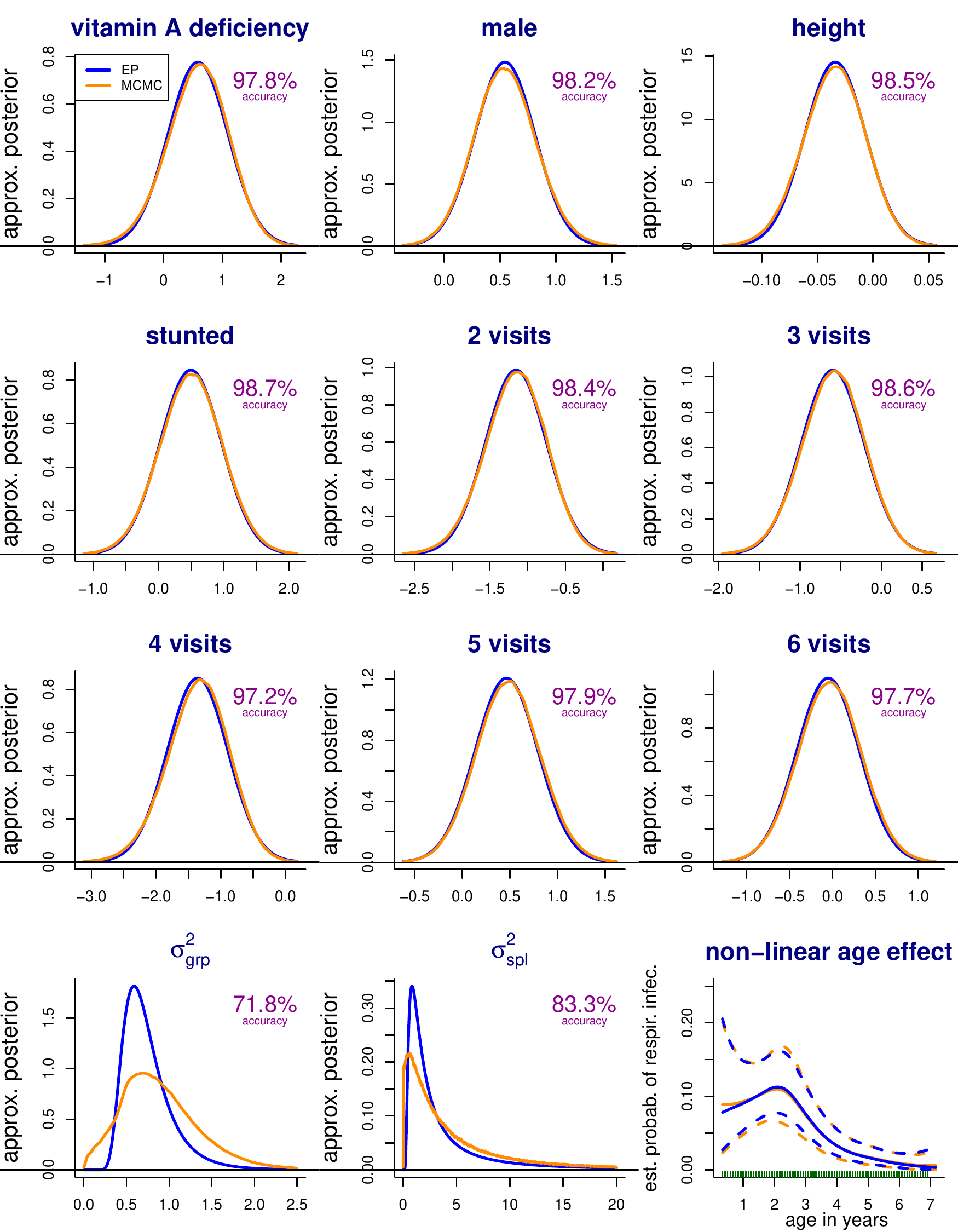}}
\caption{\it Comparison of two approximate Bayesian inference methods,
expectation propagation and Markov chain Monte Carlo,
for model (\ref{eq:indonModel}) applied to the Indonesian Children's Health Study Data.
The first three rows compares approximate posterior density functions for the
fixed effects parameters. The heading at the top of the panel 
is the corresponding predictor. The first two panels in the fourth row
compares approximate posterior density functions for the two variance parameters.
The accuracy percentages correspond to the definition at (\ref{eq:accDefn}).
The bottom right panel compares the low-rank smoothing spline fits for the
non-linear age effect on the probability of respiratory infection 
with all other predictors set to their averages. In this panel, 
the dashed curves indicate pointwise 95\% credible intervals and the 
tick marks show the age data. 
}
\label{fig:indonRespirCompar} 
\end{figure}
}
{
\begin{figure}[!ht]
\centering
{\includegraphics[width=\textwidth]{indonRespirComparBaW.pdf}}
\caption{\it Comparison of two approximate Bayesian inference methods,
expectation propagation and Markov chain Monte Carlo,
for model (\ref{eq:indonModel}) applied to the Indonesian Children's Health Study Data.
The first three rows compares approximate posterior density functions for the
fixed effects parameters. The heading at the top of the panel 
is the corresponding predictor. The first two panels in the fourth row
compares approximate posterior density functions for the two variance parameters.
The accuracy percentages correspond to the definition at (\ref{eq:accDefn}).
The bottom right panel compares the low-rank smoothing spline fits for the
non-linear age effect, with the dashed curves indicating pointwise 95\%
credible intervals and the tick marks showing the age data. 
}
\label{fig:indonRespirCompar} 
\end{figure}
}

We fit (\ref{eq:indonModel}) using 1,000 iterations of
expectation propagation message passing on the 
factor graph of Figure \ref{fig:indonRespirFacGraph}.
We also conducted Markov chain Monte Carlo fitting via the 
function \texttt{stan()} in the \textsf{R} package \textsf{rstan}
(Guo, Gabry \myand Goodrich, 2017), which interfaces the \textsf{Stan} language
(Carpenter \textit{et al.}, 2017), with a warmup size of 50,000 and a
retained sample size of 1,000,000.
The hyperparameters were set to 
$\bmu_{\bbeta}=\bzero$, $\bSigma_{\bbeta}=10^{10}\bI$, 
$\sigmagrp=\sigmaspl=10^5$ with continuous variables standardized for the 
analyses and then results transformed to correspond to the original units.
Figure \ref{fig:indonRespirCompar} compares the Bayesian inference 
arising from the two approaches. The first three rows compare
the expectation and Markov chain Monte Carlo approximate
posterior density functions for the fixed effects parameters.
The last row contains similar comparisons for the variance
parameters and the low-rank smoothing spline fits for the
non-linear age effect. The estimated probability functions
are such that all other predictors are set at their average
values, and are accompanied by pointwise 95\% credible intervals.

The posterior density function comparisons are accompanied by
accuracy percentages. For a generic parameter $\theta$, the
accuracy of the approximation $q(\theta)$ to the posterior
density function $p(\theta|\by)$ is given by
\begin{equation}
\mbox{accuracy}\equiv\,
100\left\{1-\smhalf\infint\big|q(\theta)-p(\theta|\by)\big|\,d\theta\right\}\%.
\label{eq:accDefn}
\end{equation}
The Markov chain Monte Carlo-based posterior density functions, as well
as the accuracy percentages on which they depend, are 
binned kernel density estimates obtained using the \textsf{R} function
\texttt{bkde()} in the package \textsf{KernSmooth} (Wand \myand Ripley, 2015)
with direct plug-in bandwidth selection via the function \texttt{dpik()}.
The density estimates should be very close to the actual posterior density
functions since they are  based on one million posterior draws.

We see from Figure \ref{fig:indonRespirCompar} that expectation propagation
achieves excellent accuracy for the fixed effect parameters, in keeping
with the simulation studies of Kim \myand Wand (2017). The variance
parameter posterior density estimates are not as good for this particular
example with accuracy scores of about 72\% and 83\%. Such mediocre accuracy
was not apparent in the Kim \myand Wand (2017) simulations although their
Figures 9 and 11 show accuracies for variance parameters being substantially 
lower than that those for fixed effect parameters. We ran the code that produced Figure 
\ref{fig:indonRespirCompar} on some simulated data and got accuracy
scores in the 85\%-95\% range for the variance parameters. Further
research is needed to gain a fuller understanding of the 
accuracy of expectation propagation in the generalized additive mixed model
context corresponding to this example.
 
\section{More Elaborate Expectation Propagation Fragments}\label{sec:elaborate}

The fragments listed in Table \ref{tab:GLMMfrags} and covered in Section \ref{sec:fragGLMM}
are the most fundamental ones for generalized, linear and mixed models. Whilst these fragments
support expectation propagation fitting of a wide range of models, additional 
fragments are needed for various elaborations. We now illustrate this fact by
investigating fragments needed for (a) the extension to multivariate random
effects, and (b) models where the response variable is modeled according to the $t$ distribution.
As we will see, expectation propagation is quite numerically challenging for 
such extensions.

\subsection{Multivariate Random Effects}

The fragments in Table \ref{tab:GLMMfrags} can handle the univariate random effects
structure 
$$u|\sigma^2\sim N(0,\sigma^2)$$
but they do not cover the \emph{multivariate} random effects extension:
$$\bu|\bSigma\sim N(\bzero,\bSigma)$$
where $\bSigma$ is a unstructured $\dbu\times\dbu$ matrix. 

\begin{figure}[h!]
\begin{center}
\includegraphics[width=0.3\textwidth]{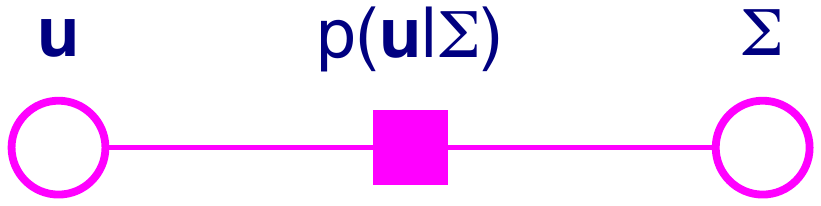}   
\caption{\it The factor graph fragment corresponding the factor $p(\bu|\bSigma)$.}
\label{fig:multRanEffFrag}
\end{center}
\end{figure}

The fragment corresponding to the factor 
$$p(\bu|\bSigma)=(2\pi)^{-\dbu/2}|\bSigma|^{-1/2}\exp(-\smhalf\,\bu^T\bSigma^{-1}\bu)$$
is shown in Figure \ref{fig:multRanEffFrag}.
%

Under the usual conjugacy constraints, the message from $p(\bu|\bSigma)$ to $\bSigma$ is 
\begin{equation}
\mSUBpuSigmaTOSigma=\frac{\projIW\big[\mSUBSigmaTOpuSigma\int_{\realnos^{\dbu}}p(\bu|\bSigma)
\mSUBuTOpuSigma\,d\bu/Z\big]}{\mSUBSigmaTOpuSigma}
\label{eq:IWmsg}
\end{equation}
where $\projIW$ denotes projection onto the $\dbu$-dimensional Inverse Wishart family.
The messages on the right-hand side of (\ref{eq:IWmsg}) have the form
$$\mSUBSigmaTOpuSigma=\exp\left\{
\left[
\begin{array}{c}
\log|\bSigma|\\[1ex]
\vecof(\bSigma^{-1})
\end{array}
\right]^T\etaSUBSigmaTOpuSigma
\right\}$$
and
$$\mSUBuTOpuSigma=\exp\left\{
\left[
\begin{array}{c}
\bu\\[1ex]
\vecof(\bu\bu^T)
\end{array}
\right]^T\etaSUBuTOpuSigma
\right\}
$$
Introducing the shorthand 
$$\bdeta^{\heartsuit}\equiv\etaSUBSigmaTOpuSigma\quad\mbox{and}\quad
\bdeta^{\diamondsuit}\equiv\etaSUBuTOpuSigma,
$$
arguments analogous to those given in Appendix \ref{alg:GaussFrag} lead to the function
of $\bSigma$ inside the $\projIW[\ \cdot\ ]$ in (\ref{eq:IWmsg}) being proportional to 
\begin{equation}
\begin{array}{rcl}
&&
|\bSigma|^{\eta_1^{\heartsuit}}|2\bSigma\vecof^{-1}(\bdeta_2^{\diamondsuit})-\bI|^{-1/2}
\mbox{tr}\{\bSigma^{-1}\vecof^{-1}(\bdeta_2^{\heartsuit})\}\\[1ex]
&&\qquad\times
\exp\Big[-\quarter(\bdeta_1^{\diamondsuit})^T\{2\bSigma\vecof^{-1}(\bdeta_2^{\diamondsuit})-\bI\}   
\{\vecof^{-1}(\bdeta_2^{\diamondsuit})\}^{-1}\bdeta_1^{\diamondsuit}\Big].
\end{array}
\label{eq:kernelSigma}
\end{equation}
The next step is to compute $E\{\log|\bSigma|\}$ and $E(\bSigma^{-1})$ with expectation
with respect to the density function obtained by normalizing (\ref{eq:kernelSigma}).
This is a particularly challenging numerical problem since it involves numerical
integration of the cone of $\dbu\times\dbu$ symmetric positive definite matrices.
Then 
\begin{equation}
\etaSUBpuSigmaTOSigma=(\nabla A)_{\tiny{\mbox{IW}}}^{-1}\left(
\left[\begin{array}{c}         
E\{\log|\bSigma|\}\\[2ex]
E\{\vech(\bSigma^{-1})\}
\end{array}
\right]\right).
\label{eq:etaTOSigma}
\end{equation}
Note that the function $(\nabla A)_{\tiny{\mbox{IW}}}$ admits the explicit form
$$(\nabla A)_{\tiny{\mbox{IW}}}\left(
\left[   
\begin{array}{c}
\eta_1\\
\bdeta_2
\end{array}
\right]\right)
=\left[   
\begin{array}{c}
\log\Big|-\vech^{-1}(\bdeta_2)\Big|-{\displaystyle\sum_{j=1}^{\dbu}}
\,\mbox{digamma}\big(-\eta_1-\smhalf(\dbu+j)\big)\\[4ex]
\{\eta_1+\smhalf(\dbu+1)\}\vech[\{\vech^{-1}(\bdeta_2)\}^{-1}]
\end{array}
\right]
$$
where $[\eta_1\ \bdeta_2^T]^T$ is the partition of the natural parameter vector 
into the first entry ($\eta_1$) and the remaining $\smhalf\dbu(\dbu+1)$ entries ($\bdeta_2$).
However, evaluation of (\ref{eq:etaTOSigma}) involves numerical inversion of 
$(\nabla A)_{\tiny{\mbox{IW}}}$ in $\{1+\smhalf\dbu(\dbu+1)\}$-dimensional space.

In conclusion, literal application of expectation propagation for multivariate
random effects is quite daunting and effective implementation for even $2\le\dbu\le5$
is a very challenging numerical problem.

\subsection{$t$ Likelihood}

The Gaussian fragment, treated in Section \ref{sec:GaussLik}, corresponds to the 
specification $y|\alpha,\sigma^2\sim N(\alpha,\sigma^2)$. Now consider the
extension to the $t$ distribution:
\begin{equation}
y|\alpha,\sigma^2,\nu\sim t(\alpha,\sigma^2,\nu)
\label{eq:tLik}
\end{equation}
where $\nu>0$ is the degrees of freedom parameter. Low values of $\nu$ correspond to heavy-tailed 
distributions. The Gaussian likelihood is the $\nu\to\infty$ limiting case. The density function
corresponding to (\ref{eq:tLik}) is
$$p(y|\alpha,\sigma^2,\nu)=\displaystyle{\frac{\Gamma\left(\frac{\nu+1}{2}\right)}
{\sqrt{\pi\nu}\Gamma(\nu/2)\{1+(y-\alpha)^2/(\nu\sigma^2)\}^{\frac{\nu+1}{2}}}}.$$
One could work with this density function in the expectation propagation message equations
(\ref{eq:stochToFac}) and (\ref{eq:facToStoch}), but trivariate numerical integration
is required. In other Bayesian computation contexts such 
as Markov chain Monte Carlo (e.g. Verdinelli \myand Wasserman, 1991)
and variational message passing (e.g. McLean \myand Wand, 2018) it is common to replace 
(\ref{eq:tLik}) by the auxiliary variable representation
\begin{equation}
y|\alpha,\sigma^2,a\sim N(\alpha,a\sigma^2),\quad a|\nu\sim\mbox{Inverse-$\chi^2$}(\nu,\nu)
\label{eq:tAuxRepn}
\end{equation}
to aid tractability. Expectation propagation also benefits from this representation of the $t$-likelihood
specification. The fragments corresponding to the factor product 
\begin{equation}
p(y|\alpha,\sigma^2,a)\,p(a|\nu)\,p(\nu)
\label{eq:tFragsProducForm}
\end{equation}
are shown in Figure \ref{fig:tFrags}.
%
\ifthenelse{\boolean{ColourVersion}}
{
\begin{figure}[h!]
\begin{center}
\includegraphics[width=0.90\textwidth]{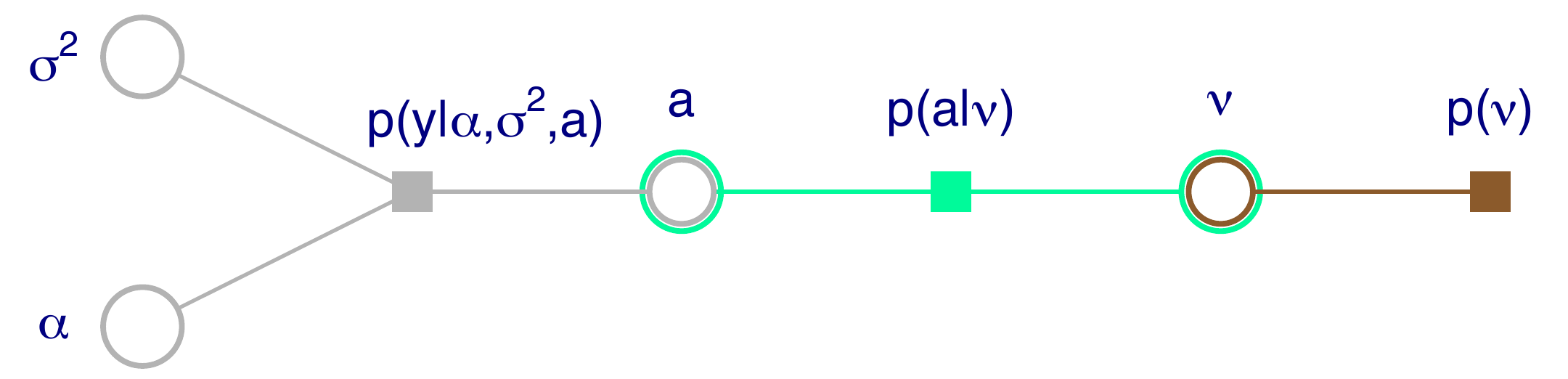}   
\caption{\it 
Color-coded fragments corresponding to the factors $p(y|\alpha,\sigma^2,a)$, 
$p(a|\nu)$ and $p(\nu)$ appearing in (\ref{eq:tFragsProducForm}).}
\label{fig:tFrags}
\end{center}
\end{figure}
}
{
\begin{figure}[h!]
\begin{center}
\includegraphics[width=0.90\textwidth]{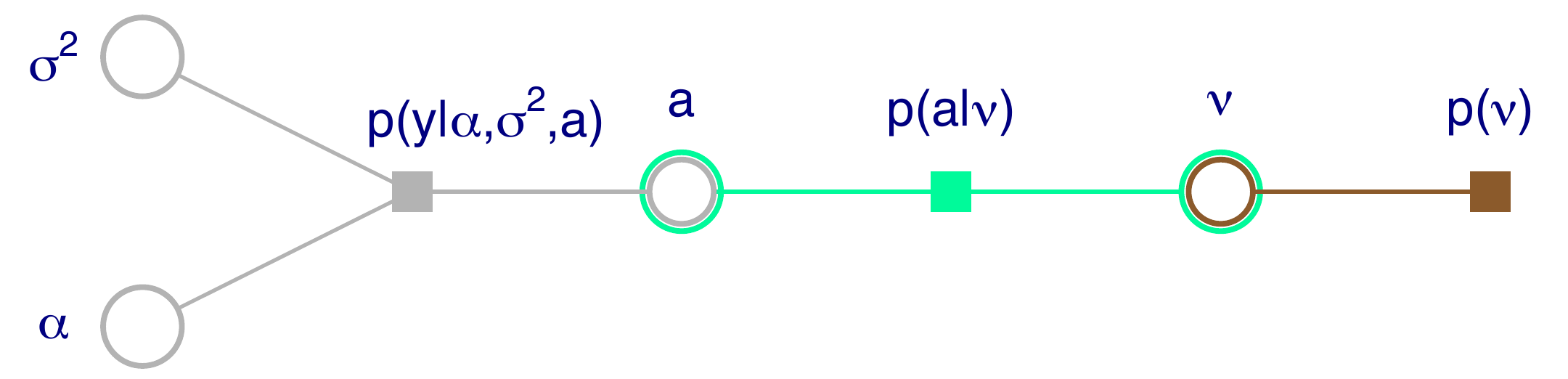}   
\caption{\it 
Fragments corresponding to the factors $p(y|\alpha,\sigma^2,a)$, 
$p(a|\nu)$ and $p(\nu)$ appearing in (\ref{eq:tFragsProducForm}).}
\label{fig:tFrags}
\end{center}
\end{figure}
}

None of these fragments are among those treated in Section \ref{sec:fragGLMM}.
Therefore, extension to $t$ likelihood models requires expectation propagation updates
for these three new fragments. Unfortunately, as we will see, difficult numerical challenges
arise for these updates. We now focus on each one in turn.

\subsubsection{The $p(y|\alpha,\sigma^2,a)$ Fragment}

The factor for this fragment is 
$$p(y|\alpha,\sigma^2,a)=(2\pi\,a\sigma^2)^{-1/2}\exp\left\{-\frac{(y-\alpha)^2}{2\,a\sigma^2}\right\}.$$
Conjugacy considerations dictate the assumption:
\begin{equation}
\begin{array}{l}
\mbox{all messages passed to $\alpha$ from factors outside of the}\\[0.1ex] 
\mbox{fragment are in the Univariate Normal family and all}\\[0.1ex] 
\mbox{messages passed to either $a$ of $\sigma^2$ from factors outside}\\[0.1ex] 
\mbox{of the fragment are in the Inverse Chi-Squared family.} 
\end{array}
\label{eq:conjugacytLik}
\end{equation}
This leads to the factor to stochastic node messages taking the forms: 
$$\mSUBpyalphasigsqaTOalpha=\exp
\left\{
\left[
\begin{array}{c}
\alpha\\
\alpha^2
\end{array}
\right]^T\etaSUBpyalphasigsqaTOalpha
\right\},
$$
$$\mSUBpyalphasigsqaTOsigsq=\exp
\left\{
\left[
\begin{array}{c}
\log(\sigma^2)\\
1/\sigma^2
\end{array}
\right]^T\etaSUBpyalphasigsqaTOsigsq
\right\}
$$
and
$$
\mSUBpyalphasigsqaTOa=\exp
\left\{
\left[
\begin{array}{c}
\log(a)\\
1/a
\end{array}
\right]^T\etaSUBpyalphasigsqaTOa
\right\}.
$$
The derivations of the natural parameter updates are similar in nature
to those given in Appendix \ref{sec:GaussLikFragDeriv} for the Gaussian
fragment. However, the form $a\sigma^2$ (rather than $\sigma^2$) in the
variance means that the natural parameter updates require evaluation of 
the bivariate integral-defined function
\begin{eqnarray*}
&&\Bsc_2(p,q_1,q_2,r_1,r_2,s,t,u)\equiv\\[1ex]
&&\qquad\quad\int_{-\infty}^{\infty}\int_{-\infty}^{\infty}
\frac{x_1^p\exp\{q_1x_1+q_2x_2-r_1e^{x_1}-r_2e^{x_2}-se^{x_1+x_2}/(t+e^{x_1+x_2})\}}
{(t+e^{x_1+x_2})^u}\,dx_1\,dx_2
\end{eqnarray*}
for
$$p\ge 0,\ q_1,q_2\in\realnos,\ r_1,r_2>0,\ s\ge0,\ t>0,\ u>0$$
rather than the univariate integral-defined function $\Bsc(p,q,r,s,t,u)$
given by (\ref{eq:AscAndBsc}). 
 
\subsubsection{The $p(a|\nu)$ Fragment}

\def\halfnu{\upsilon}

\begin{figure}[!ht]
\centering
{\includegraphics[width=0.98\textwidth]{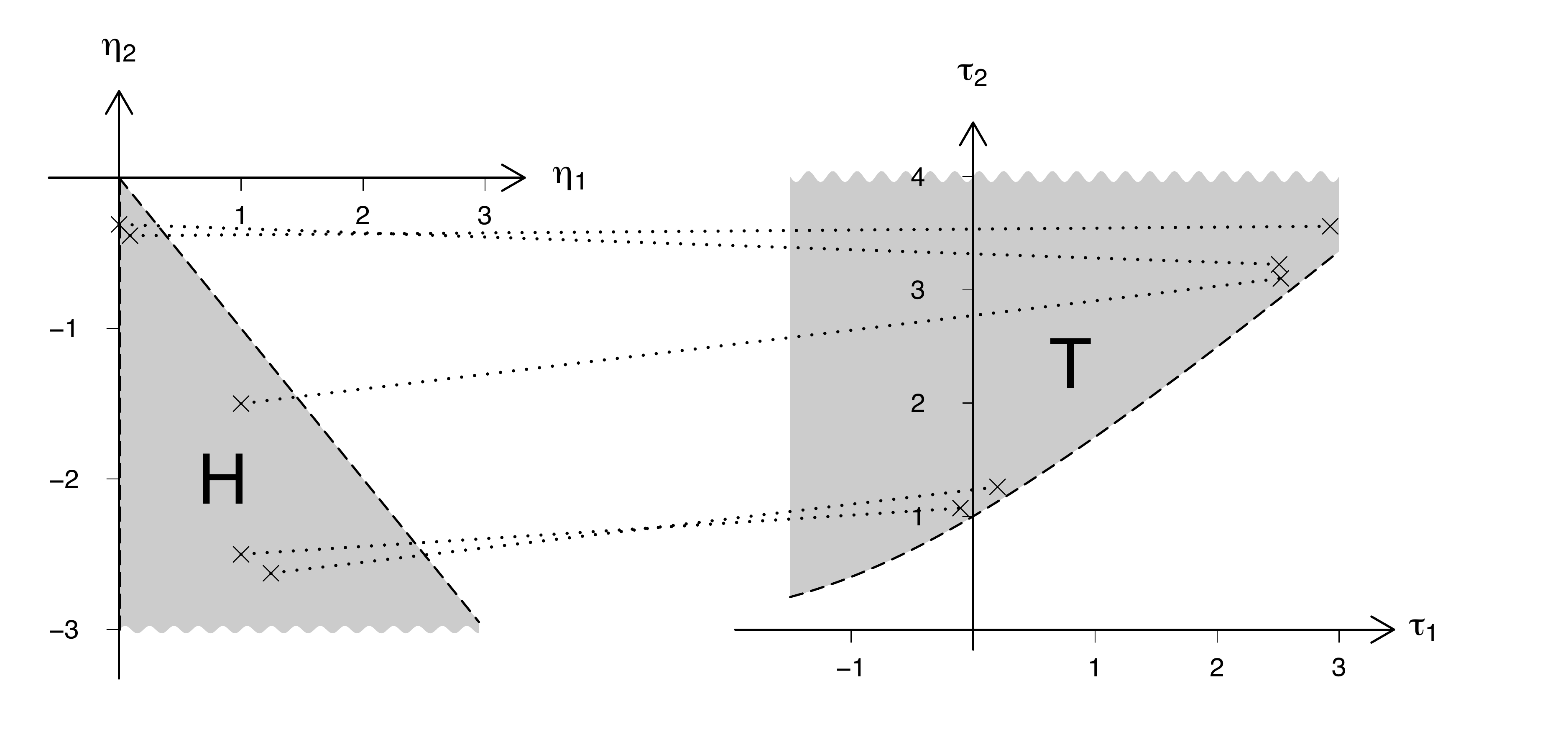}}
\caption{\it Illustration of the bijective maps between $H$ and $T$ for
the Moon Rock exponential family. The crosses and dotted lines depict five example $\bdeta\in H$ and 
$\btau=\nabla\AMR(\bdeta)\in T$ pairs. Since $\nabla\AMR$ is a bijective map, the crosses
and dotted lines equivalently depict five example $\btau\in T$ and $\bdeta=(\nabla\AMR)^{-1}(\btau)\in H$
pairs.}
\label{fig:mapPlotMoonRock} 
\end{figure}

The relevant factor is 
$$p(a|\nu)=\frac{(\nu/2)^{\nu/2}}{\Gamma(\nu/2)}\,a^{-(\nu/2)-1}\,\exp\{-(\nu/2)/a\},\quad a,\nu>0.$$
Let $\halfnu\equiv\nu/2$ be a simple linear transformation of $\nu$. For the remainder of this section
we work with $\halfnu$, rather than $\nu$, since it leads to a simpler exposition.
Now note that
$$p(a|\halfnu)\propto\left\{
\begin{array}{lcl}
\exp\left\{
\left[\begin{array}{c}      
\log(a)\\[2ex]
1/a
\end{array}
\right]^T
\left[\begin{array}{c}      
-\halfnu\\[2ex]
-\halfnu-1
\end{array}
\right]
\right\}&\quad&\mbox{as a function of $a$},\\[8ex]
\exp\left\{
\left[\begin{array}{c}      
\halfnu\log(\halfnu)-\log\{\Gamma(\halfnu)\}\\[2ex]
\halfnu
\end{array}
\right]^T
\left[\begin{array}{c}      
1\\[2ex]
-1/a-\log(a)
\end{array}
\right]
\right\}&\quad&\mbox{as a function of $\halfnu$.}
\end{array}
\right.$$
To ensure conjugacy we should then impose the restriction:
\begin{equation}
\begin{array}{l}
\mbox{all messages passed to $a$ from factors outside of the}\\[0.1ex] 
\mbox{fragment are in the Inverse Chi-Squared family and all}\\[0.1ex] 
\mbox{messages passed to either $\halfnu$ from factors outside}\\[0.1ex] 
\mbox{of the fragment are in the Moon Rock family.} 
\end{array}
\label{eq:conjugacytAux}
\end{equation}
The definition of the Moon Rock family is given in Table \ref{tab:exponFam}.
The messages passed from $p(a|\halfnu)$ are then of the form
$$\mSUBpahalfnuTOa=
\exp\left\{
\left[
\begin{array}{c}      
\log(a)\\[2ex]
1/a
\end{array}
\right]^T\etaSUBpahalfnuTOa\right\}
$$
and
$$
\mSUBpahalfnuTOhalfnu=\exp\left\{
\left[
\begin{array}{c}      
\halfnu\log(\halfnu)-\log\{\Gamma(\halfnu)\}\\[2ex]
\halfnu
\end{array}
\right]^T
\etaSUBpahalfnuTOhalfnu
\right\}.
$$
\noindent
The message $\mSUBpahalfnuTOa$ has a treatment similar to that for $\mSUBpsigsqaTOsigsq$
and $\mSUBpsigsqaTOa$ in Appendix \ref{sec:iterInvChisqFragDeriv} and 
$\mSUBpyalphasigsqTOsigsq$ in Appendix \ref{sec:GaussLikFragDeriv}
with projection onto the Inverse Chi-Squared family, although  bivariate
numerical integration is required. On the other hand,
$$\mSUBpahalfnuTOhalfnu=\frac{\projMR\Big[\mSUBhalfnuTOpahalfnu
\int_0^{\infty}p(a|\halfnu)\,\mSUBpahalfnuTOa\,da\big/Z\Big]}
{\mSUBhalfnuTOpahalfnu}.
$$
where $\projMR$ denotes projection onto the Moon Rock family.
The function of $\halfnu$ inside the $\projMR[\ ]$ is proportional
to
$$h(\halfnu)\equiv\{\halfnu^{\halfnu}\big/\Gamma(\halfnu)\}^{\eta_1^{\flat}+1}
e^{\eta_2^{\flat}\halfnu}\Gamma(\halfnu-\eta_1^{\sharp})/(\halfnu+\eta_2^{\sharp})^{\halfnu-\eta_1^{\sharp}}
$$
where 
$$\bdeta^{\sharp}\equiv\etaSUBpahalfnuTOa\quad\mbox{and}\quad
\bdeta^{\flat}\equiv\etaSUBhalfnuTOpahalfnu.
$$
Then
$$\etaSUBpahalfnuTOhalfnu=(\nabla\AMR)^{-1}\left(
\left[   
\begin{array}{c}
\int_0^{\infty}\,\{\halfnu\,\log(\halfnu)-\log\Gamma(\halfnu)\}
h(\halfnu)\,d\halfnu\big/\int_0^{\infty}h(\halfnu)\,d\halfnu\\[1ex]
\int_0^{\infty}\,\halfnu\,h(\halfnu)\,d\halfnu\big/\int_0^{\infty}h(\halfnu)\,d\halfnu
\end{array}
\right]\right)
$$
where
$$\AMR(\bdeta)\equiv\log\left[\int_0^{\infty}\{t^t/\Gamma(t)\}^{\eta_1}
\exp(\eta_2\,t)\,dt\right]$$
is the log-partition function of the Moon Rock exponential family.
This implies that
$$(\nabla\AMR)(\bdeta)=
\left[
\begin{array}{c}
\int_0^{\infty}\{t\log(t)-\log\Gamma(t)\}\{t^t/\Gamma(t)\}^{\eta_1}\exp(\eta_2\,t)\,dt
\Big/\int_0^{\infty}\{t^t/\Gamma(t)\}^{\eta_1}\exp(\eta_2\,t)\,dt\\[1ex]
\int_0^{\infty}\,t\{t^t/\Gamma(t)\}^{\eta_1}\exp(\eta_2\,t)\,dt
\Big/\int_0^{\infty}\{t^t/\Gamma(t)\}^{\eta_1}\exp(\eta_2\,t)\,dt
\end{array}
\right].
$$
This particular exponential family is not well-studied and we are not aware
of any published theory concerning the properties of $\nabla\AMR$ and $(\nabla\AMR)^{-1}$.
Standard analytic arguments can be used to show that the domain of $\nabla\AMR$ is 
$$H=\left\{\left[\begin{array}{c}
\eta_1\\
\eta_2 
\end{array}\right]:\eta_1\ge0,\ \eta_1+\eta_2<0 \right\}.
$$
It is conjectured that the image of $H$ under $\nabla\AMR$ is 
$$T=\left\{\left[\begin{array}{c}
\tau_1\\
\tau_2 
\end{array}\right]:\tau_1<\tau_2\log(\tau_2)-\log\Gamma(\tau_2) 
\right\}.
$$
Figure \ref{fig:mapPlotMoonRock} shows the domain of $\nabla\AMR$
and the conjectured domain of $(\nabla\AMR)^{-1}$ as well as some 
example mappings between the two spaces.

Evaluation of $(\nabla\AMR)^{-1}$ is a non-trivial problem. It requires numerical inversion
techniques such as Newton-Raphson iteration. Moreover, each of the iterative updates
involves evaluation of $\nabla\AMR$ and, possibly, its first partial derivatives.
None of these functions are available in closed form and require numerical integration.

\subsubsection{The $p(\nu)$ Fragment}

This simple fragment has the factor to stochastic node message
$$\mSUBpnuTOnu\propto p(\nu)$$
corresponding to the prior distribution on $\nu$. The conjugate family of prior density
functions is 
$$p(\nu)\propto \{(\nu/2)^{\nu/2}\big/\Gamma(\nu/2)\}^{A_{\nu}}\,\exp(-\smhalf\,B_{\nu}\nu),\quad\nu>0.$$
for hyperparameters $A_{\nu}\ge0$ and $B_{\nu}>A_{\nu}$.

In terms of $\upsilon=\nu/2$, the relevant message is 
$$\mSUBphalfnuTOhalfnu=\exp\left\{
\left[
\begin{array}{c}      
\halfnu\log(\halfnu)-\log\{\Gamma(\halfnu)\}\\[2ex]
\halfnu
\end{array}
\right]^T
\left[
\begin{array}{c}      
A_{\nu}\\[2ex]
-B_{\nu}
\end{array}
\right]
\right\}.
$$

\subsection{Summary of Numerical Challenges}

The previous two subsections make it clear that elaborations such 
as multivariate random effects and fancier likelihoods involve 
profound numerical challenges for the expectation propagation
paradigm. Table \ref{tab:numerDemands} summarizes the numerical
challenges of all of the non-trivial fragments treated in
this article.

\begin{table}[!ht]
\begin{center}
\begin{tabular}{lll}
\hline
fragment name &  numeric. integrat. demands & Kull.-Leib. projec. demands \\[0.1ex]
\hline\\[-0.9ex]
Gaussian prior          & none                   & none                            \\[0.74ex]
Inverse Wishart prior   & none                   & none                            \\[0.74ex]
Moon Rock prior         & none                  & none\\[0.74ex]
Iterated Inverse Chi Squared  & univariate quadrature  &  inversion of $\log-\mbox{digamma}$ \\[0.74ex]
Linear comb. deriv. var. & none                   &  none     \\[0.74ex]
Multiv. lin. comb. deriv. var. & none                   &  none     \\[0.74ex]
Gaussian                & univariate quadrature   &  inversion of $\log-\mbox{digamma}$ \\[0.74ex]
Logistic likelihood     & univariate quadrature   &  trivial \\[0.74ex]
Probit likelihood       & none                  &  trivial \\[0.74ex]
Poisson likelihood      & univariate quadrature &  trivial \\[0.74ex]
Multiple random effects & multivariate quadrature  & inversion of a multivariate \\
                        &                          &function     \\[0.74ex]
$t$ likelihood direct   & trivariate quadrature   & inversion of $\log-\mbox{digamma}$ \\
                        &                        &  and a non-explicit bivariate   \\
                        &                        &  function         \\[0.74ex]
$t$ likelihood aux. var.  & bivariate quadrature   &  inversion of $\log-\mbox{digamma}$\\
                        &                        &  and a non-explicit bivariate   \\
                        &                        &  function         \\[0.74ex]
\hline
\end{tabular}
\caption{\textit{The numerical integration and Kullback-Leibler projection demands of the
non-trivial fragments discussed in this article.}}
\label{tab:numerDemands} 
\end{center}
\end{table}

The first ten fragments in Table \ref{tab:numerDemands} have the attraction
of requiring only numerical evaluation of univariate integral within the families
given by (\ref{eq:AscAndBsc}) and (\ref{eq:Csc}).
The probit likelihood fragment stands out as a special case of a likelihood that
does not require any numerical methods for expectation propagation message passing.

The last three fragments of Table \ref{tab:numerDemands} are considerably 
more demanding in terms of numerical analysis. In a recent article, 
Gelman \textit{et al.} (2017) discuss the possibility of adopting Monte Carlo 
methods to deal with difficult computational problems in expectation propagation,
but we are not aware of any existing methodology of this type.

\section*{Acknowledgments}

\ifthenelse{\boolean{UnBlinded}}{
This research was supported by Australian Research Council
Discovery Project DP180100597.}{\emph{Blinded text place-holder.}}

\section*{References}

\bib
Azzalini, A. (2017). The \textsf{R} package \texttt{sn}: The skew-normal and skew-t
 distributions (version 1.5). \texttt{http://azzalini.stat.unipd.it/SN}

\bib
Carpenter, B., Gelman, A., Hoffman, M.D., Lee, D., Goodrich, B., Betancourt, M., 
Brubaker, M., Guo, J., Li, P. and Riddell, A. (2017). 
Stan: A probabilistic programming language. \textit{Journal of Statistical Software},
{\bf 76}, Issue 1, 1--32.

\bib
Gelman, A., Vehtari, A., Jyl\"anki, P., Sivula, T., Tran, D., 
Suhai, S., Blomstedt, P. Cunningham, J.P., Schiminovic, D. and Robert, C. (2017). 
Expectation propagation as a way of life: A framework for Bayesian inference
on partitioned data. Unpublished manuscript (\texttt{arXiv:1412.4869v2}).

\bib
Guo, B.-N. and Qi, F. (2013). Refinements of lower bounds for
polygamma functions. \textit{Proceedings of the American 
Mathematical Society}, {\bf 141}, 1007--1015.

\bib
Guo, J., Gabry, J. and Goodrich, B. (2017). 
The \textsf{R} package \textsf{rstan}: 
\textsf{R} interface to \textsf{Stan}.
\textsf{R} package (version 2.17.2). \texttt{http://mc-stan.org}.

\bib
Harville, D.A. (2008). \textit{Matrix Algebra from a Statistician's
Perspective}. New York: Springer.

\bib
Kim, A.S.I. and Wand, M.P. (2016).
The explicit form of expectation propagation for 
a simple statistical model. 
\textit{Electronic Journal of Statistics},
{\bf 10}, 550--581.

\bib
Kim, A.S.I. and Wand, M.P. (2017).
On expectation propagation for generalized, linear and mixed models. 
\textit{Australian and New Zealand Journal of Statistics},
{\bf 59}, in press.

\bib
Minka, T. (2005). 
Divergence measures and message passing.
\textit{Microsoft Research Technical Report Series}, 
{\bf MSR-TR-2005-173}, 1--17.

\bib
McLean, M.W. and Wand, M.P. (2018).
Variational message passing for elaborate response regression
models. \textit{Bayesian Analysis}, under revision.

\bib
Nolan, T.H. and Wand, M.P. (2017).
Accurate logistic variational message passing: 
algebraic and numerical details. \textit{Stat}, {\bf 6},
102--112.

\bib
Smyth, G. (2015). The \textsf{R} package \texttt{statmod}: Statistical
modeling (version 1.4).\\
\texttt{http://http://cran.r-project.org}

\bib
Sommer, A. (1982). {\it Nutritional Blindness}.
New York: Oxford University Press.

\bib
Wainwright, M.J. and Jordan, M.I. (2008).
Graphical models, exponential families, and variational inference.
\textit{Foundations and Trends in Machine Learning}, {\bf 1}, 1--305.

\bib
Wand, M.P. (2017).
Fast approximate inference for arbitrarily large semiparametric
regression models via message passing (with discussion).
\textit{Journal of the American Statistical Association}, 
{\bf 112}, 137--168.

\bib
Wand, M.P. and Jones, M.C. (1993). Comparison of smoothing
parameterizations in bivariate density estimation.
{\it Journal of the American Statistical
Association}, {\bf 88}, 520--528.

\bib
Wand, M.P. and Ripley, B.D. (2015).
The \textsf{R} package \texttt{KernSmooth}.
Functions for kernel smoothing supporting Wand \myand Jones (1995)
(version 2.23).\\ 
\texttt{https://cran.R-project.org}.

\null\vfill\eject
\vfill\eject
%
%
\renewcommand{\theequation}{S.\arabic{equation}}
\renewcommand{\thesection}{S.\arabic{section}}
\renewcommand{\thetable}{S.\arabic{table}}
\setcounter{equation}{0}
\setcounter{table}{0}
\setcounter{section}{0}
\setcounter{page}{1}
\setcounter{footnote}{0}

\begin{center}

{\Large Supplement for:}
\vskip3mm

\centerline{\LARGE\bf Factor Graph Fragmentization of Expectation Propagation}
\vskip7mm
\ifthenelse{\boolean{UnBlinded}}{\centerline{\normalsize\sc By Wilson Y. Chen and Matt P. Wand}
\vskip5mm
\centerline{\textit{University of Technology Sydney}}
\vskip6mm}{\null}

\end{center}

\section{Function Definitions}

Expectation propagation algorithms that are based on the fragments in
Section \ref{sec:fragGLMM} have straightforward implementation
once a few key functions are identified. Many of the functions
are simple but long-winded. However, they only have to be implemented
once and after that all fragment updates are simple.

The functions can be divided into three types:
\begin{enumerate}
\item functions defined via non-analytic integral families.
\item a function defined via inversion of an established function
\item functions that are explicit given function types 1. and 2.
\end{enumerate}
We now give details of each of these types in turn.

\subsection{Functions Defined via Non-Analytic Integral Families}\label{sec:funcNonAna}

Two fundamental families of integrals for expectation propagation
in linear model contexts are:
\begin{equation}
\begin{array}{rcl}
\Asc(p,q,r,s,t,u)&\equiv&\displaystyle{
\infint\frac{x^p\,\exp(qx-rx^2)\,dx}{(x^2+sx+t)^u}},\\[2ex]
&&\quad
p\ge0,\ q\in\realnos, r>0, s\in\realnos, t>\quarter\,s^2, u>0\\[3ex]
\mbox{and}\quad\Bsc(p,q,r,s,t,u)&\equiv&\displaystyle{
\infint\frac{x^p\,\exp\{qx-re^x-se^x/(t+e^x)\}\,dx}{(t+e^x)^u}},\\[2ex]
&&\quad
p\ge0,\ q\in\realnos, r>0, s\ge0, t>0, u>0.
\end{array}
\label{eq:AscAndBsc}
\end{equation}
An additional family of non-analytic functions that we need is:
\begin{equation}
\Csc_b(p,q,r)\equiv\infint x^p\,\exp\{qx-rx^2-b(x)\}\,dx,
\label{eq:Csc}
\end{equation}
where $q\in\realnos$, $r>0$ and $b:\realnos\to\realnos$ is any function for 
which $\Csc_b(p,q,r)$ exists.

To avoid underflow and overflow working with logarithms and
suitably modified integrands is recommended. For example, 
$$\Csc(0,q,r)=e^M\infint\exp\{qx-rx^2-b(x)-M\}\,dx$$
where $M\equiv\sup\{x\in\realnos:qx-rx^2-b(x)\}$ implies that
$$\log\{\Csc(0,q,r)\}=M+\log\left[\infint\exp\{qx-rx^2-b(x)-M\}\,dx\right]$$
Sine the last-written integrand has a maximum of $1$, its
values are not overly large or small.

\subsection{Function Defined via Inversion}\label{sec:funcInver}

Theorem 1 of Kim \myand Wand (2016) asserts that the function 
$\log-\mbox{digamma}$ is a bijective mapping from $\realnos_+$ onto
$\realnos_+$. Therefore its inverse 
$$(\log-\mbox{digamma})^{-1}:\realnos_+\to\realnos_+$$
is well-defined. Bounds given in Guo \myand Qi (2013) imply that 
$$\frac{1}{2x}<(\log-\mbox{digamma})^{-1}(x)<\frac{1}{x}\quad\mbox{for all}\ x>0.$$
Therefore, 
$$(\log-\mbox{digamma})^{-1}(x)\approx\mbox{geometric mean of $\displaystyle{\frac{1}{2x}}$ 
and $\displaystyle{\frac{1}{x}}$}
=\frac{1}{x\sqrt{2}}$$
and provides useful starting values for iterative inversion of $\log-\mbox{digamma}$.

In addition, care is required for evaluation of $(\log-\mbox{digamma})(x)$ for
large $x$ since direct computation round-off error can lead to an erroneous answer of zero.
Software such as the function \texttt{logmdigamma()} in the \textsf{R} package
\textsf{statmod} (Smyth, 2015).

\subsection{Explicit Functions}\label{sec:explicitFns}

The $N(0,1)$ density function and cumulative distribution functions are 
denoted by 
$$\phi(x)\equiv(2\pi)^{-1}e^{-x^2/2}\quad\mbox{and}\quad\Phi(x)\equiv\int_{-\infty}^x\phi(t)\,dt.$$
We also define
$$\zeta(x)\equiv\log\{2\Phi(x)\}\quad\mbox{so that}\quad\zeta'(x)\equiv\phi(x)/\Phi(x).$$
Stable computation of $\zeta'$ is available from, for example, the function 
\texttt{zeta()} in the \textsf{R} package $\textsf{sn}$
(Azzalini, 2017).

The functions $G^{\mbox{\tiny N}}$ and $G^{\mbox{\tiny IG1}}$ 
defined in Kim \myand Wand (2016) are also needed here. 
The function $G^{\mbox{\tiny IG2}}$ from Kim \myand Wand (2016) requires
generalization to handle Half-$t$ priors on standard deviation parameters
with arbitrary degrees of freedom. The generalization is denoted
by $G^{\mbox{\tiny IG3}}$. Each of  $G^{\mbox{\tiny N}}$, $G^{\mbox{\tiny IG1}}$ 
and $G^{\mbox{\tiny IG3}}$ depend on the functions defined in 
Appendices \ref{sec:funcNonAna} and \ref{sec:funcInver}
but otherwise ther  are simple, albeit long-winded, functions
with multiple vector arguments. Their definitions are given 
in Section A.4 of Kim \myand Wand (2016) and are repeated here
for convenience. We also define the explicit functions 
$\Hprobit$, $\Hlogistic$ and $\HPoisson$.
First set:
\begin{eqnarray*}
&&\alpha\left(k,\left[\begin{array}{c}   
a_1\\
a_2
\end{array}\right],
\left[\begin{array}{c}   
b_1\\
b_2
\end{array}\right],
\left[\begin{array}{c}   
c_1\\
c_2\\
c_3
\end{array}\right]
\right)
= 
\Asc\left(k,a_1,-a_2,\frac{-2c_2}{c_1},\frac{c_3-2b_2}{c_1},
\frac{c_1-2b_1-2}{2}\right)\\[1ex]
\mbox{and}&&\\[1ex]
&&\beta\left(k,\ell,v,w,\left[\begin{array}{c}   
a_1\\
a_2
\end{array}\right],
\left[\begin{array}{c}   
b_1\\
b_2
\end{array}\right],
\left[\begin{array}{c}   
c_1\\
c_2\\
c_3
\end{array}\right]
\right)
= \\
&&\qquad\qquad\qquad
\Bsc\left(k,\frac{\ell+c_1-1}{2}-a_1,\frac{c_1c_3-c_2^2}{2c_1}-a_2,
-b_2\left(\frac{c_2}{c_1}+\frac{b_1}{2b_2}\right)^2,\,v,\,w\right).
\end{eqnarray*}
Next, define
\begin{eqnarray*}
&&g(\ell,v,w,\ba,\bb,\bc)=(\log-\mbox{digamma})^{-1}\left(
\log\Bigg\{\frac{\beta(0,\ell+1,v,w,\ba,\bb,\bc)}
{\beta(0,\ell-1,v,w,\ba,\bb,\bc)}\right\}\,\\[2ex]
&&\qquad\qquad\qquad\qquad\qquad\qquad\qquad\qquad\qquad\qquad\qquad\qquad
-\,\frac{\beta(1,\ell-1,v,w,\ba,\bb,\bc)}{\beta(0,\ell-1,v,w,\ba,\bb,\bc)}\Bigg).
\end{eqnarray*}
Now we are ready to give the expressions
for $G^{\mbox{\tiny N}}$, $G^{\mbox{\tiny IG1}}$,  $G^{\mbox{\tiny IG2}}$,
and $G^{\mbox{\tiny IG3}}$:
$$
G^N(\ba,\bb;\bc)= 
\left[
\frac{\alpha(2,\ba,\bb,\bc)}{\alpha(0,\ba,\bb,\bc)}
-\left\{\frac{\alpha(1,\ba,\bb,\bc)}{\alpha(0,\ba,\bb,\bc)}\right\}^2\right]^{-1}
\left[
\begin{array}{c}
\alpha(1,\ba,\bb,\bc)/\alpha(0,\ba,\bb,\bc)\\
-1/2
\end{array}
\right]-\ba,
$$
$$
G^{\tiny\mbox{IG1}}\left(
\ba,\left[\begin{array}{c}   
b_1\\
b_2
\end{array}\right];
\left[\begin{array}{c}   
c_1\\
c_2\\
c_3
\end{array}\right]
\right)
=
\left[
\begin{array}{c}
-1-g(0,-2b_2/c_1,\smhalf,\ba,\bb,\bc)\\[1ex]
\displaystyle{
\frac{-g(0,-2b_2/c_1,\smhalf,\ba,\bb,\bc)
\,\beta(0,-1,-2b_2/c_1,\smhalf,\ba,\bb,\bc)}{\beta(0,1,-2b_2/c_1,\smhalf,\ba,\bb,\bc)}}
\end{array}
\right]-\ba
$$
and
\begin{eqnarray*}
&&G^{\tiny\mbox{IG3}}\left(\ba,\left[\begin{array}{c}   
b_1\\
b_2
\end{array}\right];k,\ell\right)=\\[2ex]
&&\qquad\qquad\qquad\qquad
\left[
\begin{array}{c}
-1-g\left(k-2,-b_2/\ell,\ell-k/2-b_1,\ba,\zerobtwo,\twozerozero\right)\\[6ex]
\displaystyle{
\frac{\left\{
\begin{array}{c}
-g\left(k-2,-b_2/\ell,\ell-k/2-b_1,\ba,\zerobtwo,\twozerozero\right)\\[1ex]
\times\beta\left(0,k-3,-b_2/\ell,\ell-k/2-b_1,\ba,\zerobtwo,\twozerozero\right)
\end{array}
\right\}
}
{
\beta\left(0,k-1,-b_2/\ell,\ell-k/2-b_1,\ba,\zerobtwo,\twozerozero\right)
}}
\end{array}
\right]-\ba.
\end{eqnarray*}
This definition is a generalization of the function $G^{\tiny\mbox{IG2}}$
given in Kim \myand Wand (2016, 2017) and is such that 
$$G^{\tiny\mbox{IG3}}\left(\ba,\left[\begin{array}{c}   
b_1\\
b_2
\end{array}\right];k,1\right)
=G^{\tiny\mbox{IG2}}\left(\ba,\left[\begin{array}{c}   
b_1\\
b_2
\end{array}\right];k\right).
$$
Put
$$
H_b\left(\left[\begin{array}{c}   
a_1\\
a_2
\end{array}
\right];y\right)
=
\left[
\begin{array}{c}
\displaystyle{
\frac{\Csc_b(1,a_1+y,-a_2)/\Csc_b(0,a_1+y,-a_2)}
{\displaystyle{\frac{\Csc_b(2,a_1+y,-a_2)}{\Csc_b(0,a_1+y,-a_2)}}- 
\displaystyle{\frac{\Csc_b(1,a_1+y,-a_2)^2}{\Csc_b(0,a_1+y,-a_2)^2}}}}\\[7ex]
\displaystyle{
\frac{-1/2}
{\displaystyle{\frac{\Csc_b(2,a_1+y,-a_2)}{\Csc_b(0,a_1+y,-a_2)}}- 
\displaystyle{\frac{\Csc_b(1,a_1+y,-a_2)^2}{\Csc_b(0,a_1+y,-a_2)^2}}}}
\end{array}
\right]
-\left[\begin{array}{c}
a_1\\
a_2
\end{array}\right]
$$
for any $a_1\in\realnos$, $a_2<0$ and $y\in\realnos$ and then let
\begin{equation}
{\setlength{\arraycolsep}{1pt}
\begin{array}{rcl}
\Hlogistic\left(\left[\begin{array}{c}   
a_1\\
a_2
\end{array}
\right];y\right)&\equiv&
H_b\left(\left[\begin{array}{c}   
a_1\\
a_2
\end{array}
\right];y\right)\quad\mbox{for $\quad b(x)=\log(1+e^x)$}\\[2ex]
\mbox{and}\quad\HPoisson\left(\left[\begin{array}{c}   
a_1\\
a_2
\end{array}
\right];y\right)&\equiv&
H_b\left(\left[\begin{array}{c}   
a_1\\
a_2
\end{array}
\right];y\right)\quad\mbox{for $\quad b(x)=e^x$.}
\end{array}
}
\label{eq:HlogPoiDefns}
\end{equation}
Lastly,
$$
\Hprobit\left(\left[\begin{array}{c}   
a_1\\
a_2
\end{array}
\right];y\right)
\equiv
\frac{1}
{1-2a_2-\zeta'(r)\big\{r+\zeta'(r)\big\}}
\left[\setlength\arraycolsep{0.5pt}{
\begin{array}{c}   
a_1(1-2a_2)\\[0.5ex]
+(2y-1)\zeta'(r)\sqrt{2a_2(2a_2-1)}\\[3ex]
a_2(1-2a_2)
\end{array}}
\right]
-
\left[
\setlength\arraycolsep{0.5pt}{
\begin{array}{c}   
a_1\\[6ex]
a_2
\end{array}}
\right]
$$
where $r\equiv(2y-1)a_1/\sqrt{2a_2(2a_2-1)}$.

\section{Derivations}\label{sec:derivations}

We now provide derivations of each of algorithms given in Section \ref{sec:fragGLMM}.
Definitions used in the derivations are:
$$\AN(\bdeta)\equiv\,-\quarter(\eta_1^2/\eta_2)-\smhalf\log(-2\eta_2)\quad
\mbox{and}\quad
\AICS(\bdeta)\equiv(\eta_1+1)\log(-\eta_2)+\log\Gamma(-\eta_1-1)$$
for the log-partition functions of the Normal and Inverse Chi-Squared families
respectively. In a similar vein, $\projN[\ \cdot\ ]$
denotes Kullback-Leibler projection onto the (possibly Multivariate) 
Normal family of density functions and $\projICS[\ \cdot\ ]$ denotes Kullback-Leibler 
projection onto the Inverse Chi-Squared family of density functions. We use $Z$ to 
denote the normalizing factor of the function inside a Kullback-Leibler projection 
operator.

\subsection{Derivation of Algorithm \ref{alg:GaussianPriorFrag}}\label{sec:GaussPriorFragDeriv}

Plugging into (\ref{eq:facToStoch}) we get
$$
\mSUBpthetaTOtheta=\frac{\projN\Big[\mSUBthetaTOptheta\,p(\btheta)\big/Z\Big]}{\mSUBthetaTOptheta}
=\frac{\mSUBthetaTOptheta\,p(\btheta)}{\mSUBthetaTOptheta}=p(\theta).
$$
The second equality follows from the fact that $\mSUBthetaTOptheta$ is a Multivariate
Normal density function, which is a consequence of (\ref{eq:conjugacyGaussPrior}).
Since 
$$p(\btheta)\propto
\exp\big\{-\smhalf(\btheta-\bmu_{\btheta})^T\bSigma_{\btheta}^{-1}(\btheta-\bmu_{\btheta})\big\}
\propto
\exp\left\{
\left[
\begin{array}{c}
\btheta\\[1ex]
\vecof(\btheta\btheta^T)
\end{array}
\right]^T
\left[
\begin{array}{c}
\bSigma_{\btheta}^{-1}\bmu_{\btheta}\\[1ex]
-\smhalf\vecof(\bSigma_{\btheta}^{-1})
\end{array}
\right]
\right\}
$$
we have
$$\mSUBpthetaTOtheta=\exp\left\{
\left[
\begin{array}{c}
\btheta\\[1ex]
\vecof(\btheta\btheta^T)
\end{array}
\right]^T
\etaSUBpthetaTOtheta
\right\}
$$
where 
$$\etaSUBpthetaTOtheta=\left[
\begin{array}{c}
\bSigma_{\btheta}^{-1}\bmu_{\btheta}\\[1ex]
-\smhalf\vecof(\bSigma_{\btheta}^{-1})
\end{array}
\right].$$

\subsection{Derivation of Algorithm \ref{alg:InvWishartPriorFrag}}\label{sec:InvWishartFragDeriv}

Using arguments similar to those given in Section \ref{sec:GaussPriorFragDeriv} for
the Gaussian prior fragment lead to
\begin{eqnarray*}
\mSUBpThetaTOTheta&\propto&
p(\bTheta)\propto\,|\bTheta|^{-(\kappaTheta+\dTheta+1)/2}\exp\{-\smhalf\tr(\LambdaTheta\bTheta^{-1})\}\\[2ex]
&=&
\exp\left\{
\left[
\begin{array}{c}
\log|\bTheta|\\[1ex]
\vecof(\bTheta^{-1})
\end{array}
\right]^T
\left[
\begin{array}{c}
-\smhalf(\kappaTheta+\dTheta+1)\\[1ex]
-\smhalf\vecof(\LambdaTheta^{-1})
\end{array}
\right]
\right\}.
\end{eqnarray*}
Hence
$$\mSUBpThetaTOTheta=\exp\left\{
\left[
\begin{array}{c}
\log|\bTheta|\\[1ex]
\vecof(\bTheta^{-1})
\end{array}
\right]^T
\etaSUBpThetaTOTheta
\right\}
$$
where 
$$\etaSUBpThetaTOTheta=\left[
\begin{array}{c}
-\smhalf(\kappaTheta+\dTheta+1)\\[1ex]
-\smhalf\vecof(\LambdaTheta^{-1})
\end{array}
\right].$$

\subsection{Derivation of Algorithm \ref{alg:iterInvChisqFrag}}\label{sec:iterInvChisqFragDeriv}

The message from $p(\sigma^2|a)$ to $\sigma^2$ is 
\begin{equation}
\mSUBpsigsqaTOsigsq\thickarrow\frac{\projICS\big[\mSUBsigsqTOpsigsqa
\int_0^{\infty}p(\sigma^2|a)\,\mSUBaTOpsigsqa\,da\big/Z\big]}
{\mSUBsigsqTOpsigsqa}.
\label{eq:mpsigsqaTOsigsq}
\end{equation}
Assumption (\ref{eq:conjugacyIterIG}) implies that
$$\mSUBaTOpsigsqa=\exp\left\{
\left[\begin{array}{c}
\log(a)\\
1/a
\end{array}\right]^T 
\etaSUBaTOpsigsqa            
\right\}=a^{\eta_1^{\clubsuit}}\exp(\eta_2^{\clubsuit}/a)
$$
and
$$\mSUBsigsqTOpsigsqa=\exp\left\{
\left[\begin{array}{c}
\log(\sigma^2)\\
1/\sigma^2
\end{array}\right]^T 
\etaSUBsigsqTOpsigsqa            
\right\}=(\sigma^2)^{\eta_1^{\spadesuit}}\exp(\eta_2^{\spadesuit}/\sigma^2)
$$
where 
$$\bdeta^{\clubsuit}\equiv\etaSUBaTOpsigsqa\quad\mbox{and}\quad
\bdeta^{\spadesuit}\equiv \etaSUBsigsqTOpsigsqa.$$ 
As a function of $\sigma^2$, the integral in (\ref{eq:mpsigsqaTOsigsq}) is 
\begin{eqnarray*}
&&\int_0^{\infty}\frac{(\nu/a)^{\nu/2}}{\Gamma(\nu/2)}\,(\sigma^2)^{-(\nu/2)-1}\exp\{-\nu/(\sigma^2a)\}\,
a^{\eta_1^{\clubsuit}}\exp(\eta_2^{\clubsuit}/a)\,da\\[1ex]
&&\qquad\qquad\propto(\sigma^2)^{-(\nu/2)-1}
\int_0^{\infty}\,a^{-\{-\eta_1^\clubsuit+(\nu/2)-1\}-1}\exp[-\{(\nu/\sigma^2)-\eta_2^{\clubsuit}\}/a]\,da\\[1ex]
&&\qquad\qquad\propto(\sigma^2)^{-(\nu/2)-1}\{(\nu/\sigma^2)-\eta_2^{\clubsuit}\}^{\eta_1^\clubsuit-(\nu/2)+1}.
\end{eqnarray*}
Therefore, the density function inside the $\projICS[\ \cdot\ ]$ is proportional to
$$(\sigma^2)^{\eta_1^{\spadesuit}-(\nu/2)-1}
\{(\nu/\sigma^2)-\eta_2^{\clubsuit}\}^{\eta_1^\clubsuit-(\nu/2)+1}\exp(\eta_2^{\spadesuit}/\sigma^2)
$$
and the numerator of (\ref{eq:mpsigsqaTOsigsq}) is proportional to 
\def\bdetanum{\bdeta_{\mbox{\tiny numer}}}
$$
\exp\left\{
\left[\begin{array}{c}
\log(\sigma^2)\\
1/\sigma^2
\end{array}\right]^T 
\bdetanum
\right\}
$$
where
$$\bdetanum\equiv(\nabla\AICS)^{-1}\left(
\left[
\begin{array}{c}
\frac{\int_0^{\infty}\log(\sigma^2)(\sigma^2)^{\eta_1^{\spadesuit}-(\nu/2)-1}
\{(\nu/\sigma^2)-\eta_2^{\clubsuit}\}^{\eta_1^\clubsuit-(\nu/2)+1}
\exp(\eta_2^{\spadesuit}/\sigma^2)\,d\sigma^2}
{\int_0^{\infty}(\sigma^2)^{\eta_1^{\spadesuit}-(\nu/2)-1}
\{(\nu/\sigma^2)-\eta_2^{\clubsuit}\}^{\eta_1^\clubsuit-(\nu/2)+1}
\exp(\eta_2^{\spadesuit}/\sigma^2)\,d\sigma^2}\\[4ex]
\frac{\int_0^{\infty}(1/\sigma^2)(\sigma^2)^{\eta_1^{\spadesuit}-(\nu/2)-1}
\{(\nu/\sigma^2)-\eta_2^{\clubsuit}\}^{\eta_1^\clubsuit-(\nu/2)+1}
\exp(\eta_2^{\spadesuit}/\sigma^2)\,d\sigma^2}
{\int_0^{\infty}(\sigma^2)^{\eta_1^{\spadesuit}-(\nu/2)-1}
\{(\nu/\sigma^2)-\eta_2^{\clubsuit}\}^{\eta_1^\clubsuit-(\nu/2)+1}
\exp(\eta_2^{\spadesuit}/\sigma^2)\,d\sigma^2}
\\[2ex]
\end{array}
\right]
\right).$$
Steps analogous to those given in Appendix A.5.4 of Kim \myand Wand (2016)
can then be used to derive the $\etaSUBpsigsqaTOsigsq$ update. 
However, note that Algorithm \ref{alg:iterInvChisqFrag} supports
general $\mbox{Half-t}(A,\nu)$ prior distributions and Kim \myand Wand (2016)
only deal with the $\nu=1$ (Half-Cauchy) special case. 
Hence, the function $G^{\mbox{\tiny IG2}}$ in Kim \myand Wand (2016)
has to be generalized to the $G^{\mbox{\tiny IG3}}$ function.

Also, note that the $\etaSUBpsigsqaTOsigsq$ update has $\thickarroweps$
rather than $\thickarrow$ to allow for the damping adjustment 
defined by (\ref{eq:epsAdjust}). The same adjustment 
applies to the remainder of the derivations given in Section \ref{sec:derivations}.

The message from $p(\sigma^2|a)$ to $a$ has a similar form
and arguments analogous to those just given for the message
from $p(\sigma^2|a)$ to $\sigma^2$ lead to the update for 
$\etaSUBpsigsqaTOa$ given in Algorithm \ref{alg:iterInvChisqFrag}.

\subsection{Derivation of Algorithm \ref{alg:LinCombFrag}}\label{sec:LinCombFragDeriv}

Algorithm \ref{alg:LinCombFrag} is the $\dtheta=1$ special case of Algorithm \ref{alg:MultLinCombFrag}
and therefore its derivation is covered by the general $\dtheta$ case given next
in Section \ref{sec:MultLinCombFragDeriv}.

\subsection{Derivation of Algorithm \ref{alg:MultLinCombFrag}}\label{sec:MultLinCombFragDeriv}

Algorithm \ref{alg:MultLinCombFrag} depends on Result 2 below,
which extends Theorem 1 of Kim \myand Wand (2017) to multivariate linear combination
derived variables.

Let 
$$p_{\smallN_d}(\bx;\bmu,\bSigma)=(2\pi)^{-d/2}|\bSigma|^{-1/2}
\exp\{-\smhalf(\bx-\bmu)^T\bSigma^{-1}(\bx-\bmu)\}
$$
denote the density function of a $d$-variate $\text{N}(\bmu,\bSigma)$ 
random vector. Then we have:

\vskip2mm
\noindent
{\bf Result 2.} 
{\sl For all $d\times d'$ matrices $\coeffmat$ such $\coeffmat^T\bSigma\coeffmat$
is positive definite and $d'\times1$ vectors $\bv$:} 
\begin{equation}
\int_{\realnos^d} \,p_{\smallN_d}
(\bx;\bmu,\bSigma)\,\delta\big(\bv-\coeffmat^T\bx\big)\,d\,\bx
=p_{\smallN_{d'}}(\bv;\coeffmat^T\bmu,\coeffmat^T\bSigma\coeffmat).
\label{eq:proposition}
\end{equation}

\vskip2mm
\noindent
{\bf Derivation of Result 2.}

Via the change of variable $\bz=\bSigma^{-1/2}(\bx-\bmu)$, the 
left-hand side of (\ref{eq:proposition}) is
$$\int_{\realnos^d} \,p_{\smallN_d}
(\bz;\bzero,\bI)\,\delta\big(\bv-\bL^T\bmu-\coeffmat^T\bSigma^{1/2}\bz\big)\,d\,\bz
=
\int_{\realnos^d} \,p_{\smallN_d}
(\bz;\bzero,\bI)\,\delta\big(\bvdagg-(\coeffmatdagg)^T\bz\big)\,d\,\bz
$$
where 
$$\bvdagg\equiv\bv-\bL^T\bmu\quad\mbox{and}\quad
\coeffmatdagg\equiv\bSigma^{1/2}\bL.
$$
Next note that
\setlength\arraycolsep{2pt}{
\begin{eqnarray*}
&&\int_{\realnos^d} \,p_{\smallN_d}
(\bx;\bzero,\bI)\,\delta\big(\bvdagg-(\coeffmatdagg)^T\bz\big)\,d\,\bz=
\int_{\realnos^d}\lim_{\varepsilon\to 0}\,p_{\smallN_d}
(\bz;\bzero,\bI)
\,p_{\smallN_{d'}}\big(\bvdagg-(\coeffmatdagg)^T\bz;
\bzero,\varepsilon\,\bI\big)\,d\,\bz\\
&&\qquad\qquad=(2\pi)^{-d/2}\int_{\realnos^d}\lim_{\varepsilon\to 0}
\big[(2\pi\varepsilon)^{-d'/2}\exp\{r(\bz,\bvdagg,\coeffmatdagg,\varepsilon)\}\big]\,d\bz
\end{eqnarray*}
}
where
\setlength\arraycolsep{2pt}{
\begin{eqnarray*}
r(\bz,\bvdagg,\coeffmatdagg,\varepsilon)&=&-\smhalf\bz^T\bz
-\textstyle{\frac{1}{2\varepsilon}}\{\bvdagg-(\coeffmatdagg)^T\bz\}^T
\{\bvdagg-(\coeffmatdagg)^T\bz\}\\[1ex]
&=&-\smhalf(\bz-\bm)^T\{\bI+\textstyle{\frac{1}{\varepsilon}}
\coeffmatdagg(\coeffmatdagg)^T\}(\bz-\bm)
-\textstyle{\frac{1}{2\varepsilon}}(\bvdagg)^T\bvdagg\\
&&\qquad\qquad+\textstyle{\frac{1}{2\varepsilon^2}}(\coeffmatdagg\bvdagg)^T
\{\bI+\textstyle{\frac{1}{\varepsilon}}\coeffmatdagg(\coeffmatdagg)^T\}^{-1}
\coeffmatdagg\bvdagg
\end{eqnarray*}
}
with $\bm=
\bm(\bvdagg,\coeffmatdagg,\varepsilon)
\equiv\textstyle{\frac{1}{\varepsilon}}
(\bI+\textstyle{\frac{1}{\varepsilon}}\bL\bL^T)^{-1}(\coeffmatdagg)^T
\bvdagg$.
We then have
\begin{equation}
\setlength\arraycolsep{2pt}{
\begin{array}{rcl}
&&\int_{\realnos^d} \,p_{\smallN_d}
(\bz;\bzero,\bI)\,\delta\big(\bvdagg-(\coeffmatdagg)^T\bz\big)\,d\,\bz
=\lim_{\varepsilon\to 0}\Big[ (2\pi\varepsilon)^{-d'/2}
|\bI+\textstyle{\frac{1}{\varepsilon}}\coeffmatdagg(\coeffmatdagg)^T|^{-1/2}\\
&&\qquad\qquad\qquad\times
\exp\left\{\textstyle{\frac{1}{2\varepsilon^2}}(\coeffmatdagg\bvdagg)^T
\{\bI+\textstyle{\frac{1}{\varepsilon}}\coeffmatdagg(\coeffmatdagg)^T\}^{-1}
\coeffmatdagg\bvdagg
-\textstyle{\frac{1}{2\varepsilon}}(\bvdagg)^T\bvdagg\right\}\Big]
\end{array}
}
\label{eq:firstLim}
\end{equation}
where we have used the fact 
$$\int_{\realnos^d}(2\pi)^{-d/2}
|\bI+\textstyle{\frac{1}{\varepsilon}}\coeffmatdagg(\coeffmatdagg)^T|^{1/2}
\exp\Big\{-\smhalf(\bz-\bm)^T\big\{\bI
+\textstyle{\frac{1}{\varepsilon}}\coeffmatdagg(\coeffmatdagg)^T\big\}
(\bz-\bm)\Big\}\,d\bz=1$$
and assumed that the integrand possesses properties that are sufficient to justify
interchanging the limit as $\varepsilon\to0$ and the integral over $\realnos^d$.
Using Theorem 18.1.1 of Harville (2008), the determinant in 
(\ref{eq:firstLim}) is
\begin{equation}
|\bI+\textstyle{\frac{1}{\varepsilon}}\coeffmatdagg(\coeffmatdagg)^T|
=\varepsilon^{-d'}|\varepsilon\,\bI+(\coeffmatdagg)^T\coeffmatdagg|
=\varepsilon^{-d'}|\varepsilon\,\bI+\bL^T\bSigma\bL|.
\label{eq:detResult}
\end{equation}
Next, application of Theorem 18.2.8 of Harville (2008) gives
\begin{equation}
\{\bI+\textstyle{\frac{1}{\varepsilon}}\coeffmatdagg(\coeffmatdagg)^T\}^{-1}
=\bI-\coeffmatdagg\big\{\varepsilon\bI+(\coeffmatdagg)^T\coeffmatdagg\big\}^{-1}
(\coeffmatdagg)^T
\label{eq:invResult}
\end{equation}
which implies that
\setlength\arraycolsep{1pt}{
\begin{eqnarray*}
(\coeffmatdagg\bvdagg)^T
\{\bI+\textstyle{\frac{1}{\varepsilon}}\coeffmatdagg(\coeffmatdagg)^T\}^{-1}
\coeffmatdagg\bvdagg
&=&(\bvdagg)^T(\coeffmatdagg)^T\coeffmatdagg\bvdagg
-(\coeffmatdagg\bvdagg)^T
\coeffmatdagg\big\{\varepsilon\bI+(\coeffmatdagg)^T\coeffmatdagg\big\}^{-1}
(\coeffmatdagg)^T\coeffmatdagg\bvdagg\\[1ex]
&=&(\bvdagg)^T\bL^T\bSigma\bL\,\bvdagg
-(\bvdagg)^T\coeffmat^T\bSigma\coeffmat
\big(\varepsilon\bI+\coeffmat^T\bSigma\coeffmat\big)^{-1}
\coeffmat^T\bSigma\coeffmat\bvdagg.
\end{eqnarray*}
}
The exponent in the expression on the right-hand side of (\ref{eq:firstLim})
is then
\setlength\arraycolsep{1pt}{
\begin{eqnarray*}
&&\textstyle{\frac{1}{2\varepsilon^2}}
\left\{(\bvdagg)^T\bL^T\bSigma\bL\,\bvdagg
-(\bvdagg)^T\coeffmat^T\bSigma\coeffmat
\big(\varepsilon\bI+\coeffmat^T\bSigma\coeffmat\big)^{-1}
\coeffmat^T\bSigma\coeffmat\bvdagg
-\varepsilon(\bvdagg)^T\bvdagg
\right\}\\[2ex]
&&\quad=-\smhalf(\bvdagg)^T
\big(\varepsilon\bI+\coeffmat^T\bSigma\coeffmat\big)^{-1}\bvdagg
=-\smhalf(\bv-\bL^T\bmu)^T
\big(\varepsilon\bI+\coeffmat^T\bSigma\coeffmat\big)^{-1}
(\bv-\bL^T\bmu).
\end{eqnarray*}
}
Substitution of this result and (\ref{eq:detResult}) into 
(\ref{eq:firstLim}) then gives
\begin{eqnarray*}
&&\int_{\realnos^d} \,p_{\smallN_d}
(\bz;\bzero,\bI)\,\delta\big(\bvdagg-(\coeffmatdagg)^T\bz\big)\,d\,\bz\\[1ex]
&&\quad
=\lim_{\varepsilon\to0}\big[(2\pi)^{-d'/2}|\varepsilon\bI+\bL^T\bSigma\bL|^{-1/2}  
\exp\{-\smhalf(\bv-\bL^T\bmu)^{-1}(\varepsilon\bI+\bL^T\bSigma\bL)^{-1}
(\bv-\bL^T\bmu)\}\big].\\[1ex]
&&\quad =p_{\smallN_{d'}}(\bv;\coeffmat^T\bmu,\coeffmat^T\bSigma\coeffmat)
\end{eqnarray*}
and the result follows.

\rightline{\endproof}

\vskip10mm
From (\ref{eq:facToStoch}) we have
\begin{equation}
\mSUBbdeltaTObalpha\thickarrow\frac{\projN\Big[\mSUBbalphaTObdelta
{\displaystyle\int_{\realnos^{\dtheta}}}\delta(\balpha-\bA^T\btheta)
\mSUBthetaTObdelta\,d\btheta\big/Z\Big]}{\mSUBbalphaTObdelta}
\label{eq:MSGdeltaTOalpha}
\end{equation}
where $\dtheta$ is the dimension of $\btheta$. It follows from assumption 
(\ref{eq:conjugacyLinComb}) and (\ref{eq:stochToFac}) that
\begin{equation}
\mSUBbalphaTObdelta=\exp\left\{\left[\begin{array}{c}      
\balpha\\
\vecof(\balpha\balpha^T)
\end{array}\right]^T\etaSUBbalphaTObdelta\right\}
\end{equation}
and
\begin{eqnarray*}
\mSUBthetaTObdelta&=&\exp\left\{
\left[\begin{array}{c}
\btheta\\
\vecof(\btheta\btheta^T)
\end{array}
\right]^T\etaSUBthetaTObdelta
\right\}\\[1ex]
&\propto&(2\pi)^{-\dtheta/2}|\bSigma_{\odot}|^{-1/2}\exp\{
-\smhalf(\btheta-\bmu_{\odot})^T\bSigma_{\odot}^{-1}(\btheta-\bmu_{\odot})\}
\end{eqnarray*}
where 
$$\bmu_{\odot}\equiv -\smhalf\Big\{\vecof^{-1}\big(\etaSUBthetaTObdelta\big)_2\Big\}^{-1}
\Big(\etaSUBthetaTObdelta\Big)_1$$
and
$$\bSigma_{\odot}\equiv -\smhalf\Big\{\vecof^{-1}\big(\etaSUBthetaTObdelta\big)_2\Big\}^{-1}$$
are the common parameters matching $\etaSUBthetaTObdelta$. 
From Result 2 in Appendix \ref{sec:MultLinCombFragDeriv},
\noindent
\begin{eqnarray*}
&&\int_{\realnos^{\dtheta}}\delta(\balpha-\bA^T\btheta)
(2\pi)^{-\dtheta/2}|\bSigma_{\odot}|^{-1/2}\exp\{
-\smhalf(\btheta-\bmu_{\odot})^T\bSigma_{\odot}^{-1}(\btheta-\bmu_{\odot})\}\,d\btheta\\[1ex]
&&\qquad\qquad=|2\pi\,\bA^T\bSigma_{\odot}\bA|^{-1/2}
\exp\Big\{-\smhalf(\balpha-\bA^T\bmu_{\odot})^T(\bA^T\bSigma_{\odot}\bA)^{-1}
(\balpha-\bA^T\bmu_{\odot})\Big\}.
\end{eqnarray*}
Substitution into (\ref{eq:MSGdeltaTOalpha}) then leads to 
{\setlength{\arraycolsep}{1pt}
\begin{eqnarray*}
&&\mSUBbdeltaTObalpha=\\
&&\ \frac{\projN\left[
\exp\left\{\left[\begin{array}{c}      
\balpha\\
\vecof(\balpha\balpha^T)
\end{array}\right]\etaSUBbalphaTObdelta\right\}
\exp\left\{\left[\begin{array}{c}      
\balpha\\
\vecof(\balpha\balpha^T)
\end{array}
\right]^T
\left[\begin{array}{c}      
(\bA^T\bSigma_{\odot}\bA)^{-1}\bA^T\bmu_{\odot}\\[1ex]
-\smhalf\vecof\big((\bA^T\bSigma_{\odot}\bA)^{-1}\big)
\end{array}
\right]
\right\}\Bigg/Z\right]
}{\exp\left\{\left[\begin{array}{c}      
\balpha\\
\vecof(\balpha\balpha^T)
\end{array}\right]\etaSUBbalphaTObdelta\right\}}\\[1ex]
&=&\exp\left\{\left[\begin{array}{c}      
\balpha\\
\vecof(\balpha\balpha^T)
\end{array}
\right]^T
\left[\begin{array}{c}      
(\bA^T\bSigma_{\odot}\bA)^{-1}\bA^T\bmu_{\odot}\\[1ex]
-\smhalf\vecof\big((\bA^T\bSigma_{\odot}\bA)^{-1}\big)
\end{array}
\right]
\right\}.
\end{eqnarray*}
}
Therefore, setting $\bOmega\thickarrow-2\bSigma_{\odot}\bA$, we get 
$$\etaSUBbdeltaTObalpha\thickarrow\left[\begin{array}{c}      
(\bOmega^T\bA)^{-1}\bOmega^T\Big(\etaSUBthetaTObdelta\Big)_1\\[2ex]
\vecof\big((\bOmega^T\bA)^{-1}\big)
\end{array}
\right]
$$
which are the first two updates of Algorithm \ref{alg:LinCombFrag}.

For the third update, note that (\ref{eq:facToStoch}) gives
{\setlength{\arraycolsep}{1pt}
\begin{eqnarray*}
&&\mSUBbdeltaTOtheta\\[1ex]
&&\quad\thickarrow
\frac{\projN\Big[\mSUBthetaTObdelta
\displaystyle{\int_{\realnos^{d^{\balpha}}}\delta(\balpha-\bA^T\btheta)\mSUBbalphaTObdelta\,d\balpha\big/Z\Big]}}
{\mSUBthetaTObdelta}\\[1ex]
&&\quad=\frac{\projN\Big[\mSUBthetaTOdelta\,
\biggerm_{\mbox{\footnotesize$\balpha\to\delta(\balpha-\bA^T\btheta)$}}(\bA^T\btheta)/Z\Big]}{\mSUBthetaTOdelta}\\[1ex]
&&\quad=\frac{\projN\left[\exp\left\{
\left[\begin{array}{c}
\btheta\\
\vecof(\btheta\btheta^T)
\end{array}
\right]^T\etaSUBthetaTObdelta
\right\}
\exp\left\{
\left[
\begin{array}{c}
\bA^T\btheta\\
\vecof\big((\bA^T\btheta)(\bA^T\btheta)^T\big)
\end{array}
\right]^T
\etaSUBbalphaTObdelta
\right\}\Bigg/Z
\right]}
{\exp\left\{
\left[\begin{array}{c}
\btheta\\
\vecof(\btheta\btheta^T)
\end{array}
\right]^T\etaSUBthetaTObdelta
\right\}}\\[1ex]
&&\quad=
\exp\left\{
\left[
\begin{array}{c}
\btheta\\
\vecof(\btheta\btheta^T)
\end{array}
\right]^T
\left[
\begin{array}{c}
\bA\Big(\etaSUBalphaTOdelta\Big)_1\\
(\bA\otimes\bA)\Big(\etaSUBalphaTOdelta\Big)_2
\end{array}
\right]
\right\}.
\end{eqnarray*}
}
The last step uses the identity $\vecof(\bB\bC\bD)=(\bD^T\otimes\bB)\vecof(\bC)$
for any compatible matrices $\bB$, $\bC$ and $\bD$.
The third update follows immediately.

\subsection{Derivation of Algorithm \ref{alg:GaussFrag}}\label{sec:GaussLikFragDeriv}

The message from $p(y|\alpha,\sigma^2)$ to $\alpha$ is, from (\ref{eq:facToStoch}),
\begin{equation}
\mSUBpyalphasigsqTOalpha=\frac{\projN\big[\mSUBalphaTOpyalphasigsq\,\int_0^{\infty}p(y|\alpha,\sigma^2)
\,\mSUBsigsqTOpyalphasigsq\,d\sigma^2
\big]}{\mSUBalphaTOpyalphasigsq}.
\label{eq:mLIKtosigsq}
\end{equation}
It follows from (\ref{eq:stochToFac}) and (\ref{eq:conjugacyGaussLik}) that 
$$\mSUBsigsqTOpyalphasigsq=\exp\left\{          
\left[
\begin{array}{c}
\log(\sigma^2)\\
1/\sigma^2
\end{array}
\right]^T
\etaSUBsigsqTOpyalphasigsq
\right\}
=(\sigma^2)^{\eta^{\boxplus}_1}\exp(\eta^{\boxplus}_2/\sigma^2).
$$
where $\eta^{\boxplus}\equiv\etaSUBsigsqTOpyalphasigsq$.
Hence 
\setlength\arraycolsep{2pt}
{
\begin{eqnarray*}
\int_0^{\infty}p(y|\alpha,\sigma^2)
\,\mSUBsigsqTOpyalphasigsq\,d\sigma^2&=&(2\pi)^{-1/2}\int_0^{\infty}(\sigma^2)^{\eta^{\boxplus}_1-1/2}
\exp[\{\eta^{\boxplus}_2-\smhalf(y-\alpha)^2\}/\sigma^2)]\,d\sigma^2\\[1ex]
&\propto&\big\{\smhalf(y-\alpha)^2-\eta^{\boxplus}_2\big\}^{\eta^{\boxplus}_1+\smhalf}.
\end{eqnarray*}
}
Therefore, letting $\eta^{\StarOfDavid}\equiv\etaSUBalphaTOpyalphasigsq$, 
the numerator of the right-hand side of (\ref{eq:mLIKtosigsq}) is
\begin{eqnarray*}
&&\projN\left[\frac{\exp(\alpha\eta^{\StarOfDavid}_1 + \alpha^2\eta^{\StarOfDavid}_1)}
{\big\{(y-\alpha)^2-2\eta^{\boxplus}_2\big\}^{-\eta^{\boxplus}_1-\smhalf}}\right]\\[1ex]
&&\quad=\exp\left\{
\left[
\begin{array}{c}
\alpha\\[1ex]
\alpha^2
\end{array}
\right]^T
(\nabla\AN)^{-1}\left(
\left[
\begin{array}{c}
\frac{\infint\alpha\exp(\alpha\eta^{\StarOfDavid}_1 + \alpha^2\eta^{\StarOfDavid}_2)\big/
\big\{(y-\alpha)^2-2\eta^{\boxplus}_2\big\}^{-\eta^{\boxplus}_1-\smhalf}\,d\alpha}
{\infint\exp(\alpha\eta^{\StarOfDavid}_1 + \alpha^2\eta^{\StarOfDavid}_2)\big/
\big\{(y-\alpha)^2-2\eta^{\boxplus}_2\big\}^{-\eta^{\boxplus}_1-\smhalf}\,d\alpha}\\[5ex]
\frac{\infint\alpha^2\exp(\alpha\eta^{\StarOfDavid}_2 + \alpha^2\eta^{\StarOfDavid}_2)\big/
\big\{(y-\alpha)^2-2\eta^{\boxplus}_2\big\}^{-\eta^{\boxplus}_1-\smhalf}\,d\alpha}
{\infint\exp(\alpha\eta^{\StarOfDavid}_1 + \alpha^2\eta^{\StarOfDavid}_2)\big/
\big\{(y-\alpha)^2-2\eta^{\boxplus}_2\big\}^{-\eta^{\boxplus}_1-\smhalf}\,d\alpha}
\end{array}
\right]
\right)
\right\}\\[1ex]
&&\quad=\exp\left\{
\left[
\begin{array}{c}
\alpha\\[1ex]
\alpha^2
\end{array}
\right]^T
(\nabla A_N)^{-1}\left(
\left[
\begin{array}{c}
\frac{\Asc(1,\eta^{\StarOfDavid}_1,-\eta^{\StarOfDavid}_1,-2y,y^2-2\eta^{\boxplus}_2,-\eta^{\boxplus}_1-\smhalf)}
{\Asc(0,\eta^{\StarOfDavid}_1,-\eta^{\StarOfDavid}_2,-2y,y^2-2\eta^{\boxplus}_2,-\eta^{\boxplus}_1-\smhalf)}\\[2ex]
\frac{\Asc(2,\eta^{\StarOfDavid}_1,-\eta^{\StarOfDavid}_1,-2y,y^2-2\eta^{\boxplus}_2,-\eta^{\boxplus}_1-\smhalf)}
{\Asc(0,\eta^{\StarOfDavid}_1,-\eta^{\StarOfDavid}_2,-2y,y^2-2\eta^{\boxplus}_2,-\eta^{\boxplus}_1-\smhalf)}
\end{array}
\right]
\right)
\right\}.
\end{eqnarray*}
Hence 
$$\etaSUBpyalphasigsqTOalpha\thickarrow
(\nabla A_N)^{-1}\left(
\left[
\begin{array}{c}
\frac{\Asc(1,\eta^{\StarOfDavid}_1,-\eta^{\StarOfDavid}_1,-2y,y^2-2\eta^{\boxplus}_2,-\eta^{\boxplus}_1-\smhalf)}
{\Asc(0,\eta^{\StarOfDavid}_1,-\eta^{\StarOfDavid}_2,-2y,y^2-2\eta^{\boxplus}_2,-\eta^{\boxplus}_1-\smhalf)}\\[2ex]
\frac{\Asc(2,\eta^{\StarOfDavid}_1,-\eta^{\StarOfDavid}_1,-2y,y^2-2\eta^{\boxplus}_2,-\eta^{\boxplus}_1-\smhalf)}
{\Asc(0,\eta^{\StarOfDavid}_1,-\eta^{\StarOfDavid}_2,-2y,y^2-2\eta^{\boxplus}_2,-\eta^{\boxplus}_1-\smhalf)}
\end{array}
\right]
\right)-\etaSUBalphaTOpyalphasigsq
$$
and the first update in Algorithm \ref{alg:GaussFrag} follows from 
the definition of $G^{\mbox{\tiny N}}$ given in 
Section \ref{sec:explicitFns}.

The message $p(y|\alpha,\sigma^2)$ to $\sigma^2$ is
\begin{equation}
\mSUBpyalphasigsqTOsigsq=\frac{\projICS\big[\mSUBsigsqTOpyalphasigsq\,\int_{-\infty}^{\infty}p(y|\alpha,\sigma^2)
\,\mSUBalphaTOpyalphasigsq\,d\alpha
\big]}{\mSUBsigsqTOpyalphasigsq}.
\label{eq:mLIKtoalpha}
\end{equation}
It follows from (\ref{eq:stochToFac}) and (\ref{eq:conjugacyGaussLik}) that 
$$
\mSUBalphaTOpyalphasigsq
=\exp\left\{\left[\begin{array}{c}   
\alpha\\
\alpha^2
\end{array}
\right]^T\etaSUBalphaTOpyalphasigsq\right\}
=\exp\big(\alpha\,\eta_1^{\oplus}+\alpha^2\eta_2^{\oplus}\big)
$$
where $\bdeta^{\oplus}\equiv \etaSUBalphaTOpyalphasigsq$. The integral in (\ref{eq:mLIKtoalpha})
is
\begin{eqnarray*}
&&\infint(2\pi\sigma^2)^{-1/2}\exp\left\{\frac{-(y-\alpha)^2}{2\sigma^2}\right\}
\exp\big(\alpha\,\eta_1^{\oplus}+\alpha^2\eta_2^{\oplus}\big)\,d\alpha\\[1ex]
&&\qquad\qquad\propto\infint(2\pi\sigma^2)^{-1/2}\exp\left\{\frac{-(y-\alpha)^2}{2\sigma^2}\right\}
\{2\pi(\sigma^{\oplus})^2\}^{-1/2}\exp\left\{\frac{-(y-\mu^{\oplus})^2}{2(\sigma^{\oplus})^2}\right\}\,d\alpha
\end{eqnarray*}
where
$$\left[
\begin{array}{c}
\mu^{\oplus}\\[1ex]
(\sigma^{\oplus})^2
\end{array}
\right]\equiv
\left[
\begin{array}{c}
-\eta_1^{\oplus}/(2\eta_2^{\oplus})\\[1ex]
-1/(2\eta_2^{\oplus})
\end{array}
\right]
$$
is the common parameter vector corresponding to $\bdeta^{\oplus}$.
Using (A.2) of Wand and Jones (1993), the last-written integral is
$$\{2\pi(\sigma^2+(\sigma^{\oplus})^2\}^{-1/2}\exp\left\{\frac{-(y-\mu^{\oplus})^2}
{2\{\sigma^2+(\sigma^{\oplus})^2\}}\right\}.$$
Therefore, the function inside the $\projICS\big[\ \cdot\ ]$ in (\ref{eq:mLIKtoalpha}) is
$$
(\sigma^2)^{\eta^{\boxplus}_1}\exp(\eta^{\boxplus}_2/\sigma^2)
[2\pi\{\sigma^2-1/(2\eta_2^{\oplus})\}]^{-1/2}\exp\left[\frac{-\{y+\eta_1^{\oplus}/(2\eta_2^{\oplus})\}^2}
{2\{\sigma^2-1/(2\eta_2^{\oplus})\}}\right].
$$
Plugging into (\ref{eq:mLIKtoalpha}) we then have 
$$\mSUBpyalphasigsqTOsigsq=\exp\left\{
\left[
\begin{array}{c}
\log(\sigma^2)\\
1/\sigma^2
\end{array}
\right]^T\etaSUBpyalphasigsqTOsigsq
\right\}
$$
where
$$\etaSUBpyalphasigsqTOsigsq\thickarrow (\nabla\AICS)^{-1}
\left(
\left[
\begin{array}{c}
\frac{\int_0^{\infty}\log(\sigma^2)(\sigma^2)^{\eta^{\boxplus}_1}
\{\sigma^2-1/(2\eta_2^{\oplus})\}^{-1/2}\exp\left[\frac{\eta^{\boxplus}_2}{\sigma^2}-\frac{\{y+\eta_1^{\oplus}/(2\eta_2^{\oplus})\,\}^2}
{2\{\sigma^2-1/(2\eta_2^{\oplus})\}}\right]\,d\sigma^2}
{\int_0^{\infty}(\sigma^2)^{\eta^{\boxplus}_1}
\{\sigma^2-1/(2\eta_2^{\oplus})\}^{-1/2}\exp\left[\frac{\eta^{\boxplus}_2}{\sigma^2}-\frac{\{y+\eta_1^{\oplus}/(2\eta_2^{\oplus})\,\}^2}
{2\{\sigma^2-1/(2\eta_2^{\oplus})\}}\right]\,d\sigma^2}\\[4ex]
\frac{\int_0^{\infty}(1/\sigma^2)(\sigma^2)^{\eta^{\boxplus}_1}
\{\sigma^2-1/(2\eta_2^{\oplus})\}^{-1/2}\exp\left[\frac{\eta^{\boxplus}_2}{\sigma^2}-\frac{\{y+\eta_1^{\oplus}/(2\eta_2^{\oplus})\,\}^2}
{2\{\sigma^2-1/(2\eta_2^{\oplus})\}}\right]\,d\sigma^2}
{\int_0^{\infty}(\sigma^2)^{\eta^{\boxplus}_1}
\{\sigma^2-1/(2\eta_2^{\oplus})\}^{-1/2}\exp\left[\frac{\eta^{\boxplus}_2}{\sigma^2}-\frac{\{y+\eta_1^{\oplus}/(2\eta_2^{\oplus})\,\}^2}
{2\{\sigma^2-1/(2\eta_2^{\oplus})\}}\right]\,d\sigma^2}
\end{array}
\right]
\right)-\bdeta^{\boxplus}.
$$
The change of variable $\sigma^2=e^{-x}$ and some simple algebra then leads to 
$$\etaSUBpyalphasigsqTOsigsq\thickarrow (\nabla\AICS)^{-1}
\left(
\left[
\begin{array}{c}
\frac{-\Bsc(1,-\eta_1^{\boxplus},-\eta_2^{\boxplus},-2\eta_2^{\oplus}\{y+\eta_1^{\oplus}/(2\eta_2^{\oplus})\,\}^2,
-2\eta_2^{\oplus},\smhalf)}
{\Bsc(0,-\eta_1^{\boxplus},-\eta_2^{\boxplus},-2\eta_2^{\oplus}
\{y+\eta_1^{\oplus}/(2\eta_2^{\oplus})\,\}^2,-2\eta_2^{\oplus},\smhalf)}\\[2ex]
\frac{\Bsc(0,1-\eta_1^{\boxplus},-\eta_2^{\boxplus},-2\eta_2^{\oplus}
\{y+\eta_1^{\oplus}/(2\eta_2^{\oplus})\,\}^2,-2\eta_2^{\oplus},\smhalf)}
{\Bsc(0,-\eta_1^{\boxplus},-\eta_2^{\boxplus},-2\eta_2^{\oplus}
\{y+\eta_1^{\oplus}/(2\eta_2^{\oplus})\,\}^2,-2\eta_2^{\oplus},\smhalf)}
\end{array}
\right]
\right)-\etaSUBsigsqTOpyalphasigsq.
$$
The second update in Algorithm \ref{alg:GaussFrag} follows from 
the definition of $G^{\mbox{\tiny IG1}}$ given in Section \ref{sec:explicitFns}.

\subsection{Derivation of Algorithm \ref{alg:logisticFrag}}\label{sec:logisticFragDeriv}

The only factor to stochastic node message for the logistic fragment is,
from (\ref{eq:facToStoch}):
$$\mSUBpyalphaTOalpha=\frac{\proj[\mSUBalphaTOpyalpha\,p(y|\alpha)\big/Z]}{\mSUBalphaTOpyalpha}$$
with projection onto an appropriate exponential family.
Assumption (\ref{eq:logistConjugAss}) and conjugacy considerations implies 
projection onto the  Univariate Normal family.
Setting 
$$\bdeta^{\#}\equiv\etaSUBalphaTOpyalpha,$$
we then have
$$\mSUBpyalphaTOalpha=\frac{\projN[\exp(\eta^{\#}_1\alpha+\eta^{\#}_2\alpha^2)\,\exp\{y\alpha-\log(1+e^{\alpha})\}]}
{\exp(\eta^{\#}_1\alpha+\eta^{\#}_2\alpha^2)}.$$
The numerator is 
\begin{eqnarray*}
&&\projN\big[\exp\{(\eta^{\#}_1+y)\alpha+\eta^{\#}_2\alpha^2-\log(1+e^{\alpha})\}\big]\\[1ex]
&&\qquad\qquad\propto
\exp\left\{
\left[
\begin{array}{c}
\alpha\\
\alpha^2
\end{array}
\right]^T
(\nabla A_N)^{-1}\left(
\left[  
\begin{array}{c}
\frac{\infint\alpha\exp\{(\eta^{\#}_1+y)\alpha+\eta^{\#}_2\alpha^2-\log(1+e^{\alpha})\}\,d\alpha}
{\infint\exp\{(\eta^{\#}_1+y)\alpha+\eta^{\#}_2\alpha^2-\log(1+e^{\alpha})\}\,d\alpha}\\[2ex]
\frac{\infint\alpha^2\exp\{(\eta^{\#}_1+y)\alpha+\eta^{\#}_2\alpha^2-\log(1+e^{\alpha})\}\,d\alpha}
{\infint\exp\{(\eta^{\#}_1+y)\alpha+\eta^{\#}_2\alpha^2-\log(1+e^{\alpha})\}\,d\alpha}
\end{array}
\right]\right)
\right\}\\[2ex]
&&\qquad\qquad\propto
\exp\left\{
\left[
\begin{array}{c}
\alpha\\
\alpha^2
\end{array}
\right]^T
(\nabla A_N)^{-1}\left(
\left[  
\begin{array}{c}
\frac{\Csc(1,\eta^{\#}_1+y,\eta^{\#}_1)}
{\Csc(0,\eta^{\#}_1+y,\eta^{\#}_1)}\\[2ex]
\frac{\Csc(2,\eta^{\#}_1+y,\eta^{\#}_1)}
{\Csc(0,\eta^{\#}_1+y,\eta^{\#}_1)}
\end{array}
\right]\right)
\right\}.
\end{eqnarray*}
Hence
$$\etaSUBpyalphaTOalpha\thickarrow
(\nabla A_N)^{-1}\left(
\left[  
\begin{array}{c}
\frac{\Csc(1,\eta^{\#}_1+y,\eta^{\#}_1)}
{\Csc(0,\eta^{\#}_1+y,\eta^{\#}_1)}\\[2ex]
\frac{\Csc(2,\eta^{\#}_1+y,\eta^{\#}_1)}
{\Csc(0,\eta^{\#}_1+y,\eta^{\#}_1)}
\end{array}
\right]
\right)-\etaSUBalphaTOpyalpha
$$
and the update in Algorithm \ref{alg:logisticFrag} follows from 
the definition of $\Hlogistic$ given in Section \ref{sec:explicitFns}.

\subsection{Derivation of Algorithm \ref{alg:probitFrag}}\label{sec:probitFragDeriv}

Via arguments analogous to those given in Section \ref{sec:logisticFragDeriv}, the 
factor to stochastic node natural parameter update is 
$$\etaSUBpyalphaTOalpha\thickarrow
(\nabla A_N)^{-1}\left(
\left[  
\begin{array}{c}
\frac{\infint\alpha\Phi((2y-1)\alpha)\exp\{\eta^{\#}_1\alpha+\eta^{\#}_2\alpha^2\}\,d\alpha}
{\infint\Phi((2y-1)\alpha)\exp\{\eta^{\#}_1\alpha+\eta^{\#}_2\alpha^2\}\,d\alpha}\\[2ex]
\frac{\infint\alpha^2\Phi((2y-1)\alpha)\exp\{\eta^{\#}_1\alpha+\eta^{\#}_2\alpha^2\}\,d\alpha}
{\infint\Phi((2y-1)\alpha)\exp\{\eta^{\#}_1\alpha+\eta^{\#}_2\alpha^2\}\,d\alpha}
\end{array}
\right]\right)-\etaSUBalphaTOpyalpha
$$
where, as before, $\bdeta^{\#}\equiv\etaSUBalphaTOpyalpha$. The integral results
\begin{equation}
\setlength\arraycolsep{2pt}
{
\begin{array}{rcl}
{\displaystyle\infint}\Phi(a+bx)\phi(x)\,dx&=&
\Phi\left(\displaystyle{\frac{a}{\sqrt{b^2+1}}}\right),\\[3ex]
{\displaystyle\infint}x\,\Phi(a+bx)\phi(x)\,dx&=&\displaystyle{\frac{b}{\sqrt{b^2+1}}}\,
\phi\left(\displaystyle{\frac{a}{\sqrt{b^2+1}}}\right)\\[3ex]
\mbox{and}\quad
{\displaystyle\infint}x^2\,\Phi(a+bx)\phi(x)\,dx&=&
\Phi\left(\displaystyle{\frac{a}{\sqrt{b^2+1}}}\right)-
\displaystyle{\frac{ab^2}{\sqrt{(b^2+1)^3}}}\,
\phi\left(\displaystyle{\frac{a}{\sqrt{b^2+1}}}\right),
\end{array}
}
\label{eq:NormIntResults}
\end{equation}
and standard algebraic manipulations lead to 
$$\etaSUBpyalphaTOalpha\thickarrow\Hprobit\left(\etaSUBalphaTOpyalpha\right)$$
where $\Hprobit$ is defined in Section \ref{sec:explicitFns}.

\subsection{Derivation of Algorithm \ref{alg:PoissonFrag}}\label{sec:PoissonFragDeriv}

The Poisson likelihood fragment derivation is very similar to that given
in Section \ref{sec:logisticFragDeriv}. The only change is that 
$$\exp\{y\alpha-\log(1+e^{\alpha})\}\quad\mbox{is replaced by}\quad
\exp(y\alpha-e^{\alpha})$$
which leads to the $\HPoisson$ function appearing in the factor to stochastic
node update rather than the $\Hlogistic$ function.
Section \ref{sec:explicitFns} provides the definition of 
$\HPoisson$.


\end{document}